\documentclass[a4paper,11pt]{article}
\pdfoutput=1
\usepackage{jheppub}

\usepackage{amssymb,amsmath}
\usepackage[normalem]{ulem}
\usepackage[utf8x]{inputenc}
\usepackage{slashed}
\usepackage{graphicx}

\usepackage{multirow}
\usepackage{diagbox}

\usepackage{midpage}

\usepackage{csquotes} 
\usepackage{comment}
\usepackage{mathrsfs}
\usepackage{float}

\usepackage{listings} 

\usepackage{color} 

\def\lsim{\mathrel{\rlap{\lower4pt\hbox{\hskip1pt$\sim$}}
    \raise1pt\hbox{$<$}}}         
\def\gsim{\mathrel{\rlap{\lower4pt\hbox{\hskip1pt$\sim$}}
    \raise1pt\hbox{$>$}}}         

\newcommand{\greekii}{I\hspace{-.1em}I}

\numberwithin{equation}{section}

\preprint{
\begin{minipage}{5cm}
\small
\flushright
EPHOU-23-020\\KEK-TH-2577\\KYUSHU-HET-274
\end{minipage}}

\title{Flux Landscape with Enhanced Symmetry Not on $SL(2, \mathbb{Z})$ Elliptic Points}

\author{Keiya Ishiguro$^{1}$,} 
\author{Takafumi Kai$^{2}$,}  
\author{Tatsuo Kobayashi$^{3}$ and} 
\author{Hajime Otsuka$^{2}$} 
\affiliation{
$^1$Graduate University for Advanced Studies (Sokendai), 1-1 Oho, Tsukuba, Ibaraki 305-0801, Japan\\
$^2$Department of Physics, Kyushu University, 744 Motooka, Nishi-ku, Fukuoka 819-0395, Japan\\
$^3$Department of Physics, Hokkaido University, Sapporo 060-0810, Japan
}
\emailAdd{ishigu@post.kek.jp}
\emailAdd{kai.takafumi.606@s.kyushu-u.ac.jp}
\emailAdd{kobayashi@particle.sci.hokudai.ac.jp}
\emailAdd{otsuka.hajime@phys.kyushu-u.ac.jp}

\abstract{
We study structures of solutions for SUSY Minkowski F-term equations on two toroidal orientifolds with $h^{2, 1} = 1$. Following our previous study \cite{Ishiguro:2020tmo}, with fixed upper bounds of a flux D3-brane charge $N_{\rm flux}$, we obtain a whole Landscape and a distribution of degeneracies of physically-distinct solutions for each case. In contrast to our previous study, we consider a non-factorizable toroidal orientifold and its Landscape on which $SL(2, \mathbb{Z})$ is violated into a certain congruence subgroup, as it had been known in past studies. We find that it is not the entire duality group that a complex-structure modulus $U$ enjoys but its outer semi-direct product with a "scaling" outer automorphism group. The fundamental region is enlarged to include the $|U| < 1$ region. In addition, we find that high degeneracy is observed at an elliptic point, not of $SL(2, Z)$ but of the outer automorphism group. 
Furthermore, $\mathbb{Z}_2$-enhanced symmetry is realized on the elliptic point. The outer automorphism group is exceptional in the sense that it is consistent with a symplectic basis transformation of background three-cycles, as opposed to the outer automorphism group of $SL(2, \mathbb{Z})$. We also compare this result with Landscape of another factorizable toroidal orientifold.
}
\makeatletter
\gdef\@fpheader{}
\makeatother

\begin{document}

\maketitle

\section{Introduction}
\label{sec:intro}
Landscape of Type IIB superstring theory (for reviews, see \cite{Blumenhagen:2006ci, Grana:2005jc, McAllister:2023vgy}) has been investigated for decades based on theoretical and phenomenological interests, as one of the most controllable low energy effective theory originated from string theory.
In areas of near purely theoretical interest, the Swampland program (for reviews, see \cite{Palti:2019pca}) has now been intensively studied.
It originally aims to provide constraints on four-dimensional (4D) theories to ensure that they are consistent with string theory, and many relevant studies on some fundamental and mathematical properties of string compactifications and the Landscape, which is defined as a set of 4D theories consistent with string theory, have been conducted.
However, it is fair to say that some problems in the Swampland program were initially raised from phenomenological perspectives, and those two aspects are intrinsically inseparable.
One of the intersections where the two perspectives meet is the moduli stabilization. 

Moduli are degrees of freedom in deformations of background six-dimensional (6D) space which is typically taken to be toroidal orientifolds and Calabi-Yau (CY) orientifolds.
In string compactifications, they enter the bosonic part of low-energy effective theory as massless scalar fields (see, e.g., for the type IIB superstring theory \cite{Grimm:2004uq, Grimm:2005fa}).
Since such fields may mediate a fifth force at low energy, one has to stabilize them appropriately (for instance, a review is provided by \cite{Adelberger:2003zx}) to give vacuum expectation values (VEVs) and masses with a phenomenological spirit.
The type IIB flux compactification is one of the most plausible mechanisms to achieve it, which generates a scalar potential for moduli by turning on background three-form fluxes.
In appropriate flux compactifications, supersymmetry (SUSY) can remain after the stabilization.
However, stabilizing all moduli is already a hard problem. As it had been pointed out in long-standing studies, and recently in the Swampland program as the Tadpole Conjecture \cite{Bena:2020xrh}, it usually needs large flux tadpole charge $N_{\rm flux}$ to fix all complex-structure moduli. 
Such large flux tadpole charge is usually inconsistent with the tadpole cancellation condition, and thus full moduli stabilization is difficult at the theoretical level.
Furthermore, one has to introduce gauge and matter sectors in theory, including the Standard Model (SM) gauge symmetries and matters.
It has also been pointed out that it is not easy to manage both moduli stabilization and introducing the SM, since the tadpole cancellation condition relates both sectors \cite{Blumenhagen:2007sm}.\footnote{The correlation between the tadpole charge and the generation number of chiral matters is recently discussed in \cite{Ishiguro:2023jjc}.}
Unfortunately, these are difficulties in the context of string Landscape before we can reproduce the observed values such as masses and mixing angles of the SM matter fields.

In this situation, it is still interesting to analyze a typical but controllable Landscape to thoroughly clarify structures in the effective theory.
We again choose type IIB flux Landscape as an example of Landscape, which is defined as a set of flux vacua with at least several moduli stabilized.
Then, the structures in the flux Landscape that we should clarify include reminiscences of the tadpole cancellation condition and the number of flux vacua with possible dualities.
The latter, the number of flux vacua, is indeed important phenomenologically since it relates to the concept of (stringy) naturalness \cite{Douglas:2003um}. 
The naturalness should be considered due to the fact flux quanta can only be taken as free parameters at this stage.
If one would like to predict some phenomenology via flux Landscape, then it encounters immediately a problem why a certain, possibly phenomenological vacuum, is selected from among the vast Landscape.
Finiteness of Landscape seems to be a necessary condition for string phenomenology in this regard\footnote{The finiteness is often argued after specifying a compactification background, but it is true that we do not even know whether candidates of the backgrounds are finite or not.
For Calabi-Yau threefolds, there is evidence for the finiteness. See \cite{yau1993open, Jejjala:2022lxh} and references therein.} as long as we stick to perturbative theory.
When the finiteness is ensured, one can associate probabilities with each vacuum.
A phenomenological vacuum becomes more natural if it has a high probability in the Landscape.
In the context of string Landscape, there have been many studies on the statistics that count flux vacua to discuss the naturalness.
Considering only the Ramond-Ramond (R-R) and Neveu Schwarz-Neveu Schwarz (NS-NS) fluxes\footnote{Note that it was pointed out that Landscape may be infinite with the non-geometric $Q$-flux \cite{Shelton:2006fd, Betzler:2019kon}, while we will only consider the R-R and NS-NS flux in the following.}, the number of "physical" vacua with a certain bound on fluxes is shown to be finite by the density method which treats fluxes as continuous variables and is reliable in large flux region \cite{Ashok:2003gk}, and many concrete examples with no approximation can be found in \cite{Giryavets:2004zr, DeWolfe:2004ns,  Conlon:2004ds, Plauschinn:2023hjw, Kachru:2002he} for instance. Furthermore, the number of SUSY vacua is proved to be finite with the D3 brane charge of fluxes bounded \cite{Grimm:2020cda}.
However, a protocol to find the whole finite vacua and its distribution should be explored concretely on each background.

Dualities\footnote{It does not necessarily mean the action must be invariant under the dualities, but we use the term duality to represent a map that relates solutions of F-term equations throughout this paper. $GL(2, \mathbb{Z})$ transformation of the axio-dilaton and rescaling of the three-form fluxes are examples. } under which relevant equations in moduli stabilization are invariant and different sets of flux quanta and VEVs are identified, is essential to ensure the finiteness of Landscape.
While the type IIB flux compactification manifestly has the S-duality in the axio-dilaton, a duality may also exist in complex-structure moduli.
The $SL(2, \mathbb{Z})$ duality on tori is the most simple case. 
Indeed, $SL(2, \mathbb{Z})$ itself, its generalization, or its subgroups are known to be realized on some toroidal orientifolds and Calabi-Yau orientifolds.
Then, all one has to do is to count "physically-distinct" flux vacua, which are modded by the dualities, correctly.
However, there remain some vacua to which we have to pay more attention because of discrete symmetries on specific points, which is called enhanced symmetries \cite{DeWolfe:2004ns}.

Counting vacua must be done by taking into account such enhanced symmetries.
In the $SL(2, \mathbb{Z})$ case, there are two enhanced symmetries: $\mathbb{Z}_3$ and $\mathbb{Z}_2$.
On the other hand, in general backgrounds, we do not know what kind of enhanced symmetries may appear in their Landscape and how large fractions of the Landscape enjoy such symmetries.

Our aim in this paper is to analyze the Landscape of other toroidal orientifolds that are not $T^6/{\mathbb{Z}_2 \times \mathbb{Z}_2'}$.
Due to the symmetric structure of $T^6/{\mathbb{Z}_2 \times \mathbb{Z}_2'}$, the Landscape may also be special compared with that of other toroidal orientifolds.
Although flux compactifications on the toroidal orientifolds had already been studied a long time ago, we will obtain the whole flux solutions with fixed $N_{\rm flux}$ explicitly and clarify structures of the Landscape by focusing on contributions of the tadpole cancellation condition and that of dualities, especially enhanced symmetries.
In this paper, we study the $T^6/{\mathbb{Z}_{6 - \rm{\greekii}}}$ and $T^6/{\mathbb{Z}_{2} \times \mathbb{Z}_4}$ orientifolds with certain lattices \cite{Markushevich:1986za, Ibanez:1987pj, Katsuki:1989bf, Kobayashi:1991rp, Lust:2005dy, Lust:2006zg}.
They have $h^{2, 1} = 1$, thus they have the simplest non-trivial complex-structure modulus spaces at least among the toroidal orientifolds.
Nevertheless, the former one is non-factorizable while the latter one is factorizable.
Comparing two resultant distributions will illustrate new structures of the Landscape.
The finiteness of flux Landscape on them is shown to be easily established while the usual $SL(2, \mathbb{Z})$ duality is partially broken on the $T^6/{\mathbb{Z}_{6 - \rm{\greekii}}}$ orientifold.
Therefore clarifying the Landscape on those orientifolds is a good starting point to accomplish the goals mentioned above.
We will study the Landscape in a manner of \cite{Betzler:2019kon, Ishiguro:2020tmo}, which explores structures of the F-term equations while the tadpole cancellation condition is generically violated.
We simply ignore the condition and assume the existence of uplifting to F-theory or those of some other justifications.
Nevertheless, $N_{\rm flux}$ enters in moduli stabilization, in particular, it is related to VEVs of the axio-dilaton and complex-structure moduli.
Indeed, it will prevent moduli VEVs from being in arbitrary regions and determine the structures.
In future work, we intend to analyze various orientifolds in the same manner, but systematically. In particular, it seems that both the tadpole cancellation condition and the existence of dualities are well determined by background geometry.
This paper will be a first step in this direction.

Finally, we believe that our studies revealing explicit distributions of degeneracies in moduli space also have some implications on the framework of modular flavor symmetry (see \cite{Feruglio:2017spp, Kobayashi:2018vbk, Penedo:2018nmg, Criado:2018thu, Kobayashi:2018scp, Novichkov:2018ovf, Novichkov:2018nkm, deAnda:2018ecu, Okada:2018yrn, Novichkov:2018yse} for earlier works and recent reviews \cite{Kobayashi:2023zzc, Ding:2023htn}).
Indeed, on general orientifolds, $SL(2, \mathbb{Z})$ may not exist as itself, but some subgroups or generalizations occur.
On CY orientifolds or toroidal orientifolds with multiple bulk moduli, the generalizations to certain symplectic groups occur and modular flavor symmetry with such symmetries has been studied in \cite{Ishiguro:2020nuf, Ishiguro:2021ccl, Ding:2020zxw, Baur:2020yjl, Nilles:2021glx, Kikuchi:2023awe, Kikuchi:2023awm}, for instance.
Furthermore, if we choose an orientifold with $h^{2, 1} = 1$ i.e., single complex-structure modulus, the superpotential will be a linear function of the modulus (couplings with axio-dilaton also exist).
This may enable us to choose either of the $\mathbb{Z}_3$ or $\mathbb{Z}_2$ (elliptic) fixed points because the VEV will be expressed by a fraction with both numerator and denominator are linear combinations of (complexified) integer flux quanta.
Thus the VEV will take its value in certain quadratic field $\mathbb{Q}(\sqrt{n})$, and it implies that two elliptic points cannot be obtained in the same Landscape simultaneously.
In this paper, we will find that an elliptic point that is not of $SL(2, \mathbb{Z})$ has high degeneracy i.e., a high probability on the $\mathbb{Z}_{6 - \rm{\greekii}}$ orientifold. 
There $SL(2, \mathbb{Z})$ is broken to its congruence subgroup, and its outer semi-direct product is the whole group which is enjoyed by the complex-structure modulus. 
The favored elliptic point turns out to be the elliptic point of the outer automorphism transformation, and $\mathbb{Z}_2$ enhanced symmetry is realized at that point.
While we will not discuss modular flavor symmetry in the following, we expect that our results are related to it in some way. 

This paper is organized as follows: In section \ref{sec:fluxcompactifications}, we briefly review the geometry of $T^6/{\mathbb{Z}_{6 - \rm{\greekii}}}$ orientifold and its flux compactification.
Then, we focus on SUSY Minkowski solutions and analyze all dualities in the F-term equations. 
We will find that a symplectic basis is consistent with a certain subgroup of $SL(2, \mathbb{Z})$ and its outer semi-direct product.
It is also found that the finiteness of Landscape is ensured by the S-duality and the group which is enjoyed by complex-structure moduli. 
In section \ref{sec:landscapeanalysis}, we illustrate the whole solutions of the F-term equations, which are modded out by the duality above.
It will turn out that VEVs with high degeneracy are changed from the previous $T^6/{\mathbb{Z}_2 \times \mathbb{Z}'_2}$ case drastically.
The VEV with the highest probability will be related to the outer semi-direct product group, and a certain enhanced symmetry will appear.
In section \ref{sec:otherorientifolds}, Landscape on the factorizable $T^6/(\mathbb{Z}_2 \times \mathbb{Z}_4)$ orientifold will be discussed as another example with $h^{2, 1} = 1$.
In this case, it is shown that the usual $SL(2, \mathbb{Z})$ is realized in the complex-structure modulus space, thus the Landscape will be shown to differ from the $\mathbb{Z}_{6 - \rm{\greekii}}$ case in several aspects.
Finally, we summarize the results of this paper and show some directions which will be studied in our future work.
We will also describe some detailed procedures to obtain the whole Landscape and the whole dualities which are consistent with the symplectic basis in Appendices \ref{app:algorhithm} and \ref{app:generalreparametrization}.
\section{Flux compactifications on $T^6/\mathbb{Z}_{6-{\rm \greekii}}$}
\label{sec:fluxcompactifications}
In the following, we consider the theory which is compactified on the $T^6/\mathbb{Z}_{6-{\rm \greekii}}$ orbifold with ${\rm SU}(6) \times {\rm SU}(2)$ lattice and its orientifold theory.
We compactify the orientifold with nontrivial flux quanta and discuss the resultant flux Landscape.
\subsection{Geometry}
\label{sec:geometry}
Let us first describe the geometry of the orientifold. Although the geometry and needed data for flux compactifications had already been studied in \cite{Kobayashi:1991rp, Font:2004cy, Lust:2005dy}, we briefly review the results for later convenience.

The Hodge numbers are $(h^{1,1}, h^{2, 1}) = (25, 1)$, and there are no twisted contribution to $h^{2, 1}$.
On the $T^6$, there are 20 real three-forms which form a third-cohomology basis:
\begin{equation}
    \begin{alignedat}{5}
        \alpha_0 &= dx^1 \wedge dx^3 \wedge dx^5, \quad& \beta^0 &= dx^2 \wedge dx^4 \wedge dx^6, \\
        \alpha_1 &= dx^2 \wedge dx^3 \wedge dx^5, \quad& \beta^1 &= -dx^1 \wedge dx^4 \wedge dx^6, \\
        \alpha_2 &= dx^1 \wedge dx^4 \wedge dx^5, \quad& \beta^2 &= -dx^2 \wedge dx^3 \wedge dx^6, \\
        \alpha_3 &= dx^1 \wedge dx^3 \wedge dx^6, \quad& \beta^3 &= -dx^2 \wedge dx^4 \wedge dx^5, \\
        \\
        \gamma_1 &= dx^1 \wedge dx^2 \wedge dx^3, \quad& \delta^1 &= - dx^4 \wedge dx^5 \wedge dx^6, \\
        \gamma_2 &= dx^1 \wedge dx^2 \wedge dx^5, \quad& \delta^2 &= - dx^3 \wedge dx^4 \wedge dx^6, \\
        \gamma_3 &= dx^1 \wedge dx^3 \wedge dx^4, \quad& \delta^3 &= - dx^2 \wedge dx^5 \wedge dx^6, \\
        \gamma_4 &= dx^3 \wedge dx^4 \wedge dx^5, \quad& \delta^4 &= - dx^1 \wedge dx^2 \wedge dx^6, \\
        \gamma_5 &= dx^1 \wedge dx^5 \wedge dx^6, \quad& \delta^5 &= - dx^2 \wedge dx^3 \wedge dx^4, \\
        \gamma_6 &= dx^3 \wedge dx^5 \wedge dx^6, \quad& \delta^6 &= - dx^1 \wedge dx^2 \wedge dx^4.
    \end{alignedat}
    \label{eq:realbasis}
\end{equation}
Choosing the orientation by $\int_{T^6} dx^1 \wedge dx^2 \wedge dx^3 \wedge dx^4 
\wedge dx^5 \wedge dx^6 = -1$, the above basis satisfies 
\begin{align}
    \int_{T^6} {\alpha_i \wedge \beta^j} = \delta^i_j, \quad \int_{T^6} {\gamma_i \wedge \delta^j} = \delta^i_j.
\end{align}
The real coordinates are defined along lattice vectors which transform under the orbifold twist $\Gamma = \mathbb{Z}_6$. Not all the three-forms (\ref{eq:realbasis}) are invariant under $\Gamma$ by themselves. On the orbifold, the orbifold twist $\Gamma$ is defined through the Coxeter element $Q$ of the corresponding Lie algebra. For the $Z_{6 - {\rm \greekii}}$ case, the action of $Q$ on the $SU(6) \times SU(2)$ root lattice is defined as
\begin{equation}
    \begin{alignedat}{2}
     Q(e_i) &= e_{i + 1} \quad (i = 1, 2, 3, 4), \\
     Q(e_5) &= - e_1 - e_2 - e_3 - e_4 - e_5, \\
     Q(e_6) &= - e_6.
    \end{alignedat}
\label{eq:twistactiononlattice}
\end{equation}
If we define a matrix representation of $Q$ by $Q(e_i) \equiv Q_{ji} e_j$, the 1-forms $dx^i$ transform as $d\mathbf{x}' = Q d \mathbf{x}$. $Q^6 = 1$ and its explicit expression is
\begin{align}
    Q = \begin{pmatrix}
        0 & 0 & 0 & 0 & -1 & 0 \\
        1 & 0 & 0 & 0 & -1 & 0 \\
        0 & 1 & 0 & 0 & -1 & 0 \\
        0 & 0 & 1 & 0 & -1 & 0 \\
        0 & 0 & 0 & 1 & -1 & 0 \\
        0 & 0 & 0 & 0 &  0 & -1 \\
    \end{pmatrix}.
    \label{eq:matrixQ}
\end{align}
We can switch to the complex coordinates on which the twist is represented diagonally:
\begin{equation}
    \begin{alignedat}{2}
    dz^1 &= dx^1 + e^{-\frac{2 \pi i}{6}} dx^2 + e^{-\frac{2 \pi i}{3}} dx^3 - dx^4 + e^{\frac{2 \pi i}{3}} dx^5, \\
    dz^2 &= dx^1 + e^{-\frac{2 \pi i}{3}} dx^2 + e^{\frac{2 \pi i}{3}} dx^3 + dx^4 + e^{-\frac{2 \pi i}{3}} dx^5, \\
    dz^3 &= \frac{1}{2\sqrt{3}} \left[ \frac{1}{3} (dx^1 - dx^2 + dx^3 - dx^4 + dx^5) + U dx^6 \right].
    \end{alignedat}
    \label{eq:complexcoordinates}
\end{equation}
In this basis, the action of the twist can be represented as $z^i \rightarrow e^{2 \pi i v^i} z^i$ with $(v^1, v^2, v^3) = \left(-\frac{1}{6}, -\frac{2}{6}, \frac{3}{6}\right)$. $U$ cannot be fixed by the orbifolding, and it is a complex-structure modulus of $T^6/\mathbb{Z}_{6-{\rm \greekii}}$. Note that the orbifold gives rise to the non-factorizable tori. 
The holomorphic three-form $\Omega \equiv dz^1 \wedge dz^2 \wedge dz^3$ with $-i \int \Omega \wedge \bar{\Omega} = 2 {\rm Im}U$ on $T^6$.\footnote{Of course, the overall numerical factor of $\Omega$ does not matter.}

The orientifold action is one for type IIB theory with O3/O7, $\Omega' (-1)^{F_L} \sigma$ where $\Omega'$ is the worldsheet parity operator, $F_L$ is the left-moving space-time fermion number, and $\sigma: \Omega \rightarrow -\Omega$ is $\mathbb{Z}_2$ isometry of the internal space. We focus on the case $\sigma \colon (z^1, z^2, z^3) \rightarrow - (z^1, z^2, z^3)$ for the toroidal orientifold compactifications in this paper. 
Under the geometric action $\sigma$, the cohomology $H^{2,3}$ are split into their even/odd parts. In the orbifold, the bulk contributions to $h^{1,1}$ and $h^{2, 1}$ are even and odd respectively. We will denote the dimensions by $h^{1,1}_{{\rm untw}., +}, h^{2,1}_{{\rm untw.}, -}$, but omit (untw., $\pm$) if there is no risk of confusion.

\subsection{Three-form flux $G_3$ and the scalar potential}
\label{sec:G3flux}

The three-form fluxes in type IIB string theory appear in the effective Lagrangian as a linear combination $G_3$ of RR- and NSNS- fluxes $F_3, H_3$. It is defined as $G_3 \equiv F_3 - S H_3$ with $S \equiv C_0 + i e^{-\phi}$ being the axio-dilaton in our notation.
The K\"{a}hler potential $K$ and the Gukov-Vafa-Witten superpotential $W$ in units of $M_{\rm Pl}=1$
\begin{align}
    K &= - \log\left(-i \left( S - \bar{S} \right)\right) - \log\left(-i \int{\Omega \wedge \bar{\Omega}}\right) - 2 \log{\mathcal{V}}, \\
    W &= \int{G_3 \wedge \Omega}
\end{align}
form the no-scale type F-term scalar potential in type IIB supergravity
\begin{align}
    V = e^K \left(G^{I J} D_I W D_{\bar{J}} \bar{W} \right).
\end{align}
Here, $G^{IJ}$ is the inverse of the K\"{a}hler metric and $D_I W = K_I W + \partial_I W$ is the covariant derivative of $W$.  
In this study, we focus on the no-scale type scalar potential where the volume contribution $- 2 \log \mathcal{V}$ in the K\"{a}hler potential cancels out the supergravity effect $-3 |W|^2$. Then $I, J$ denotes the axio-dilaton and the complex-structure modulus.
\\
As a three-form, $G_3$ can be expanded in the cohomology basis $H^3_{-}$ if the three-form fluxes are turned on in the internal space \footnote{The total parity of the three-form fluxes is even because of the intrinsic parity under $\Omega' (-1)^{F_L}$. For the toroidal orientifold compactifications, the relevant parity can be found in \cite{Kachru:2002he}.}. First, let us assume that the superpotential takes the form of
\begin{align}
    W = A + B S + U \left[ C + D S \right]. \label{eq:WGeneral}
\end{align}
It is equivalent to define $A, B, C$ and $D$ by
\begin{align}
    A + C U \equiv \int{F_3 \wedge \Omega}, \quad - (B + D U) \equiv \int{H_3 \wedge \Omega}.
\end{align}
$C$ and $D$ themselves are respectively defined as the derivatives of $\int{F_3 \wedge \Omega}$ and $-\int{H_3 \wedge \Omega}$ with respect to $U$.
Then, we obtain the Hodge decomposition of $G_3$:
\begin{align}
    G_3 = \frac{i}{2 {\rm Im} U} [ &-(\overline{A} + \overline{B} S + \overline{U} (\overline{C} + \overline{D} S)) \omega_{A_0}  \nonumber \\
    & + (A + BS + U (C + D S)) \omega_{B_0} \nonumber \\ 
    & + (\overline{A} + \overline{B} S + U (\overline{C} + \overline{D} S)) \omega_{A_3} \nonumber \\
    &  - (A + B S + \overline{U} (C + D S)) \omega_{B_3}], \label{eq:G3HodgeDecomposition}
\end{align}
where $\omega_{A_0} = dz^1 \wedge dz^2 \wedge dz^3, \omega_{A_3} = dz^1 \wedge dz^2 \wedge d\overline{z}^3$, $\omega_{B_3} = \overline{\omega_{A_3}}$ and $\omega_{B_0} = \overline{\omega_{A_0}}$ form a complex basis. Here, the superpotential is evaluated on the ambient tori, but it changes the integral only by an overall factor.
On the other hand, the fluxes are quantized on the real basis in Eq. (\ref{eq:realbasis}). We define those integral flux quanta as
\begin{equation}
    \begin{alignedat}{5}
        \int_{T^6}{(F_3, H_3) \wedge \alpha_i} &\equiv (-b_i, -d_i), \quad    &  \int_{T^6}{(F_3, H_3) \wedge \beta^i} &\equiv (a^i, c^i),  \\
    \int_{T^6}{(F_3, H_3) \wedge \gamma_i} &\equiv (-f_i, -h_i), \quad    &  \int_{T^6}{(F_3, H_3) \wedge \delta^i} &\equiv (e^i, g^i).
    \end{alignedat}
    \label{eq:realfluxquanta}
\end{equation}
These definitions are equivalently described in terms of the integrals of the three-forms on the dual cycles.
The flux quantization conditions depend on whether there are twisted $h^{2, 1}$ contributions or not, as pointed out in \cite{Blumenhagen:2003vr}. For the present case, we consider that there are no contributions from the twisted sector. However, the fluxes that are measured on the underlying tori cannot be arbitrary integers since there are fractional cycles in the orientifold. We will turn back to this point shortly.

Since there is an identification under the twist, not every flux is independent. Indeed, there are four independent singlets consisting of three-forms (\ref{eq:realbasis}). If we denote an orbit\footnote{In the physics literature, the term "orbit" sometimes means (can be sum of) a collection of $gx$ with possible duplication (of length $|G|$). Those orbits are inherited cycles from underlying tori. As formulated in \cite{Blumenhagen:2002wn, Blumenhagen:2003vr}, we may define the fractional cycles on the orbifold as the "orbit"s divided by some integers. The existence of the fractional cycles can be detected if the intersection matrix of the "orbit"s is not unimodular.
In any case, the orbits defined are the shortest candidates of the fractional cycles and we can build the cycles on the orbifold as linear combinations of our orbits.} of $x \in \mathbb{X}$ under a group $G$ by $G(x) \equiv \{g x | g \in G \}$, those singlets are given by\footnote{We choose the relative sign such that the expansion of $G_3$ agrees with that in \cite{Lust:2005dy}.}
\begin{equation}
    \begin{alignedat}{2}
    \mathbf{1}_{1} &\equiv \sum \Gamma(\alpha_0), \\
    \mathbf{1}_{2} &\equiv \sum \Gamma(\alpha_1), \\
    \mathbf{1}_{3} &\equiv - \left( \sum \Gamma(\beta^1) + \sum \Gamma(\beta^0) \right), \\
    \mathbf{1}_{4} &\equiv \sum \Gamma(\beta^0),
    \end{alignedat}
    \label{eq:singletsasorbits}
\end{equation}
where the action of $\Gamma$ on the cohomology basis is induced by the action of $Q$ (\ref{eq:matrixQ}) and satisfy 
\begin{align}
    \int_{T^6} \mathbf{1}_{1} \wedge \mathbf{1}_{4} = 6, \quad \int_{T^6} \mathbf{1}_{2} \wedge \mathbf{1}_{3} = -6
    \label{eq:integralsofsinglets}
\end{align}
with vanishing other pairings. On the orbifold, the intersection matrix of the dual cycles becomes unimodular as the integrals are divided by six.
All three-forms belonging to the same singlet must induce the same flux quantum. Expanded in the singlet basis, $G_3$ is represented as
\begin{align}
    G_3 = \frac{1}{3} (a^0 - c^0 S) \mathbf{1}_{1} + [(a^1 - c^1 S) + \frac{1}{3} (a^0 - c^0 S)] \mathbf{1}_{2} - \frac{1}{2} (b_1 - d_1 S) \mathbf{1}_{3} + (b_0 -d_0 S) \mathbf{1}_{4}. \label{eq:G3expansion}
\end{align}
The coefficients come from the fact that the lengths of the four orbits are 
\begin{align}
    |\Gamma(\alpha_0)| = 6, |\Gamma(\alpha_1)| = 3, |\Gamma(\beta^0)| = 2, |\Gamma(\beta^1)| = 6
\end{align}
and the integrals (\ref{eq:integralsofsinglets}).
In terms of the real basis, the four singlets are
\begin{equation}
    \begin{alignedat}{2}
        \mathbf{1}_1 & = 3 \alpha_0 - \alpha_1 -\alpha_2 + \beta^3 - \delta^5 + \delta^6 + \gamma_1 - \gamma_2 - \gamma_3 + \gamma_4,\\
        \mathbf{1}_2 & = \alpha_1 + \alpha_2 + \beta^3 - \delta^6 - \gamma_2 - \gamma_3, \\
        \mathbf{1}_3 & = - 2\beta^1 - \beta^2 - \delta^1 + \delta^2 + 2\delta^3 + \delta^4, \\
        \mathbf{1}_4 & = - \alpha_3 + \beta^0 + \gamma_5 - \gamma_6,
    \end{alignedat}
\end{equation}
and related to the complex basis by
\begin{equation}
    \begin{pmatrix}
        \omega_{A_0} \\ \omega_{A_3} \\ \omega_{B_3} \\ \omega_{B_0}
    \end{pmatrix}
= \frac{1}{6}
    \begin{pmatrix}
        i & \sqrt{3} & \sqrt{3} U & -3 i U \\
        i & \sqrt{3} & \sqrt{3} \overline{U} & - 3 i \overline{U} \\
        -i & \sqrt{3} & \sqrt{3} U & 3 i U \\
        -i & \sqrt{3} & \sqrt{3} \overline{U} & 3 i \overline{U}
    \end{pmatrix}
    \begin{pmatrix}
        \mathbf{1}_1 \\ \mathbf{1}_2 \\ \mathbf{1}_3 \\ \mathbf{1}_4
    \end{pmatrix}.
\end{equation}
Then we obtain expressions of $A, B, C, D$ in terms of the real fluxes:
\begin{equation}
    \begin{alignedat}{5}
        A &= - i b_0 - \frac{\sqrt{3}}{2} b_1 , \quad & C &= - i a^0 - \left( \sqrt{3} a^1 + \frac{1}{\sqrt{3}} a^0 \right), \\
        B &= i d_0 + \frac{\sqrt{3}}{2} d_1 , \quad & D &= i c^0 + \left( \sqrt{3} c^1 + \frac{1}{\sqrt{3}} c^0 \right).
    \end{alignedat}
    \label{eq:ABCDinrealfluxes}
\end{equation}

The quantization condition of the fluxes should be determined on the orbifold. As mentioned below Eq. (\ref{eq:integralsofsinglets}), the Poincar\'{e} dual cycles of the singlets (\ref{eq:singletsasorbits}) span a symplectic basis on the orbifold. Hence they are the fundamental cycles on which the fluxes are quantized. 
Following \cite{Frey:2002hf, Kachru:2002he, Cascales:2003zp, Font:2004cy}, the flux quanta is assumed even on the orbifold as a sufficient condition to exclude exotic O3-planes in this paper.
Since we defined the flux quanta (\ref{eq:realfluxquanta}) on the ambient $T^6$ cycles, the quanta obtain additional factors. The additional factor corresponding to a flux quantum is $\frac{N}{|\Gamma(\xi)|}$ where $\Gamma = \mathbb{Z}_N$ and $\xi$ is an element of the cohomology whose a Poincar\'{e} dual is the cycle on which the flux quantum is quantized.
In the present case, the quantization condition is obtained as
\begin{align}
    (b_0, d_0) \in 2 \mathbb{Z},\quad (b_1, d_1) \in 4 \mathbb{Z},\quad (a_0, c_0) \in 6 \mathbb{Z},\quad  (a_1, c_1) \in 2 \mathbb{Z}.
    \label{eq:fluxquantizationcondition}
\end{align}

\subsection{Dualities in the F-term equations}
\label{sec:dualities}
In the following, we focus on classifying dualities in the F-term equations that relate two solutions of them.
We will choose by which dualities we mod out the Landscape or not.
\paragraph{Manifest Dualities} \mbox{}\\
As pointed out in \cite{Ashok:2003gk}, there exist dualities in the F-term equations $D_I W = 0$ or the effective action level which relate solutions of the F-term equations $D_I W = 0$ and make the Landscape finite via identifications under them. First of all, the overall sign of the fluxes does not affect the solutions at all:
\begin{align}
    F_3 \rightarrow -F_3, \quad H_3 \rightarrow - H_3.
\end{align}
We mod out the Landscape by this $\mathbb{Z}_2$ symmetry that does not affect the effective action, as in \cite{Shelton:2006fd, DeWolfe:2004ns}.
The fixed locus in the flux space corresponds to turning off all the fluxes, and all valid solutions are simply doubled by this symmetry.

In type IIB flux compactifications, there is another manifest duality in the effective action\footnote{Throughout this paper, we do not consider the non-geometric fluxes in compactifications. Details of self S-dual theory after the compactifications in the presence of such non-geometric fluxes are discussed in \cite{Aldazabal:2006up}.} with respect to the axio-dilaton and the three-form fluxes, which is called the S-duality. Indeed, the K\"{a}hler function $K + \log |W|^2$ is invariant under $SL(2, \mathbb{Z})_S$ actions on the axio-dilaton and the three-form fluxes:
\begin{align}
    S \rightarrow S' = \frac{a S + b}{c S + d}, \quad \begin{pmatrix}
        F_3 \\ H_3
    \end{pmatrix} \rightarrow
    \begin{pmatrix}
        F'_3 \\ H'_3
    \end{pmatrix} =
    \begin{pmatrix}
        a & b \\
        c & d
    \end{pmatrix}
    \begin{pmatrix}
        F_3 \\ H_3
    \end{pmatrix}
\end{align}
with integers $a,b,c,d$ satisfying $ad-bc=1$. In terms of $A, B, C$ and $D$ in Eq. (\ref{eq:WGeneral}), they are rewritten as
\begin{align}
    \begin{pmatrix}
        A' \\ B'
    \end{pmatrix}
    =
    \begin{pmatrix}
        a & -b\\
        -c & d
    \end{pmatrix}
    \begin{pmatrix}
        A \\ B
    \end{pmatrix}, \quad
    \begin{pmatrix}
        C' \\ D'
    \end{pmatrix}
    =
    \begin{pmatrix}
        a & -b\\
        -c & d
    \end{pmatrix}
    \begin{pmatrix}
        C \\ D
    \end{pmatrix}.
    \label{eq:fluxesABCDtransformation-sl2z-ad}
\end{align}
\paragraph{Broken $SL(2, \mathbb{Z})$ for $U$} \mbox{}\\
For the $SL(2, \mathbb{Z})$ in the complex-structure modulus space, we can also find that non-trivial transformations of the three-form fluxes keep the F-term equation covariant. The F-term equation with respect to $U$ is 
\begin{align}
    D_U W = - \frac{1}{U - \overline{U}} \left( A + B S + \overline{U} (C + D S)\right) = 0.
\end{align}
Then, under $SL(2, \mathbb{Z})_U$ actions on the complex-structure modulus
\begin{align}
    U \rightarrow U' = \frac{a U + b}{c U + d},
\end{align}
the F-term equation remains covariant if the three-form flux quanta transform as
\begin{align}
    \begin{pmatrix}
        A' \\ C'
    \end{pmatrix}
    = 
    \begin{pmatrix}
        a & -b\\
        -c & d
    \end{pmatrix}
    \begin{pmatrix}
        A \\ C
    \end{pmatrix}, \quad
    \begin{pmatrix}
        B' \\ D'
    \end{pmatrix}
    = 
    \begin{pmatrix}
        a & -b\\
        -c & d
    \end{pmatrix}
    \begin{pmatrix}
        B \\ D
    \end{pmatrix},
    \label{eq:fluxesABCDtransformation-sl2z-cs}
\end{align}
where we expand $W \rightarrow W' = A' + B' S + U' [C' + D' S]$.
Note that this transformation is a rule defined up to a common constant factor at this level. 
If one would not like to change the Lagrangian, the constant factor is determined through a correspondent K\"{a}hler transformation of $\Omega$.
Thus, more concretely, a transformation of the flux quanta should be determined by looking at that of a period vector.

Let us summarize the flux transformations in terms of the real flux quanta (\ref{eq:realfluxquanta}). For the two generators $S, T$ \footnote{Under the identification of the overall sign of the fluxes, $S$ forms $\mathbb{Z}_2$ by itself.} of $SL(2, \mathbb{Z})_{S}$,

\begin{itemize}
    \item $S = 
        \begin{pmatrix}
            0 & -1 \\
            1 & 0
        \end{pmatrix} \in SL(2, \mathbb{Z})_S
    $\\
    \begin{equation}
        \begin{alignedat}{5}
            a^0 &\rightarrow - c^0, \quad & b_0 &\rightarrow - d_0, \\
            a^1 &\rightarrow - c^1, \quad & b_1 &\rightarrow - d_1, \\
            c^0 &\rightarrow a^0, \quad & d_0 &\rightarrow b_0, \\
            c^1 &\rightarrow a^1, \quad & d_1 &\rightarrow b_1. \\
        \end{alignedat}
    \end{equation}
\end{itemize}

\begin{itemize}
    \item $T^q = 
        \begin{pmatrix}
            1 & q \\
            0 & 1
        \end{pmatrix} \in SL(2, \mathbb{Z})_S
    $\\
    \begin{equation}
        \begin{alignedat}{5}
            a^0 &\rightarrow a^0 + q c^0, \quad & b_0 &\rightarrow b_0 + q d_0, \\
            a^1 &\rightarrow a^1 + q c^1, \quad & b_1 &\rightarrow b_1 + q d_1, \\
            c^0 &\rightarrow c^0, \quad & d_0 &\rightarrow d_0, \\
            c^1 &\rightarrow c^1, \quad & d_1 &\rightarrow d_1. \\
        \end{alignedat}
    \end{equation}
\end{itemize}
For those of $SL(2, \mathbb{Z})_U$, 
\begin{itemize}
    \item $S = 
        \begin{pmatrix}
            0 & -1 \\
            1 & 0
        \end{pmatrix} \in SL(2, \mathbb{Z})_U
    $\\
    \begin{equation}
        \begin{alignedat}{5}
            a^0 &\rightarrow - b_0, \quad & b_0 &\rightarrow a^0, \\
            a^1 &\rightarrow \frac{b_0}{3} - \frac{b_1}{2}, \quad & b_1 &\rightarrow 2 a^1 + \frac{2}{3} a_0, \\
            c^0 &\rightarrow -d_0, \quad & d_0 &\rightarrow c_0, \\
            c^1 &\rightarrow \frac{d_0}{3} - \frac{d_1}{2}, \quad & d_1 &\rightarrow \frac{2}{3}c^0 + 2 c^1. \\
        \end{alignedat}
        \label{eq:fluxtrf-S}
    \end{equation}
\end{itemize}

\begin{itemize}
    \item $T^q = 
        \begin{pmatrix}
            1 & q \\
            0 & 1
        \end{pmatrix} \in SL(2, \mathbb{Z})_U
    $\\
    \begin{equation}
        \begin{alignedat}{5}
            a^0 &\rightarrow a^0, \quad & b_0 &\rightarrow b_0 - q a^0, \\
            a^1 &\rightarrow a^1, \quad & b_1 &\rightarrow b_1 - q \left(\frac{2}{3}a^0 + 2 a^1\right), \\
            c^0 &\rightarrow c^0, \quad & d_0 &\rightarrow d_0 - q c^0, \\
            c^1 &\rightarrow c^1, \quad & d_1 &\rightarrow d_1 - q \left( \frac{2}{3} c^0 + 2 c^1 \right).  \\
        \end{alignedat}
        \label{eq:fluxtrf-T}
    \end{equation}
\end{itemize}
Notice that, the non-trivial transformations of the fluxes under $SL(2, \mathbb{Z})_U$ cannot be realized for general flux quanta.
Indeed, we see the fractional coefficients in Eqs. (\ref{eq:fluxtrf-S}) and (\ref{eq:fluxtrf-T}). Especially, the coefficients of $S$-transformation violate the quantization condition (\ref{eq:fluxquantizationcondition}). Thus, we see the $SL(2, \mathbb{Z})_U$ duality does not exist for arbitrary fluxes.

Moreover, if we consider only fluxes on which the quantization condition is not violated under $SL(2, \mathbb{Z})$, there is no $SL(2, \mathbb{Z})$ but its certain subgroup survives.
It can be understood via the $SL(2, \mathbb{Z})_U$ action on the period vector of $\Omega$.
One can check that the period vector is transformed by $M \in Sp(b_3, \mathbb{Q})$ instead of $Sp(b_3, \mathbb{Z})$.
Then the duality is realized if the period vector of $G_3$ is transformed by $M^T$, but it implies the fractional coefficients in general.
We shall demonstrate the property of the period vector via the explicit expression. Taking into account the intersection matrix (\ref{eq:integralsofsinglets}), we build the period vector as
\begin{align}
\Pi = 
    \begin{pmatrix}
        \int_{A^0} \Omega \\
        \int_{A^1} \Omega \\
        \int_{B^0} \Omega \\
        \int_{B^1} \Omega 
    \end{pmatrix}
    = \begin{pmatrix}
        \int_{T^6/{\mathbb{Z}_{6-\rm{\greekii}}}} \Omega \wedge \mathbf{1}_4 \\
        \int_{T^6/{\mathbb{Z}_{6-\rm{\greekii}}}} \Omega \wedge (-\mathbf{1}_3)\\
        \int_{T^6/{\mathbb{Z}_{6-\rm{\greekii}}}} \Omega \wedge \mathbf{1}_1\\
        \int_{T^6/{\mathbb{Z}_{6-\rm{\greekii}}}} \Omega \wedge \mathbf{1}_2
    \end{pmatrix}
    = - \frac{1}{6}
    \begin{pmatrix}
        i \\
        \sqrt{3}\\
        3 i U\\
        \sqrt{3}U
    \end{pmatrix}.
\end{align}
The overall factor can be absorbed by the K\"{a}hler transformation. A K\"{a}hler transformation and $Sp(b_3, \mathbb{Z})$ lead us to the period vector $\Pi' = (1, ~ -\sqrt{3}iU, ~ 3U, ~ \sqrt{3}i)^T$ where the derivative of the prepotential is understood as usual. Then we consider the modular transformation of $U: U \rightarrow \frac{a U + b}{c U + d}$. It can also be understood as the $Sp(b_3, \mathbb{Q})$ transformation on $\Pi'$:
\begin{align}
    \Pi' \rightarrow (c U + d)^{-1} 
    \begin{pmatrix}
        d & 0 & \frac{c}{3} & 0 \\
        0 & a & 0 & -b \\
        3b & 0 & a & 0 \\
        0 & -c & 0 & d
    \end{pmatrix} \Pi' \equiv (c U + d)^{-1} M \Pi'.
    \label{eq:period-under-sl2z}
\end{align}
This implies that there is no basis transformation of $H_3(T^6/\mathbb{Z}_{6-{\rm \greekii}}, \mathbb{Z})$ which corresponds to the general $SL(2, \mathbb{Z})$ transformation. 
Since the geometrical dualities in the complex-structure moduli sector originate from the basis transformation, it is more natural to consider the effective theory has only the duality with $c \equiv 0 ~(\text{mod}~3)$ due to the orbifolding.
The restricted transformations of $U$ can be shown to form a congruence subgroup of the modular group $SL(2, \mathbb{Z})_U$. It is called the Hecke congruence subgroup\footnote{We can find generators and presentations of the corresponding (Fuchsian) subgroup $\overline{\Gamma}_0(N)$ of $PSL(2, \mathbb{Z})$ in \cite{lascurain2002some}. \texttt{GAP} \cite{GAP4} can also give generators of congruence subgroups. 
It is observed that the Hecke congruence subgroups  or similar groups appear in partition functions on $T^6/{\mathbb{Z}_N}$ with $N=4, 6-\rm{\greekii}$ in \cite{Mayr:1993mq}, while they consider a different lattice for $\mathbb{Z}_{6-\rm{\greekii}}$.} of level 3 and denoted by $\Gamma_0(3)$:
\begin{align}
    \Gamma_0(n) \equiv \left\{\begin{pmatrix}
        a & b \\
        c & d
    \end{pmatrix} \in SL(2, \mathbb{Z}) \middle| ~ c \equiv 0 \quad \text{mod}~n
    \right\}.
\end{align}
$\Gamma_0(3)$ can be generated by two generators. One of them can be chosen to be $T$, as the $T$-transformation which is not broken by the restriction on $c$. The fundamental region of $\Gamma_0(3)$ is also known and larger than that of the $SL(2, \mathbb{Z})$.
The two generators which we choose are
\begin{align}
    T = \begin{pmatrix}
        1 & 1\\
        0 & 1\\
    \end{pmatrix}, \quad
    S' = \begin{pmatrix}
        -1 & 1\\
        -3 & 2
    \end{pmatrix},
\end{align}
where $S'$ is at order of $6$: $(S')^3 = - \mathbf{1}_{2\times2}$ and $S' = S^{-1}T^{-3}S^{-1}T^{-1}$. The fixed point under $S'$ is $U = \frac{1}{2} + i \frac{\sqrt{3}}{6}$.
Substituting $c = 3 c'~(c \in 3 \mathbb{Z})$, the transformation matrix $M$ is turned out to be an element of $Sp(b_3, \mathbb{Z})$. 
Then the period vector of $G_3$ on the basis, $\mathbb{G}'_3$, should transforms as
\begin{align}
    \mathbb{G}'_3 \rightarrow M \mathbb{G}'_3 \label{eq:symplectictrf-G3}
\end{align}
as the $\Gamma_0(3)$ duality of the effective theory. The axio-dilaton $S$ is not transformed under the duality of the complex-structure modulus. The explicit transformation of the flux quanta is obtained via Eq. (\ref{eq:G3expansion});

For the $S'$-transformation of $\Gamma_0(3)_U$, 
\begin{itemize}
    \item $S' = 
        \begin{pmatrix}
            -1 & 1 \\
            -3 & 2
        \end{pmatrix} \in \Gamma_0(3)_U
    $\\
    \begin{equation}
        \begin{alignedat}{5}
            a^0 &\rightarrow 2 a^0 + 3 b_0, \quad & b_0 &\rightarrow -b_0 - a^0, \\
            a^1 &\rightarrow 2 a^1 - b_0 + \frac{3}{2}b_1, \quad & b_1 &\rightarrow - b_1 - \frac{2}{3} a_0 - 2 a_1, \\
            c^0 &\rightarrow 2 c^0 + 3 d_0 , \quad & d_0 &\rightarrow - d_0 - c^0, \\
            c^1 &\rightarrow 2 c^1 - d_0 + \frac{3}{2} d_1, \quad & d_1 &\rightarrow -d_1 - \frac{2}{3}c^0 - 2c^1, \\
        \end{alignedat}
        \label{eq:fluxtrf-Sprime}
    \end{equation}
\end{itemize}
and the $T$-transformation of the flux quanta is same as Eq. (\ref{eq:fluxtrf-T}). 

Since $S'^3 = -1$, fluxes quanta $(a^0, a^1, b_0, b_1, c^0, c^1, d_0, d_1)$ transforms as $\rho(S')^3 = -1$ which is trivial up to the overall sign reversion. We also see that the quantization condition (\ref{eq:fluxquantizationcondition}) is conserved under the transformation which implies the duality $\Gamma_0(3)$ exists for arbitrary fluxes.

It is ensured that, under the above symplectic flux transformations which are accompanied by $SL(2, \mathbb{Z})_{S}$ and $\Gamma_0(3)$, the D3 tadpole charge
\begin{align}
    N_{\rm flux} \equiv \int H_3 \wedge F_3 > 0 \label{eq:def-of-Nflux} 
\end{align}
is invariant. The positivity is ensured by the imaginary self-dual (ISD) condition required by SUSY solutions with $DW = 0$\footnote{$N_{\rm flux} = 0$ is not allowed for Minkowski solutions as we will explicitly confirm later.}. $N_{\rm flux}$ appears in the D3 tadpole cancellation condition which generalizes the Gauss theorem in electromagnetism.
The integral is equal to
\begin{align}
    N_{\rm flux}&=\left(\int{\Omega \wedge \overline{\Omega}}\right)^{-1}\left[ \left(\overline{\int{F_3 \wedge \Omega}} \right) \left(\int{H_3 \wedge \Omega}\right) - (\partial_U \partial_{\overline{U}}K)^{-1} \left(D_{\overline{U}}\overline{\int{F_3 \wedge \Omega}}\right) \left(D_{U}\int{H_3 \wedge \Omega}\right) \right]\nonumber\\
                &= - 2 {\rm Re} (A \overline{D}) + 2 {\rm Re} (B \overline{C})\\
                &= 2 b_0 c^0 + b_1 c^0 + 3 b_1 c^1 -2 a^0 d_0 -a^0 d_1 - 3 a^1 d_1 \in 24 \mathbb{Z},
\end{align}
where $A, B, C$ and $D$ are defined in Eq. (\ref{eq:WGeneral}). Note that in the second line, the expression of $N_{\rm flux}$ is calculated with an assumption $\int{\Omega \wedge \overline{\Omega}} = U - \overline{U}$.
\paragraph{"Scaling" duality $S_{(3)} \equiv U \rightarrow - \frac{1}{3U}$}\mbox{}\\
There is another "scaling"\footnote{
We call it the "scaling" group since its isometric circle is centered at the origin.
Such outer automorphism is exceptional in the sense that an outer automorphism group $\notin GL(2, \mathbb{Z})$ but $\in SL(2, \mathbb{R})$ does not exist for the usual $SL(2, \mathbb{Z})$.} duality, which is denoted by $S_{(3)}$. It does not change $N_{\rm flux}$ with an appropriate transformation of ${\mathbb{G}}_3$. Note that, although the matrix (\ref{eq:period-under-sl2z}) is derived straightforwardly, there is another way to construct a symplectic matrix for specific $a, b, c, d$. Indeed, for $U \rightarrow -\frac{1}{3U}$, the matrix (\ref{eq:period-under-sl2z}) cannot be a symplectic matrix with integer coefficients. However, if we enter appropriate parameters in different positions, we can construct a symplectic transformation (up to a K\"{a}hler transformation):
\begin{align}
    \Pi \rightarrow (c U + d)^{-1} (i \sqrt{3}) 
    \begin{pmatrix}
    0 & 1 & 0 & 0 \\
    1 & 0 & 0 & 0 \\
    0 & 0 & 0 & 1\\
    0 & 0 & 1 & 0
    \end{pmatrix}
    \Pi'.
    \label{eq:periodtrfscaling}
\end{align}
Thus, $U \rightarrow - \frac{1}{3U} \in SL(2, \mathbb{R}) \notin SL(2, \mathbb{Z})$ 
\footnote{It is an element of $SL(2, \mathbb{R})$ with $a, d = 0$ and $b=-\frac{1}{\sqrt{3}}, c = \sqrt{3}$. 
As a linear fractional transformation on $U$, it is in the class of elliptic transformations, where $|a + d| < 2$ and $a, d \in \mathbb{R}$. 
It is isomorphic to $\mathbb{Z}_2$, as suggested by $a + d = 0$. 
Its two finite fixed points, which are called elliptic points, are $U = \pm \frac{1}{\sqrt{3}}i$.} 
can also be understood as a basis transformation of the symplectic basis or the third homology. 
In general, $SL(2, \mathbb{R})$ is not consistent with the integral symplectic structure $Sp(4, \mathbb{Z})$, but if $a, b, c, d$ can be multiplied by some factor to be integers, there is a possibility to become consistent, as this duality. 
Then, we should mod out the $\Gamma_0(3)$ Landscape by this duality. This duality can be considered as a reminiscence of the $\mathbb{Z}_{6-\rm{\greekii}}$ lattice. 
We can also show that the modular duality $\Gamma_0(3)$ and the scaling duality $S_{(3)}$ are the most general dualities which can be understood as basis transformations, as we check in Appendix \ref{app:generalreparametrization}. 
The resultant fundamental region is narrower than that of $\Gamma_0(3)$. 
We will show it in Appendix \ref{app:algorhithm}. 
Then, as a group of transformations, the complex-structure modulus $U$ enjoys the outer semi-direct product
\begin{align}
    \overline{\Gamma}_0(3) \rtimes_{\varphi(S_{(3)})} \mathbb{Z}_2,
\end{align}
where $\varphi$ is conjugation of $\overline{\Gamma}_0(3)$ by $S_{(3)}$.
Under the $S_{(3)}$, the three-form fluxes transform as
\begin{itemize}
    \item $S_{(3)} = 
        \begin{pmatrix}
            0 & -\frac{1}{\sqrt{3}} \\
            \sqrt{3} & 0
        \end{pmatrix} \in \Gamma_0(3)_U
    $\\
    \begin{equation}
        \begin{alignedat}{5}
            a^0 &\rightarrow -\frac{3}{2}b_1, \quad & b_0 &\rightarrow a^1 + \frac{1}{3}a^0, \\
            a^1 &\rightarrow b_0 + \frac{1}{2}b_1, \quad & b_1 &\rightarrow - \frac{2}{3} a_0, \\
            c^0 &\rightarrow -\frac{3}{2}d_1 , \quad & d_0 &\rightarrow c^1 + \frac{1}{3}c^0, \\
            c^1 &\rightarrow d_0 + \frac{1}{2}d_1, \quad & d_1 &\rightarrow - \frac{2}{3}c^0. \\
        \end{alignedat}
        \label{eq:fluxtrf-Sthree}
    \end{equation}
\end{itemize}
It is consistent with the quantization condition (\ref{eq:fluxquantizationcondition}).
\paragraph{Other dualities and symmetries}\mbox{}\\
In addition, there may exist several dualities and symmetries which defines relations between the solutions of the F-term equations, while we do not mod out Landscape further by them;
\begin{itemize}
    \item Orientifold action\\
    Let us briefly comment on the orientifold action $\Omega' (-1)^{F_L} \sigma$. Since $\Omega, F_3$ and $H_3$ transform as $\Omega \rightarrow - \Omega, F_3 \rightarrow F_3$ and $H_3 \rightarrow H_3$ under the total parity, the superpotential $W$ will transform as $W \rightarrow -W$ (in terms of period, $\Pi$ is invariant, while $\mathbb{G}_3$ is odd). This can be absorbed by the overall sign-flipping of the fluxes and does not affect the stabilizations at all.
    \item "pseudo-S transformation" on $U$\\
    As mentioned, under the $SL(2, \mathbb{Z})$ on $U$, the period vector $\Pi$ is transformed by $M \in Sp(4, \mathbb{Q})$ (\ref{eq:period-under-sl2z}). Then the fluxes also must be transformed by $M$, and it leads to a contradiction with the quantization condition in general. However, if the flux quanta are multiples of some integers, one may satisfy the quantization condition even after the fractional transformation, it seems that $S$-transformation which is originated from $SL(2, \mathbb{Z})$ exists. However, it still cannot be understood in terms of the basis transformation of the third homology. Thus, we do not mod out by this duality.
    \item Rescaling of the fluxes\\
    Since $W$ depends on the fluxes linearly, $(F_3, H_3) \rightarrow k (F_3, H_3)$ with $k \in \mathbb{Z}$ does not change the F-term equations. Hence, the solutions with a small $N_{\rm flux}$ are encoded in the solution space with a larger charge, $k^2 N_{\rm flux}$. This duality can be used to find the whole Landscape of $\mathbb{Z}_{6-\rm{\greekii}}$ in Appendix \ref{app:algorhithm}.
    However, physics changes as the effective action is not invariant as $N_{\rm flux} \rightarrow k^2 N_{\rm flux}$, and it cannot be understood via the basis transformation of the third homology. Thus, we do not mod out by this duality.
    \item Different transformations for each of $F_3$ and $H_3$\\
    We implicitly assume $F_3$ and $H_3$ transform with the same $M$ in Eq (\ref{eq:symplectictrf-G3}). 
    In principle, they can be transformed differently and even linearly combined into the same $M$. 
    Then, the superpotential is invariant.
    However, this transformation cannot be achieved by symplectic transformations of the basis, since under them quantities $\int{X_3} \wedge (\alpha_I, \beta^J)$ transform uniformly for arbitrary $X_3$.
    Thus, we do not mod out the Landscape by this duality.
    \item Any other transformations which cannot be interpreted as symplectic matrices\\
    As mentioned, we mod out the Landscape only by dualities which correspond to symplectic transformations of the basis. 
    There is an explicit example of such transformations under which $W$ is invariant and $N_{\rm flux}$ is also invariant. Such transformations exist if there are some symmetries in a period vector itself. We will find it in the later section for $T^6/(\mathbb{Z}_2\times\mathbb{Z}_4)$.
    \item $GL(2, \mathbb{Z})_U$ and its congruence subgroups\\
    The Landscape may be symmetrical under $U \rightarrow - \bar{U}$. Indeed, the fundamental region of $SL(2, Z)$ becomes its half if we consider $GL(2, \mathbb{Z}) \simeq SL(2, Z) \rtimes \mathbb{Z}_2$ transformation. For the congruence subgroups, we can define the corresponding groups by the semi-direct product. In any cases, there is an additional generator in $GL(2, \mathbb{Z})$:
    \begin{align}
        R = 
        \begin{pmatrix}
        1 & 0 \\
        0 & -1
        \end{pmatrix}.
    \end{align}
    $R$ forms the $\mathbb{Z}_2$ group. The period vector transforms under the $R$ with $M$ satisfying $M^T \Sigma M = - \Sigma = \Sigma^T$, and this $\mathbb{Z}_2 \ntriangleleft Sp(b_3 = 2h^{2, 1} + 2, \mathbb{Z})$. Then, we see that the group of $M$ becomes \cite{Ishiguro:2020nuf,Ishiguro:2021ccl}
    \begin{align}
        GSp(2h^{2, 1} + 2, \mathbb{Z}) = Sp(2h^{2, 1} + 2, \mathbb{Z}) \rtimes \mathbb{Z}_2.
    \end{align}
    Under the $GSp$, we can make the fluxes transform as well and $DW = 0$ be invariant. We again do not mod out the Landscape by this duality because it cannot be understood in the usual transformation of the third homology. It will be related to (generalized) $CP$-transformation \cite{Strominger:1985it, DeWolfe:2004ns, Nilles:2018wex, Novichkov:2019sqv} due to the fact that $GL(2, \mathbb{Z})$ can be introduced as the orientation reversing of the torus with $U$. Since we would like to consider several orbifold groups and lattices, we postpone studying the $CP$ for future work. 
\end{itemize}

\subsection{Moduli stabilization}
\label{sec:modulistabilization}
Due to the simple structure of $T^6/\mathbb{Z}_{6- \rm{\greekii}}$, the F-term equations can be analytically solved. We also require the F-terms of possible K\"{a}hler moduli vanish i.e., $D_{T_A} W = 0$ where $T_A$ denotes the K\"{a}hler moduli.\footnote{The stabilization of K\"{a}hler moduli will break the modular and its residual symmetries as discussed in Ref. \cite{Ishiguro:2022pde}, but we leave the detailed study for future work.} Since the superpotential does not depend on $T_A$, it simply boils down to $W = 0$. Hence we have to solve
\begin{equation}
    \left\{
    \begin{alignedat}{3}
        D_U W &= \partial_U W + K_U W &= 0\\
        D_S W & = \partial_S W + K_S W &= 0\\
        W &= 0 
    \end{alignedat}
    \right.
    \quad \Rightarrow \quad
    \left\{
    \begin{alignedat}{2}
        C + D S &= 0\\
        B + D U &= 0\\
        A + C U &= 0
    \end{alignedat}\right.
    , \label{eq:ftermeq}
\end{equation}
which is an overdetermined system. We will call the set of (fluxes, moduli VEVs) whose elements satisfy
\begin{align}
    \langle S \rangle = - \frac{C}{D}, \quad \langle U \rangle = - \frac{B}{D}, \quad A D - B C = 0
    \label{eq:ftermsolution}
\end{align}
the Landscape, where $\langle X \rangle$ denotes a VEV of $X$\footnote{We often omit this symbol in the following.}. Note that $-\bar{S}, -\bar{U}$ is also the solution. As pointed out in \cite{Ishiguro:2021csu}, the tadpole charge $N_{\rm flux}$ and VEV of the imaginary part of $S$ are related regardless of details of complex-structure moduli space. In the present case,
\begin{align}
    N_{\rm flux} = 4 {\rm Im} S {\rm Im} U |D|^2.
    \label{eq:NfluxandVEVs}
\end{align}
It implies that two VEVs ${\rm Im}S$ and ${\rm Im} U$ are correlated in the Landscape as the correlation can be observed in \cite{Betzler:2019kon, Ishiguro:2020tmo}. We also observe that rescaling of fluxes $(F_3, H_3) \rightarrow k (F_3, H_3)$ does not change the moduli VEVs.

\subsection{Identifications under the dualities}
\label{sec:identification}
Since there is no guiding principle or a mechanism that fixes specific flux quanta except for the tadpole cancellation condition, it can be assumed that the dualities in Section \ref{sec:dualities} which map different solutions of the F-term equations define physically-equivalent relations between those solutions. Namely,
\begin{align}
     ^\exists {\cal A} \in PSL(2, \mathbb{Z})_{S} \times (\overline{\Gamma}_0(3)_{U} \rtimes_{\varphi(S_{(3})} \mathbb{Z}_2) \text{ s.t. }  ({\rm fluxes', VEVs'}) = ({\cal A} ({\rm fluxes}), {\cal A} ({\rm VEVs})) \nonumber \\  \Leftrightarrow ({\rm fluxes, VEVs}) \sim ({\rm fluxes', VEVs'}),
     \label{eq:identificationmanner}
\end{align}
with denoting the non-trivial transformations of the fluxes by $\mathcal{A}$(fluxes). Note that the overall sign of the fluxes is also modded out via the definition.
This identification is consistent with the tadpole cancellation condition because the dualities do not change the value of $N_{\rm flux}$.
As a result, we can obtain the finite set of the whole physically-distinct vacua in the Landscape for the fixed value of $N_{\rm flux}$. Some technical details will be summarized in Appendix \ref{app:algorhithm}. The finiteness enables us to define probabilities that are associated with moduli VEVs.

\subsection{Tadpole cancellation condition}
\label{sec:tcc}
Here, we briefly comment that the tadpole cancellation condition which is a set of consistency conditions constraints on the three-form fluxes. Indeed, the three-form fluxes cannot be taken arbitrarily since they carry D3-brane charges (\ref{eq:def-of-Nflux}); they must satisfy restrictions which can be considered as a generalization of the Gauss law\footnote{Note that, for a fully-consistent compactification, the differential form of the tadpole cancellation condition i.e., Bianchi identity, must be satisfied.}.  In general, the tadpole cancellation condition depends on which sources are assumed in the background. The sources are classified into two general types: one is the NS-NS sources and the other is R-R sources. In this paper, we assume there are no NS-NS sources which may include NS5-branes and $5^2_2$-branes, and consider only the R-R sources which couple to the various R-R potentials. D$p$-branes and O$p$-planes are the candidates of the R-R sources.
We further assume that there is no non-geometric $Q$-flux, and consider only bulk (untwisted) cycles and correspondent cohomology which are odd under the orientifold action.
Thus, we do not consider geometric two-form flux nor non-geometric $R$-flux, which are introduced along with the $Q$-flux by considering T-duality of the NS-NS three-form flux $H_3$.

Then, the D3 tadpole cancellation condition which the three-form fluxes are explicitly constrained boils down to an integral of 
\begin{align}
    H_3 \wedge F_3 = 2\left( - N_{\rm D3} + \frac{N_{\rm O3}}{4}\right) \omega_{A_0}, \label{eq:D3tcc}
\end{align}
at the orbifold point\footnote{We ignored possible contributions from D7-branes. For smooth CY threefolds, see \cite{Plauschinn:2008yd}.}.
Here, $N_{\rm O3}$ stands for the number of the O3-planes while $N_{D3}$ stands for that of D3-branes. Since we do not consider stabilization mechanisms for open string moduli, the most reliable system here is a system with no D3-brane. 
However, it is very difficult to satisfy (\ref{eq:D3tcc}) with only the three-form fluxes, and indeed it is found impossible for some toroidal orientifolds. Nevertheless, we postpone such open moduli stabilization for future work. 
Under the type IIB orientifold action, $N_{\rm O3}$ on $T^6$ is fixed to be 64. Thus, the maximum value of $N_{\rm flux}$ (measured on the underlying tori) must be less than 32. Several O3-planes will further be identified with each other under orbifold groups, and the typical constraint will be $N_{\rm flux} \lesssim {\cal O}(10)$. The exact values were calculated for some orientifolds. Indeed, it restricts the three-form fluxes in this paper severely. Nevertheless, we simply ignore the tadpole cancellation condition to illustrate a space of the solutions.
A possible loophole to avoid the severe constraint is the F-theory uplifting.
Although there is an explicit uplifting for $T^6/{\mathbb{Z}_2 \times \mathbb{Z}'_2}$ \cite{Denef:2005mm}, it is not easy to examine that possibility on other toroidal orientifolds.
If the uplifting to F-theory is possible, there emerges a relationship between the Euler number of a CY four-fold and the contribution to the right-hand side of Eq. (\ref{eq:D3tcc}). The largest known value of the Euler number leads to a bound $N_{\rm flux} \lesssim {\cal O}(10^5)$ \cite{Candelas:1997eh, Taylor:2015xtz}, where the bound is significantly relaxed.

\section{Distributions of the moduli VEVs and degeneracies in the finite Landscape on the $T^6/{Z_{6- \rm{\greekii}}}$ orientifold}
\label{sec:landscapeanalysis}
In this section, we show the distribution of the moduli VEVs explicitly. 
The degeneracies of physically-distinct solutions are also summarized.
We will limit our discussion to the range $N_{\rm flux} \leq 24 \times 100$. 
Due to the $S_{(3)}$ duality, we can restrict ${\rm Im} U$ to be in ${\rm Im}U \geq \frac{1}{2\sqrt{3}}$, and it enables us to explore the F-term equations with large $N_{\rm flux}$ easily. 
Of course, the tadpole cancellation condition restricts the maximum value of $N_{\rm flux}$, $N_{\rm flux}^{\rm max}$, severely, and the large $N_{\rm flux}$ is not physically viable. 
For details, please see Appendix \ref{app:algorhithm}.
We indeed find whole physically-distinct SUSY Minkowski solutions on the $\mathbb{Z}_{6 - {\rm \greekii}}$ orientifold with fixed values of $N_{\rm flux}^{\rm max}$. 
As mentioned, the duality with regard to the complex-structure modulus $U$ is $\overline{\Gamma}_0(3) \rtimes_{S(3)} \mathbb{Z}_2$, and the fundamental region is widened. Concerning the context of modular flavor symmetry, we first summarize our result which is projected on the $U$-plane.

We begin with $N_{\rm flux}^{\rm max} = 24 \times 20$ case. The number of solutions is 229. On the $U$-plane and $S$-plane, as illustrated in Fig. \ref{fig:vevU_G03}, the VEVs scatter all over the fundamental region. Note that one can show that there is no solution on $U = i$ even with arbitrary $N_{\rm flux}^{\rm max}$. $U = \omega$ is realizable, and ${\rm Im} U \gg 1$ on which $q = e^{2 \pi i U}$ becomes small is also found.

\subsection{Distributions of solutions projected on the $U$-plane}
\label{sec:landscapeanalysis-on-U}

\begin{figure}[H]
\centering
\includegraphics[width = 0.8 \linewidth]{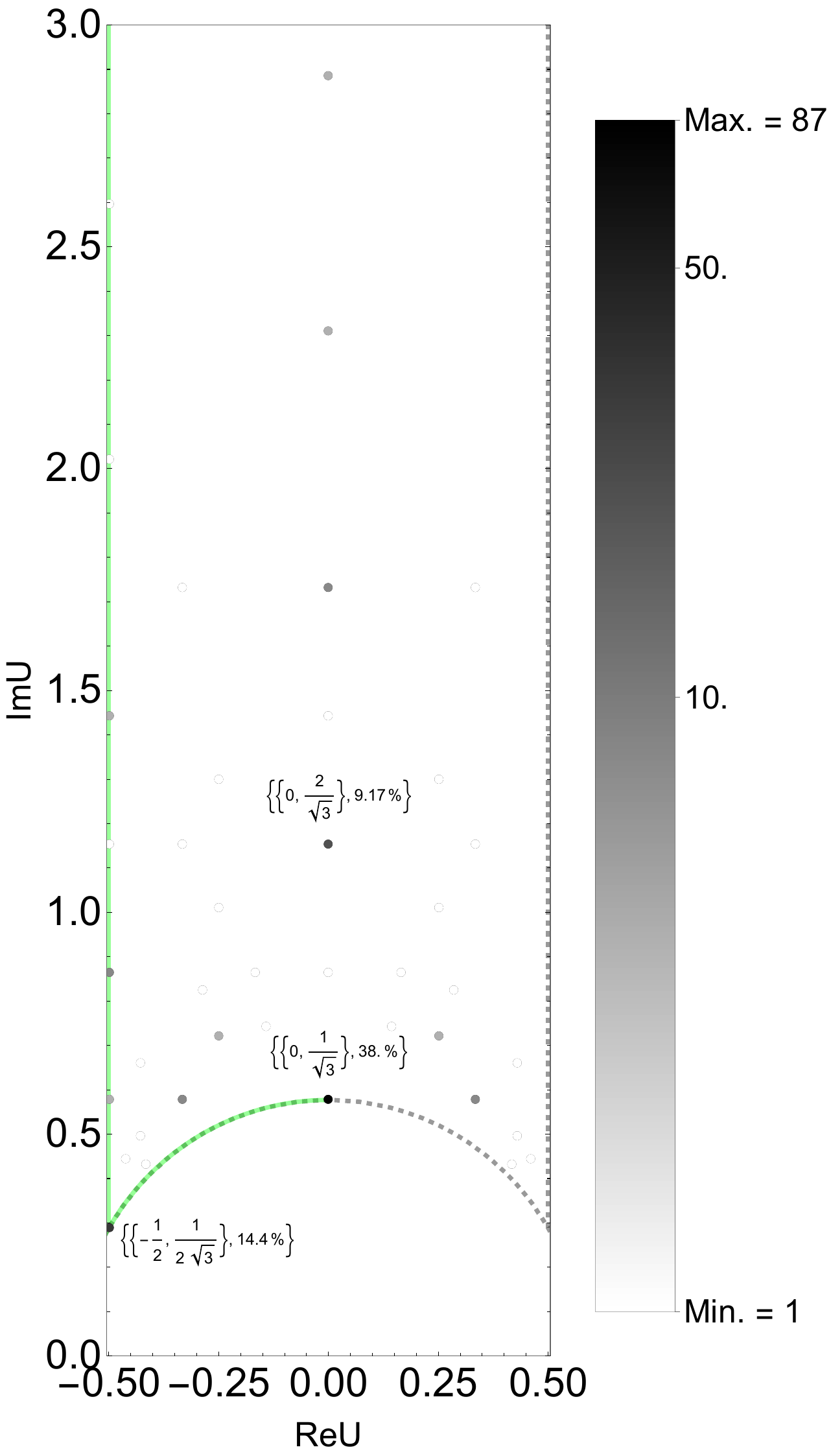}
\caption{The distribution of physically-distinct solutions on $\langle U \rangle$-plane with the figure cropped to ${\rm Im} U \leq 3.0$ is illustrated. Here, $N_{\rm flux}^{\rm max}$ is  $24 \times 20$. The colors are determined by the associated number of degeneracies among the physically-distinct solutions. Here, we additionally listed the positions and ratios of the points whose ratio $> 5.0\%$.
The dashed curves denote the (isometric) circles $|U| = \frac{1}{3}$ and $U = \pm \frac{1}{2}$. The green curves denote the boundary of the total fundamental region of $\overline{\Gamma}_0(3) \rtimes_{\varphi(S_{(3)})} \mathbb{Z}_2$.}
\label{fig:vevU_enlarge_count_G03}
\end{figure}

We also investigated the number of degeneracies on each point. Here, the solutions are projected onto the $U$-plane, and solutions on the same point may have different VEVs of $S$.
The result is summarized in Table \ref{tab:RatiovsU-NMAX20} and Fig. \ref{fig:vevU_enlarge_count_G03}. 
Surprisingly, the most favorable point, which has the largest ratio (probability), is in the region $S {\cal F}_{SL(2, \mathbb{Z})}$.
Since this is not an elliptic point of $\overline{\Gamma}_0(3)$, the result is quite different from the known $\mathbb{Z}_2 \times \mathbb{Z}_2'$ case.
Indeed, the largest probability emerges on $U = \frac{i}{\sqrt{3}}$, and it is the elliptic point of the $S_{(3)}$ duality.
The only elliptic point of $\Gamma_0(3)$, $\omega'$, has the second largest probability. Note that the probability associated with $U = \omega$ is the third largest, while the other three points have also the same probability.
Those three points and $U = \omega$ are on the circle with center at $U = \frac{2}{\sqrt{3}}i$ and radius $\frac{1}{\sqrt{3}}$.
We can also observe that ${\rm Re} U = 0$ is dominant.

\begin{figure}[H]
 \centering
  \includegraphics[width=11cm]{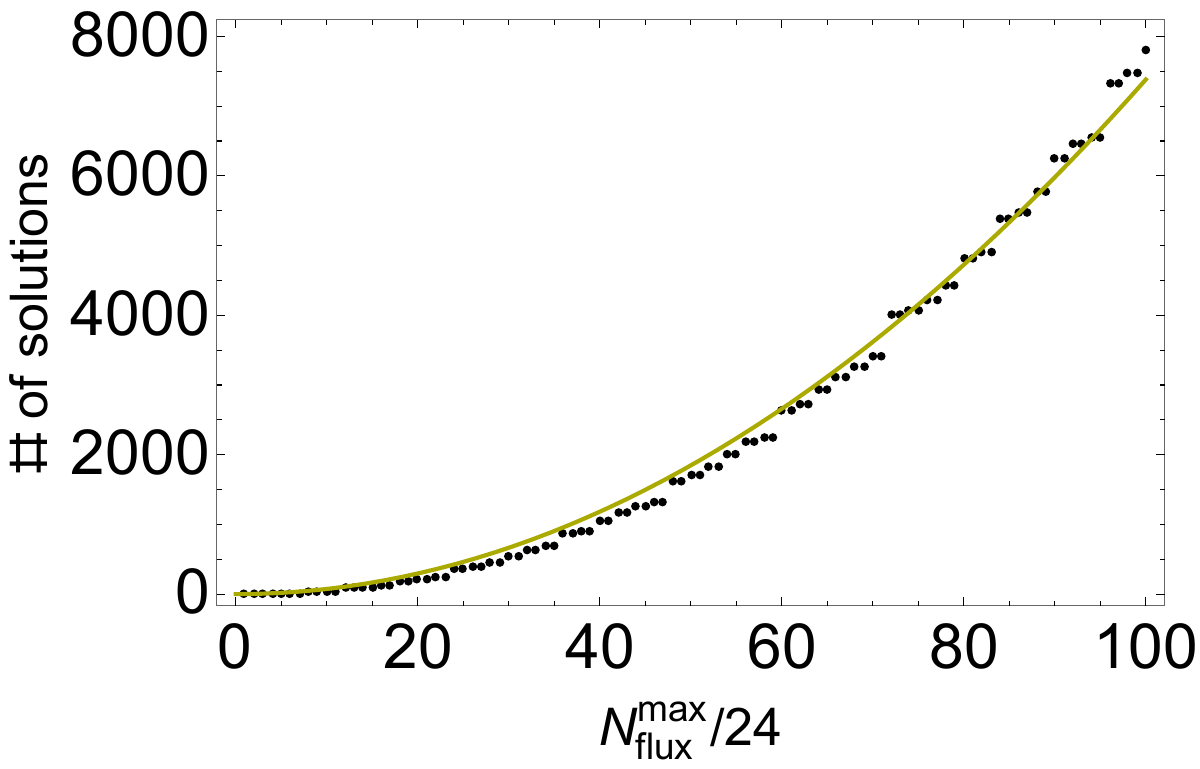}
  \caption{The number (\#) of the whole solutions with $N_{\rm flux} \leq N_{\rm flux}^{\rm max}$. The yellow line is a fitted curve with $\# = 0.738 (N_{\rm flux}^{\rm max}/24)^2.$} 
\label{fig:Nfluxvsnumsolutions}
\end{figure}

Next, we vary $N_{\rm flux}^{\rm max}$. As mentioned, small $N_{\rm flux}^{\rm max}$ is relatively reliable. 
Since small $N_{\rm flux}^{\rm max}$ makes the allowed range for fluxes very narrow, there will be only a few solutions. 
Thus, it provides us with a set of selective $U$s. The number of solutions for various $N_{\rm flux}^{\rm max}$ is summarized in Table \ref{tab:Number} and Fig. \ref{fig:Nfluxvsnumsolutions}. 
We list several such VEVs and shifts in the probabilities associated with those points as $N^{\rm max}_{\rm flux}$ increases on Table \ref{tab:Ratio-of-Dominant-VEVs}.

Thus, we conclude that on $\mathbb{Z}_{6 - \rm{\greekii}}$ with $SU(6) \times SU(2)$ lattice, there are only few solutions can be founded with small $N^{\rm max}_{\rm flux}$ and none of them are on the elliptic points of $SL(2, \mathbb{Z})$. Instead, we find $U = -\frac{1}{2} + \frac{1}{2\sqrt{3}}i$, which is only one elliptic point of $\Gamma_0(3)$, has relatively large probabilities for various $N^{\rm max}_{\rm flux}$. 
Here, we have also studied large $N^{\rm max}_{\rm flux}$ and found that $U = \frac{i}{\sqrt{3}}$ is the most dominant solution for all $N^{\rm max}_{\rm flux}$. 
Indeed, this is the elliptic point of the scaling duality $S_{(3)}: U \rightarrow -\frac{1}{3U}$.
Thus, elliptic points are still favored, but the most favored point does not need to be the usual $SL(2, \mathbb{Z})$ elliptic points.
Note that the VEVs on Table \ref{tab:Ratio-of-Dominant-VEVs} except $U = \frac{2}{\sqrt{3}}i$ are not on the fundamental region ${\cal F}$ of $SL(2, \mathbb{Z})$.\newpage
\begin{midpage}
\begin{figure}[H]
  \begin{minipage}[b]{0.45\linewidth}
 \centering
  \includegraphics[height=16cm]{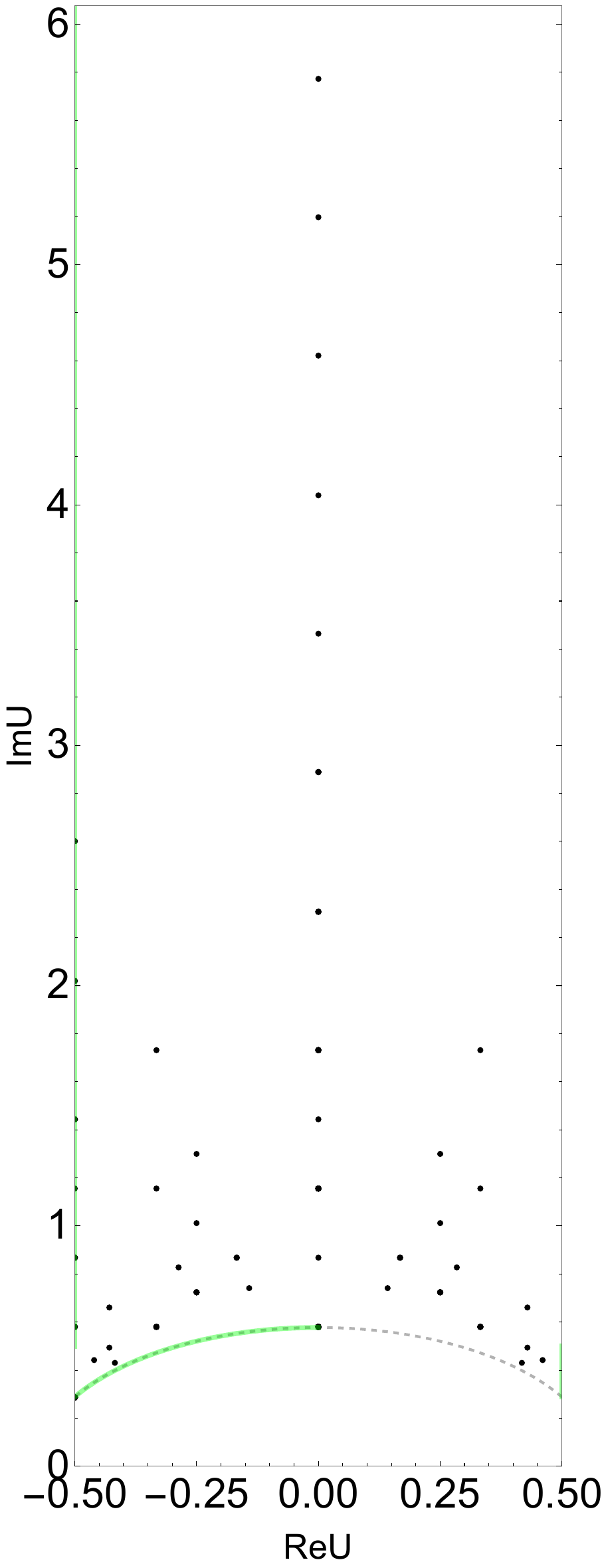}
  \end{minipage}
  \begin{minipage}[b]{0.45\linewidth}
 \centering
  \includegraphics[height=16cm]{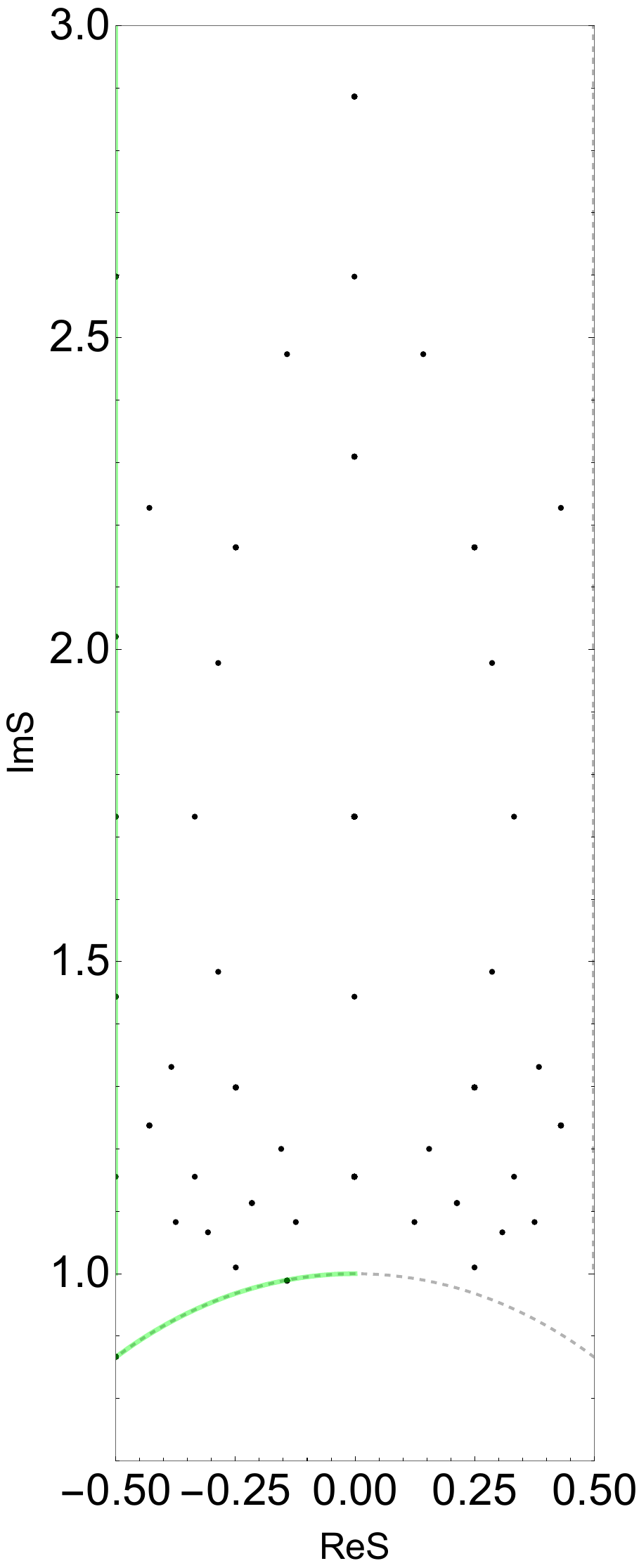}
  \end{minipage}
  \caption{In the left panel, the distribution of $\langle U \rangle$ is illustrated in the case of $N_{\rm flux}^{\rm max} = 24 \times 20$. The green curves denote the boundaries of the fundamental region of $SL(2, \mathbb{Z})$ and its images under $S, ST$ and $ST^{-1}$ while the dashed curves denote the boundary of the fundamental region of $\Gamma_0(3)$. The right panel is the enlarged view of the left panel with ${\rm Im}U \leq 1.5$.}
    \label{fig:vevU_G03}
\end{figure}
\end{midpage}
\newpage
\clearpage

\begin{table}[H]
\centering
\begin{tabular}{|c|ccccc|}
\hline
Ratio                 &$38.0\%$&$14.4\%$&$9.17\%$&$3.49\%$&$1.75\%$\\ \hline
\multirow{6}{*}{$U$} &$\frac{1}{\sqrt{3}}i$&$-\frac{1}{2} + \frac{1}{2\sqrt{3}}i$&$\frac{2}{\sqrt{3}}i$&$\frac{1}{3} + \frac{1}{\sqrt{3}}i$&$-\frac{1}{2} + \frac{1}{\sqrt{3}}i$  \\
                      &  &  & &$\sqrt{3}i$&$-\frac{1}{4} + \frac{5}{4\sqrt{3}}i$\\
                      &  &  &  &$-\frac{1}{3} + \frac{1}{\sqrt{3}}i$&$\frac{5}{\sqrt{3}}i$\\
                      &  &  &  &$-\frac{1}{2} + \frac{\sqrt{3}}{2}i$&$\frac{4}{\sqrt{3}}i$\\
                      &  &  &  & & $\frac{1}{4} + \frac{5}{4\sqrt{3}}i$\\ 
                      &  &  &  & & $-\frac{1}{2} + \frac{5}{2\sqrt{3}}i$\\ \hline
\end{tabular}
\caption{The ratios (probabilities) which are associated with each VEV on $U$-plane. $N^{\rm max}_{\rm flux} = 24 \times 20$, and we listed the ratios up to the fifth largest. } 
\label{tab:RatiovsU-NMAX20}
\end{table}

\begin{table}[H]
\centering
\begin{tabular}{|c|c c c c c c c|} \hline
$N_{\rm flux}^{\rm max} $ & $24 \times 2$  & $24 \times 4$ & $24 \times 10$ & $24 \times 20$ & $24 \times 40$ & $24 \times 80$ & $24 \times 100$ \\ \hline
$\#$ of solutions & 2 & 7 & 44 & 229 & 1053 & 4823 & 7818\\ \hline
\end{tabular}
\caption{Number of solutions for each bound $N_{\rm flux}^{\rm max}$.} 
\label{tab:Number}
\end{table}

\begin{table}[H]
\centering
\begin{tabular}{|c|c c c c c c c|} \hline
\diagbox[]{$U$}{$N_{\rm flux}^{\rm max} $} & $24 \times 2$  & $24 \times 4$ & $24 \times 10$  & $24 \times 20$ & $24 \times 40$ & $24 \times 80$ & $24 \times 100$\\ \hline
$\frac{1}{\sqrt{3}}i$ & 50\% & 57.1\% & 47.7\% & 38.0\% & 32.2\% & 27.8\%  & 26.6\% \\ 
$-\frac{1}{2} + \frac{1}{2\sqrt{3}}i$ & 50\% & 28.6\% & 20.5\% & 14.4\% & 11.5\% & 9.62\%  & 9.13 \% \\
$\frac{2}{\sqrt{3}}i$ &    & 14.3\% & 9.09\% & 9.17\% & 8.26\% & 7.03\% & 6.68 \%\\ \hline
\end{tabular}
\caption{The ratios which are associated with several dominant VEVs. Generally, the probabilities decrease with larger $N_{\rm flux}^{\rm max}$, and $U = \frac{1}{\sqrt{3}}i$ is more dominant than others. We can see that specific VEVs are exclusively selected with small $N_{\rm flux}^{\rm max}$.} 
\label{tab:Ratio-of-Dominant-VEVs}
\end{table}

\subsection{Distributions of solutions with specific $S$}
\label{sec:landscapeanalysis-on-SU}
As we can see in Eq. (\ref{eq:NfluxandVEVs}), the VEVs of $S$ and $U$ may correlate with each other. Since the theory is only evaluated perturbatively, smaller ${\rm Im}S = g_s^{-1}$ is more reliable. It may also enable us to control non-perturbative effects. In this section, we summarize the solutions by focusing only on those that are in particular regions on $S$-plane.

\begin{figure}[H]
\centering
\includegraphics[width = 0.75 \linewidth]{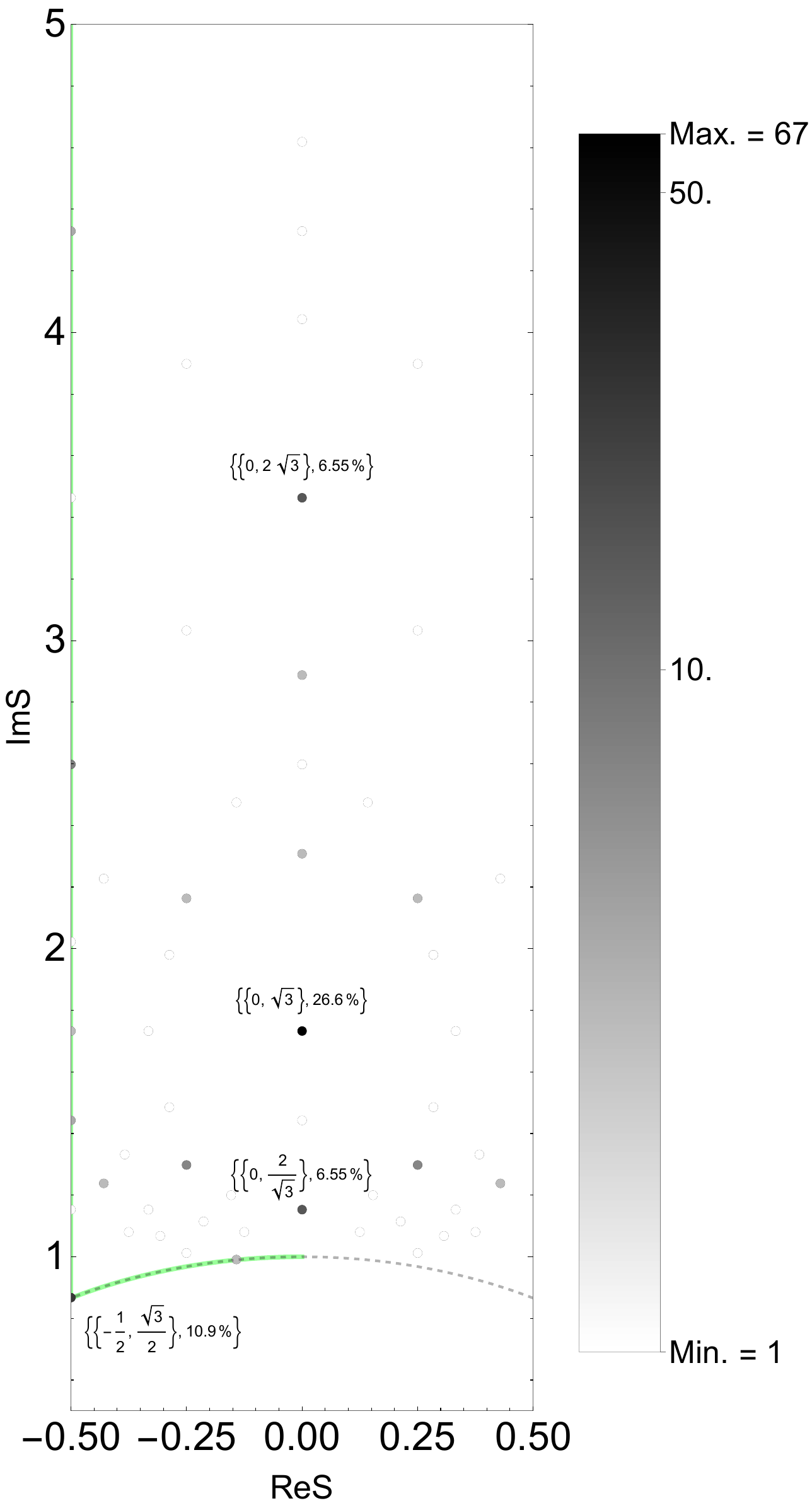}
\caption{The distribution of physically-distinct solutions on $\langle S \rangle$-plane with the figure cropped to ${\rm Im} S \leq 5.0$ is illustrated. The colors are determined by the associated number of degeneracies among the physically-distinct solutions. Here, we additionally listed the positions and ratios of the points whose ratio $> 5.0\%$.
The green curves denote the boundary of the fundamental region of $SL(2, \mathbb{Z})$. Here, $N_{\rm flux}^{\rm max}$ is  $24 \times 20$.}
\label{fig:vevS_enlarge_count_G03}
\end{figure}

First, let us illustrate the whole solutions which are projected on $S$-plane in Fig. \ref{fig:vevS_enlarge_count_G03}. We find $S = \sqrt{3}i$ is the most favored point, and $S = \omega = -\frac{1}{2} + \frac{\sqrt{3}}{2}i$ is the second one. The solutions with relatively large ${\rm Im}S$ have also comparable probabilities. On the other hand, there is no solution with $S = i$. This is due to the correlation with $U$, especially the symmetry $\Gamma_0(3)$ for $U$. One can see from the F-term equations that neither $S = i$ nor $U = i$ are allowed in the solutions. 
This is in contrast to the $T^6/(\mathbb{Z}_2 \times \mathbb{Z}'_2)$ case \cite{Ishiguro:2020tmo}, where $SL(2, \mathbb{Z})_S \times SL(2, \mathbb{Z})_U$ is the symmetry of the moduli space. 
Although the present symmetry is still a direct product of $SL(2, \mathbb{Z})_S$ and $\overline{\Gamma}_0(3) \rtimes_{\varphi(S_{(3)})} \mathbb{Z}_2$, $S$ and $U$ are correlated via the F-term equations in a non-trivial way. Note, the solutions with ${\rm Re}S = 0$ have large probabilities as well as $U$.

\begin{figure}[H]
 \centering
  \includegraphics[height=8cm]{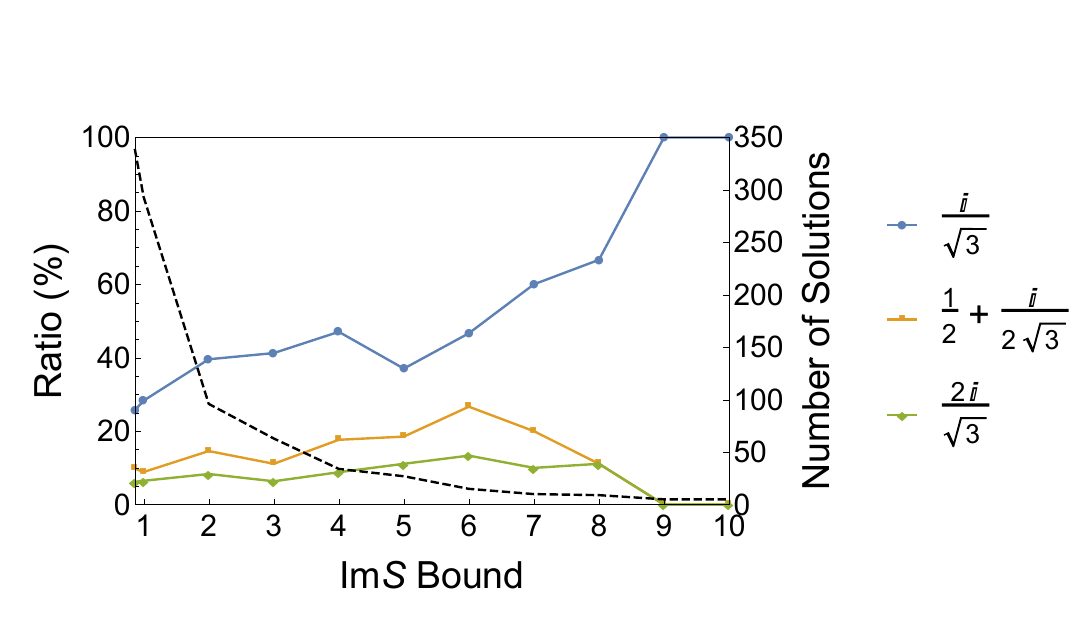}
  \caption{The solid lines present the ratios of the dominant (>5\%) VEVs of $U = \frac{i}{\sqrt{3}}, \frac{-1}{2}+\frac{1}{2\sqrt{3}}i$ and $3i$ with bounded ${\rm Im}S_{\rm bound}$ and $N_{\rm flux}^{\rm max} = 24 \times 20$. The lines start with ${\rm Im}S_{\rm bound} = \frac{\sqrt{3}}{2}$ (all region). The dashed line presents the number of solutions above certain ${\rm Im}S_{\rm bound}$.} 
\label{fig:Ratio-of-Dominant-VEVs-Bounded-Z6II}
\end{figure}

Next, we impose some bounds for ${\rm Im}S$ on the solutions to analyze the weak-coupling regime. Here, we calculate the probabilities as functions of $({\rm Re}U, {\rm Im}U)$ and $({\rm Re}S, {\rm Im}S)$. Then, the probabilities are summed up if the associated ${\rm Im}S$ satisfies the bound ${\rm Im}S_{\rm min}$. The result for dominant VEVs is summarized in Tables \ref{tab:Ratio-of-Dominant-VEVs-Bounded} and \ref{fig:Ratio-of-Dominant-VEVs-Bounded-Z6II}.
Basically, those dominant VEVs have larger probabilities with higher bounds. Although it is not exclusive for relatively small ${\rm Im}S$ but $U = \frac{i}{\sqrt{3}}$, which is not the elliptic point of $SL(2, \mathbb{Z})$ but that of $S_{(3)}$, is favored most.
From Fig. \ref{fig:Ratio-of-Dominant-VEVs-Bounded-Z6II}, we see that on solutions with ${\rm Im}S \geq 9$, only $U = \frac{i}{\sqrt{3}}$ is realized.

Due to the correlation, a VEV of $U$ can acquire the largest probability if the associated $S$ is specified $S_{\rm section}$. Indeed, it was observed in the $T^6/(\mathbb{Z}_2 \times \mathbb{Z}'_2)$ case \cite{Ishiguro:2020tmo}. If there is some mechanism to constrain $S$ to the specific value, the specific $U$ can be dominant. From this point of view, we summarized in Table \ref{tab:Ratio-of-Dominant-VEVs-Section} the probabilities as functions of $({\rm Re}U, {\rm Im}U)$ and $({\rm Re}S, {\rm Im}S)$ for several important VEVs. As a result, the exclusive correlation is not observed with favored $S$. However, besides the list, there are 41 solutions that there possible $U$ is only $U = \frac{i}{\sqrt{3}}$ with a specific $S_{\rm section}$. All of those solutions have no degeneracy and the ratio of $0.44\%$ in the whole Landscape, but the total is $17.9\%$. Thus, if there is some natural mechanism to obtain such specific $S$s, it automatically leads to $U = \frac{i}{\sqrt{3}}$. Other VEVs of $U$ are turned out not to exceed $25\%$ even if we choose a specific $S_{\rm section}$. This result is different from the $T^6/(\mathbb{Z}_2 \times \mathbb{Z}'_2)$ case, where $U = i$, which is the elliptic point of $SL(2, \mathbb{Z})$, becomes exclusive with $S = i, 2i$. 

\clearpage
\begin{table}[H]
\centering
\begin{tabular}{|c|c c c c c c |} \hline
\diagbox[]{$U$}{$ {\rm Im}S_{\rm bound} $} & $\frac{\sqrt{3}}{2}$ (All)  & $1$ & $\frac{2}{\sqrt{3}}$  & $\sqrt{3}$ & $2\sqrt{3}$ & $3\sqrt{3}$ \\ \hline
$\frac{1}{\sqrt{3}}i$ & 38.0\% & 41.3 \% & 39.3\% & 36.0\% & 52.2 \% & 47.6 \% \\ 
$\frac{-1}{2} + \frac{1}{2\sqrt{3}}i$ & 14.4\% & 12.9 \% & 12.6\% & 12.7\% & 15.2\% & 23.8\% \\
$\frac{2}{\sqrt{3}}i$ & 9.17\% &  9.45\% & 9.95\% & 9.86\% & 8.70\% & 14.3\% \\\hline
\end{tabular}
\caption{The ratios which are associated with several dominant ($>5\%$) VEVs with several bounds ${\rm Im}S_{\rm bound}$ and $N_{\rm flux}^{\rm max} = 24 \times 20$.} 
\label{tab:Ratio-of-Dominant-VEVs-Bounded}
\end{table}

\begin{table}[H]
\centering
\begin{tabular}{|c|c c c c|} \hline
\diagbox[]{$U$}{$ S_{\rm section} $} & $\sqrt{3}i$  & $-\frac{1}{2} + \frac{\sqrt{3}}{2}i$ & $ 2\sqrt{3}i$  & $\frac{2}{\sqrt{3}}i$  \\ \hline
$\frac{1}{\sqrt{3}}i$ & 13.1\% & 12\% & 33.3\% & 33.3\%  \\ 
$- \frac{1}{2} + \frac{1}{2\sqrt{3}}i$ & 4.92 \% (*)& 24 \% & 6.67 \% (*) & 6.67 \% (*) \\
$\frac{2}{\sqrt{3}}i$ & 8.20\% & 4.0 \%  (*) & 6.67 \% (*) & 6.67 \% (*) \\ 
$\omega = -\frac{1}{2} + \frac{\sqrt{3}}{2}i$ & 4.92 \% & 8.0\% & 0 \% & 0\%  \\ \hline
\end{tabular}
\caption{The ratios which are associated with several dominant ($>5\%$) VEVs with several specific $S = S_{\rm section}$ and $N_{\rm flux}^{\rm max} = 20$. The ratios are not calculated on the whole Landscape with various $S$. The solutions with $S = 2\sqrt{3}i$ and $S = \frac{2}{\sqrt{3}}i$ have also exactly equal ratios.  (*) denotes that there are other VEV(s) which have the same or higher probabilities. } 
\label{tab:Ratio-of-Dominant-VEVs-Section}
\end{table}

\begin{figure}[H]
  \begin{minipage}[b]{0.45\linewidth}
 \centering
  \includegraphics[height=14cm]{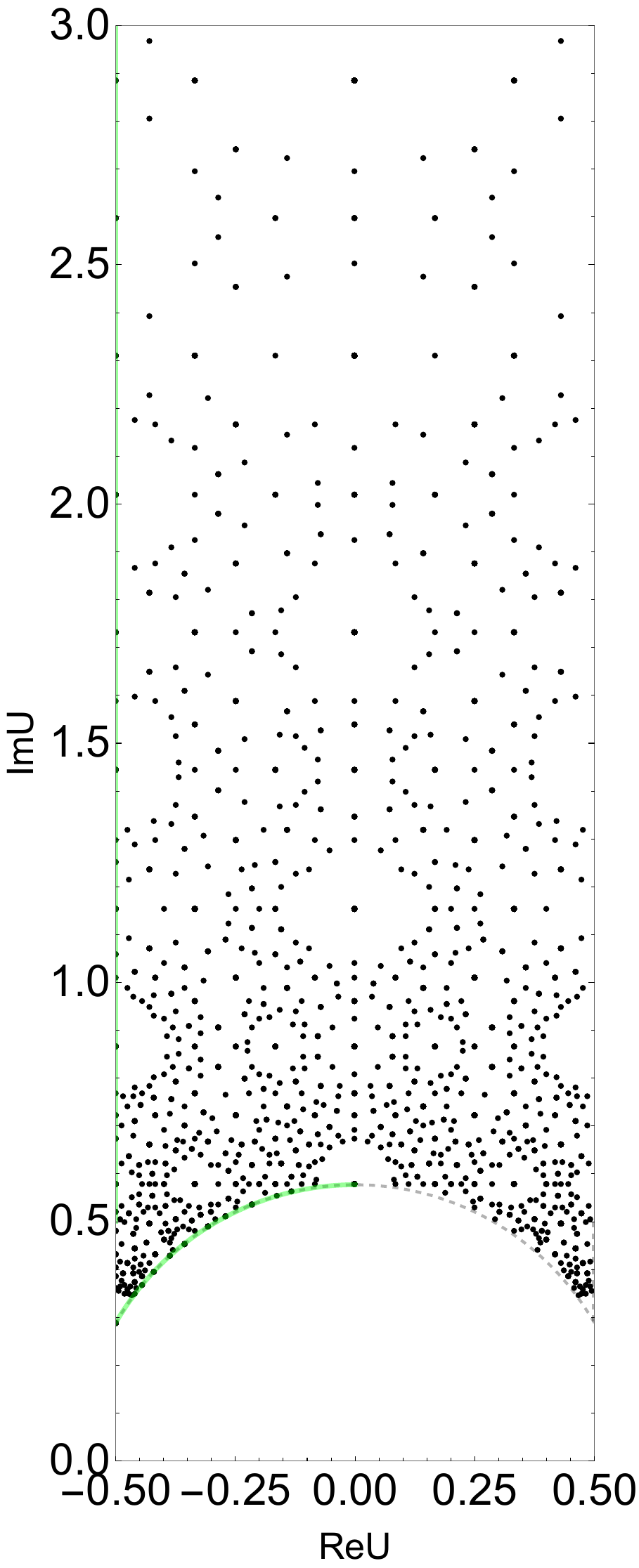}
  \end{minipage}
  \begin{minipage}[b]{0.45\linewidth}
 \centering
  \includegraphics[height=14cm]{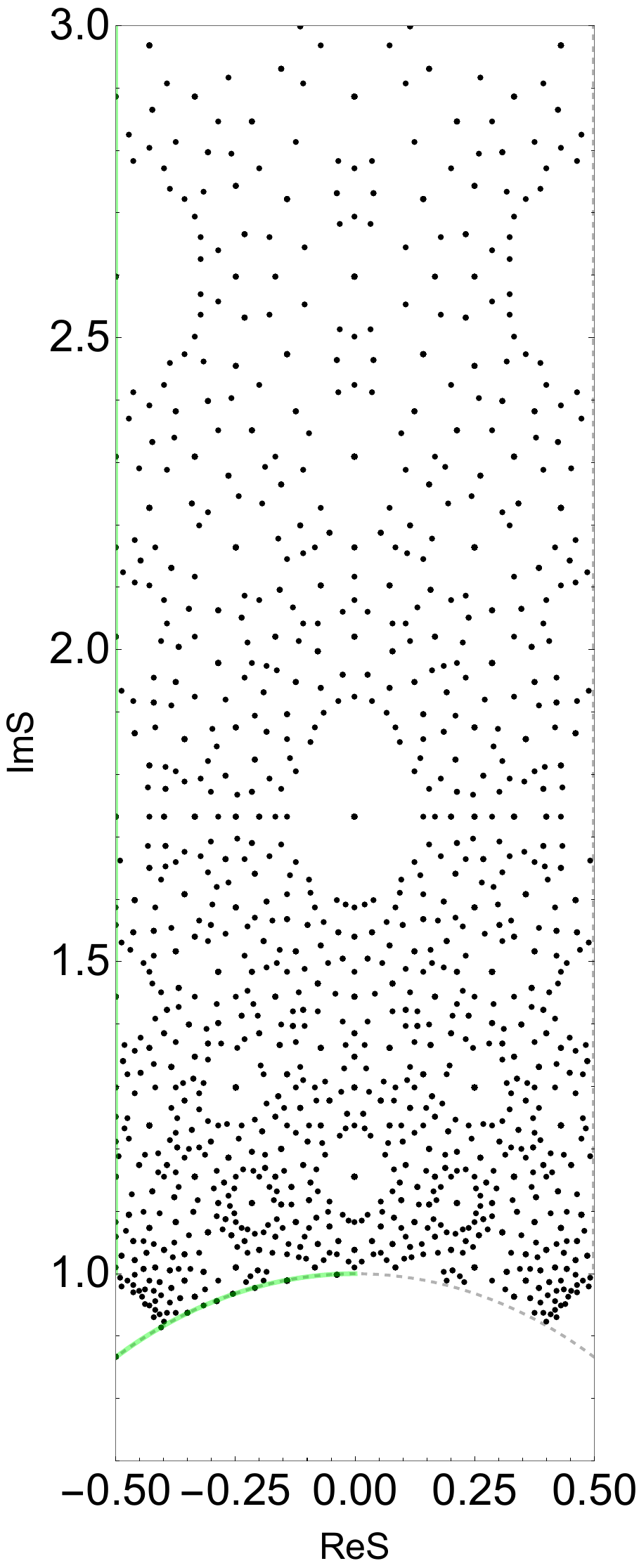}
  \end{minipage}
  \caption{The distribution of $\langle U \rangle$ and $\langle S \rangle$ are illustrated in the left panel and the right one, respectively. The figures are cropped to ${\rm Im} U \leq 3$ or ${\rm Im} S \leq 3$. The maximum value $N_{\rm flux}^{\rm max} = 24 \times 100$, and one can observe the void structure appears even where $|U|^2 < 1$.}
    \label{fig:voids}
\end{figure}

Finally, let us comment on the "void structure" in the Landscape. Since we observe several differences with the $T^6/(\mathbb{Z}_2 \times \mathbb{Z}'_2)$ case where the direct product $SL(2, \mathbb{Z}) \times SL(2, \mathbb{Z})$ is realized, the void structure may also differ in this case.
As mentioned, in the $N_{\rm flux}^{\rm max} = 24 \times 20$ case, one can see that there is a circle on which points including $U = \omega$ have the same probability and whose center is $U = \frac{2}{\sqrt{3}}$.
This is an example of the void structure. In contrast to usual toroidal orientifolds where $SL(2, \mathbb{Z})$ remains, $\omega$ may not be the center of a large circle, but it is on the large circle. We shall study this point in more detail in the future work.
We also illustrate in Fig. \ref{fig:voids} the void structure with higher $N_{\rm flux}^{\rm max}$, $24 \times 100$, where one can see that there are circles which encircle points with high degeneracies. In the orientifold, this structure exists also outside the fundamental region of $SL(2, \mathbb{Z})$.
\section{Other orientifolds: $\mathbb{Z}_2 \times \mathbb{Z}_4$ orientifold}
\label{sec:otherorientifolds}
The analysis on the $T^6/{\mathbb{Z}_{6-\greekii}}$ with ${\rm SU}(6) \times {\rm SU}(2)$ lattice can be extended to other toroidal orientifolds whose $h^{2,1}_{\rm untw.} = 1$. We still stick to orientifolds with no twisted complex-structure modulus. 
In this section, we analyze $T^6/({\mathbb{Z}_2 \times \mathbb{Z}_4})$ with ${\rm SU}(2)^2 \times {\rm SO}(5)^2$ as an example.
We use the notation as fixed already in Sec. \ref{sec:fluxcompactifications} for quantities that are defined on the underlying $T^6$.

\subsection{Flux compactification on the $T^6/(\mathbb{Z}_2 \times \mathbb{Z}_4)$ orientifold}
\label{sec:Z_2Z_4}
Henceforth, we consider the $T^6/({\mathbb{Z}_2 \times \mathbb{Z}_4)}$ with $({\rm SU}(2))^2 \times ({\rm SO}(5))^2$ lattice. The Hodge numbers are $(h^{1,1}, h^{2,1}) = (61, 1)$, and there are no twisted contributions to $h^{2,1}$. The $\mathbb{Z}_2 \times \mathbb{Z}_4$ twist lives on the lattice via the Coxeter element $Q$ which is defined as $Q = Q_1 Q_2$ with
\begin{align}
    \begin{alignedat}{2}
    Q_1(e_i) &= e_i \quad (i = 1, 2),\\
    Q_1(e_i) &= e_i \quad (i = 3, 4),\\
    Q_2(e_i) &= e_i + 2 e_{i+1} \quad (i = 3),\\
    Q_2(e_i) &= -e_{i-1} - e_i \quad (i = 4),\\
    Q_2(e_i) &= -e_i -2 e_{i+1} \quad (i = 5),\\
    Q_2(e_i) &= e_{i-1} + e_i \quad (i = 6).
    \end{alignedat}
\end{align}
The matrix representation of $Q: e_i \rightarrow Q_{ji}e_i$ is
\begin{align}
Q_{({\rm SU}(2))^2 \times ({\rm SO}(5))^2} = 
    \begin{pmatrix}
    -1  &   &   &   &   &   \\
        &-1 &   &   &   &   \\
        &   &1  &-1 &   &   \\
        &   &2  &-1 &   &   \\
        &   &   &   &-1 &1  \\
        &   &   &   &-2 &1
    \end{pmatrix},
\end{align}
where the twist is represented on the complex diagonal basis $z_i \rightarrow e^{2\pi i v_a^i} z_i ~ (a = 1, 2, \text{no sum})$ with
\begin{align}
    v_1 = \frac{1}{2}(1, 0, -1), \quad v_2 = \frac{1}{4}(0, 1, -1). 
\end{align}
The explicit form of $dz$ which we adopt is
\begin{align}
\begin{alignedat}{2}
    dz^1 &= dx^1 - U dx^2, \\
    dz^2 &= dx^3 - \frac{1}{2}(1-i) dx^4, \\
    dz^3 &=dx^5 - \frac{1}{2}(1+i) dx^6,
\end{alignedat}
\end{align}
where $\int{\Omega \wedge \overline{\Omega} = U - \overline{U}}$.
Then we find that there are four independent orbits of the real three-forms whose length is two, and their intersecting matrix takes the form of an unimodular matrix multiplied by $\frac{1}{4}$.
Thus we conclude that the fractional cycles on the orbifold are given as those orbits multiplied by two. The symplectic three-form basis on the orbifold is
\begin{align}
    \begin{alignedat}{2}
    \mathbf{1}_1 &= 2 \sum \Gamma(\beta^2), \\
    \mathbf{1}_2 &= 2 \sum \Gamma(\alpha_2), \\
    \mathbf{1}_3 &= 2 \sum \Gamma(\alpha_1) + 2 \sum \Gamma(\beta^2), \\
    \mathbf{1}_4 &= 2 \sum \Gamma(\alpha_0) = - \Gamma(\beta^1),
    \end{alignedat}
    \label{eq:singletsasorbits-Z4Z2}
\end{align}
where the orbits are defined for elements in $\Gamma = \mathbb{Z}_4 \times \mathbb{Z}_2$ and satisfy 
\begin{align}
    \int_{T^6} \mathbf{1}_1 \wedge \mathbf{1}_4 = 8, \quad \int_{T^6} \mathbf{1}_2 \wedge \mathbf{1}_3 = 8.
\end{align}
All orbits that appear in (\ref{eq:singletsasorbits-Z4Z2}) have a length of two.
The fluxes $F_3, H_3$ are quantized on the Poincar\'{e} dual cycles which span $H_3(T^6/({\mathbb{Z}_2\times \mathbb{Z}_4}), \mathbb{Z})$.
The flux quanta which are measured on the underlying $T^6$ acquire additional factors as before.
We choose $-\beta^1$ as a representative for the orbit $\mathbf{1}_4$.
Then by use of the following definitions:
\begin{align}
    \begin{alignedat}{2}
        \int_{T^6/(\mathbb{Z}_2 \times \mathbb{Z}_4)} (F_3, H_3) \wedge \mathbf{1}_1 &= \frac{1}{8} \int_{T^6} (F_3, H_3) \wedge \mathbf{1}_1 = \frac{1}{4} \sum \int_{T^6} (F_3, H_3) \wedge \Gamma(\beta^2) = \frac{1}{2} (a^2, c^2), \\
        \int_{T^6/(\mathbb{Z}_2 \times \mathbb{Z}_4)} (F_3, H_3) \wedge \mathbf{1}_2 &= \frac{1}{2}(-b_2, -d_2), \\
        \int_{T^6/(\mathbb{Z}_2 \times \mathbb{Z}_4)} (F_3, H_3) \wedge \mathbf{1}_3 &= \frac{1}{2} \left\{ (-b_1, -d_1) + (a^2, c^2) \right\}, \\
        \int_{T^6/(\mathbb{Z}_2 \times \mathbb{Z}_4)} (F_3, H_3) \wedge \mathbf{1}_4 &= \frac{1}{4} (-a^1, -c^1),
    \end{alignedat}
    \label{eq:integralsforfluxquanta-Z4Z2}
\end{align}
the three form flux $G_3$ is expanded as
\begin{align}
    G_3 = -\frac{1}{4}(a^1 - S c^1) \mathbf{1}_1 + \frac{1}{2}[(- b_1 + a^2 ) - S (-d_1 + c^2)] \mathbf{1}_2 + \frac{1}{2} (b_2 - S d_2) \mathbf{1}_3 - \frac{1}{2} (a^2 - S c^2) \mathbf{1}_4.
\end{align}
The fluxes which we choose as representatives are quantized as
\begin{align}
        (b_1, d_1) \in 4 \mathbb{Z},\quad (b_2, d_2) \in 4 \mathbb{Z},\quad (a_1, c_1) \in 8 \mathbb{Z},\quad  (a_2, c_2) \in 4 \mathbb{Z},
    \label{eq:fluxquantizationcondition-Z2Z4}
\end{align}
to realize even fluxes on the orbifold.
With this choice of the fluxes, $N_{\rm flux}$ becomes
\begin{align}
    N_{\rm flux} = a^2 c^1 - a^1 c^2 + 2 b_2 c^2 - 2 b_2 d_1 - 2 a^2 d_2 + 2 b_1 d_2 \in 32 \mathbb{Z}.
\end{align}

The period vector is given by
\begin{align}
    \Pi \equiv \begin{pmatrix}
        \int \Omega \wedge \mathbf{1}_4 \\
        \int \Omega \wedge \mathbf{1}_3 \\
        \int \Omega \wedge \mathbf{1}_1 \\
        \int \Omega \wedge \mathbf{1}_2 
    \end{pmatrix}
    = \frac{1}{8} \begin{pmatrix}
        2U \\
        2i \\
        -2+2i \\
        (2+2i)U 
    \end{pmatrix},
\end{align}
where the integrals are calculated on the $T^6/({\mathbb{Z}_2 \times \mathbb{Z}_4})$.
Then, the expansion of $W$ (\ref{eq:WGeneral}) leads
\begin{equation}
    \begin{alignedat}{5}
        A &= \frac{-1 + i}{2} \left[ a^1 + (-1 + i)b_2 \right], \quad & C &= \frac{-1 + i}{2} \left[ (-1 + i) a^2 + (-2i)b_1 \right], \\
        B &= \frac{-1 + i}{2} (-1) \left[ c^1 + (-1 + i)d_2 \right], \quad & D &= \frac{-1 + i}{2} (-1) \left[ (-1 + i)c_2 + (-2i) d_1\right].
    \end{alignedat}
    \label{eq:ABCDinrealfluxes-Z2Z4}
\end{equation}
One can notice that VEVs of $S, U$ are always rational numbers. Thus, $\langle U \rangle = \omega$ is not allowed on the $\mathbb{Z}_2 \times \mathbb{Z}_4$ orientifold with the lattice.

The modular transformation of $U: U \rightarrow \frac{a U + b}{c U + d}$ leads the symplectic transformation over $\mathbb{Z}$ as
\begin{align}
    \Pi \rightarrow (c U + d)^{-1} \begin{pmatrix}
        a & b & -b & 0 \\
        -c & d & 0 & c \\
        -2c & 0 & d & c \\
        0 & 2b & -b & a
    \end{pmatrix} \Pi \equiv (c U + d)^{-1} M \Pi,
\end{align}
where $M \in Sp(b_3, \mathbb{Z})$. Thus, the $SL(2, \mathbb{Z})_U$ duality exists. Since we do not consider the $GSp$ transformation of period vectors, it is the most general duality.  Let us summarize the flux transformations in terms of the real flux quanta. For the two generators $S, T$ of $SL(2, \mathbb{Z})_S$, 
\begin{itemize}
    \item $S = 
        \begin{pmatrix}
            0 & -1 \\
            1 & 0
        \end{pmatrix} \in SL(2, \mathbb{Z})_S
    $\\
    \begin{equation}
        \begin{alignedat}{5}
            a^1 &\rightarrow - c^1, \quad & b_0 &\rightarrow - d_1, \\
            a^2 &\rightarrow - c^2, \quad & b_1 &\rightarrow - d_2, \\
            c^1 &\rightarrow a^1, \quad & d_0 &\rightarrow b_1, \\
            c^2 &\rightarrow a^2, \quad & d_1 &\rightarrow b_2. \\
        \end{alignedat}
    \end{equation}
\end{itemize}

\begin{itemize}
    \item $T^q = 
        \begin{pmatrix}
            1 & q \\
            0 & 1
        \end{pmatrix} \in SL(2, \mathbb{Z})_S
    $\\
    \begin{equation}
        \begin{alignedat}{5}
            a^1 &\rightarrow a^1 + q c^1, \quad & b_1 &\rightarrow b_1 + q d_1, \\
            a^2 &\rightarrow a^2 + q c^2, \quad & b_2 &\rightarrow b_2 + q d_2, \\
            c^1 &\rightarrow c^1, \quad & d_1 &\rightarrow d_1, \\
            c^2 &\rightarrow c^2, \quad & d_2 &\rightarrow d_2. \\
        \end{alignedat}
    \end{equation}
\end{itemize}

For those of $SL(2, \mathbb{Z})_U$, 
\begin{itemize}
    \item $S = 
        \begin{pmatrix}
            0 & -1 \\
            1 & 0
        \end{pmatrix} \in SL(2, \mathbb{Z})_U
    $\\
    \begin{equation}
        \begin{alignedat}{5}
            a^1 &\rightarrow - 2 b_1, \quad & b_1 &\rightarrow \frac{1}{2}a^1, \\
            a^2 &\rightarrow a^1 - b_2, \quad & b_2 &\rightarrow a^2 - 2 b_1, \\
            c^1 &\rightarrow - 2 d_1, \quad & d_1 &\rightarrow \frac{1}{2}c^1, \\
            c^2 &\rightarrow c^1 - d_2, \quad & d_2 &\rightarrow c^2 - 2 d_1.
        \end{alignedat}
        \label{eq:fluxtrf-S-Z2Z4}
    \end{equation}
\end{itemize}

\begin{itemize}
    \item $T^q = 
        \begin{pmatrix}
            1 & q \\
            0 & 1
        \end{pmatrix} \in SL(2, \mathbb{Z})_U
    $\\
    \begin{equation}
        \begin{alignedat}{5}
            a^1 &\rightarrow a^1 + 2 q b_1, \quad & b_1 &\rightarrow b_1, \\
            a^2 &\rightarrow a^2, \quad & b_2 &\rightarrow b_2 - q (a^2 - 2 b_1), \\
            c^1 &\rightarrow c^1 + 2 q d_1, \quad & d_1 &\rightarrow d_1, \\
            c^2 &\rightarrow c^2, \quad & d_2 &\rightarrow d_2 - q (c^2 - 2 d_1).
        \end{alignedat}
        \label{eq:fluxtrf-T-Z2Z4}
    \end{equation}
\end{itemize}

\subsection{Distributions and degeneracies in the finite Landscape}
\label{sec:landscapeanalysis-Z2Z4}
In this section, we show the distribution of the moduli VEVs on the $\mathbb{Z}_2 \times \mathbb{Z}_4$ orientifold explicitly, along the lines of the $\mathbb{Z}_{6 - \rm{\greekii}}$ case.
The degeneracies of physically-distinct solutions are also summarized.
The procedure to obtain the whole Landscape is quite similar to the previous case in Appendix \ref{app:algorhithm}.
We will limit our discussion to SUSY Minkowski solutions with $N_{\rm flux} \leq 32 \times 100$. 
The least value of $N_{\rm flux}$ where solutions are found is $64$, which typically violates the tadpole cancellation condition.
However, we continue the discussion to explore structures of the F-term equation and its solutions.
Even if uplifting to F-theory is turned out to be possible on this background, smaller $N_{\rm flux}$ will be a more reliable region.  

\begin{figure}[H]
  \begin{minipage}[b]{0.45\linewidth}
 \centering
  \includegraphics[height=12cm]{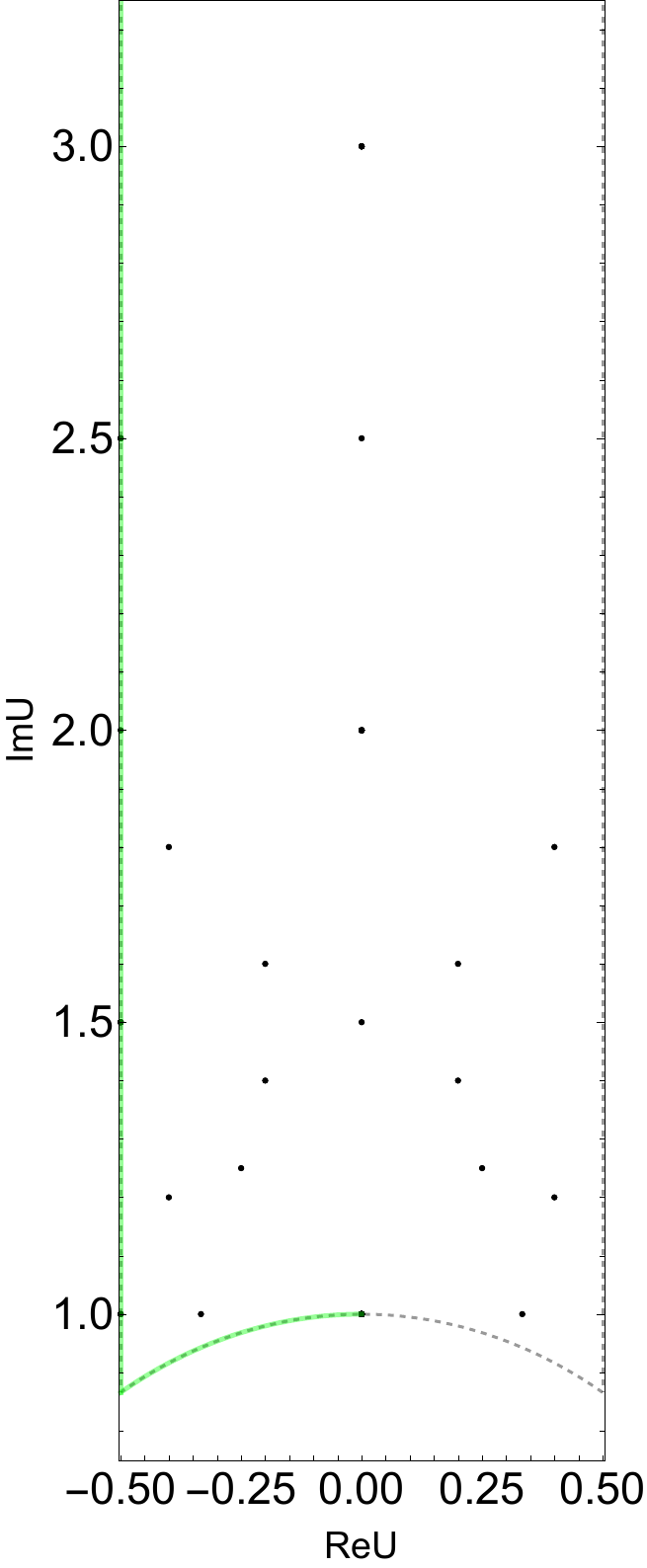}
  \end{minipage}
  \begin{minipage}[b]{0.45\linewidth}
 \centering
  \includegraphics[height=12cm]{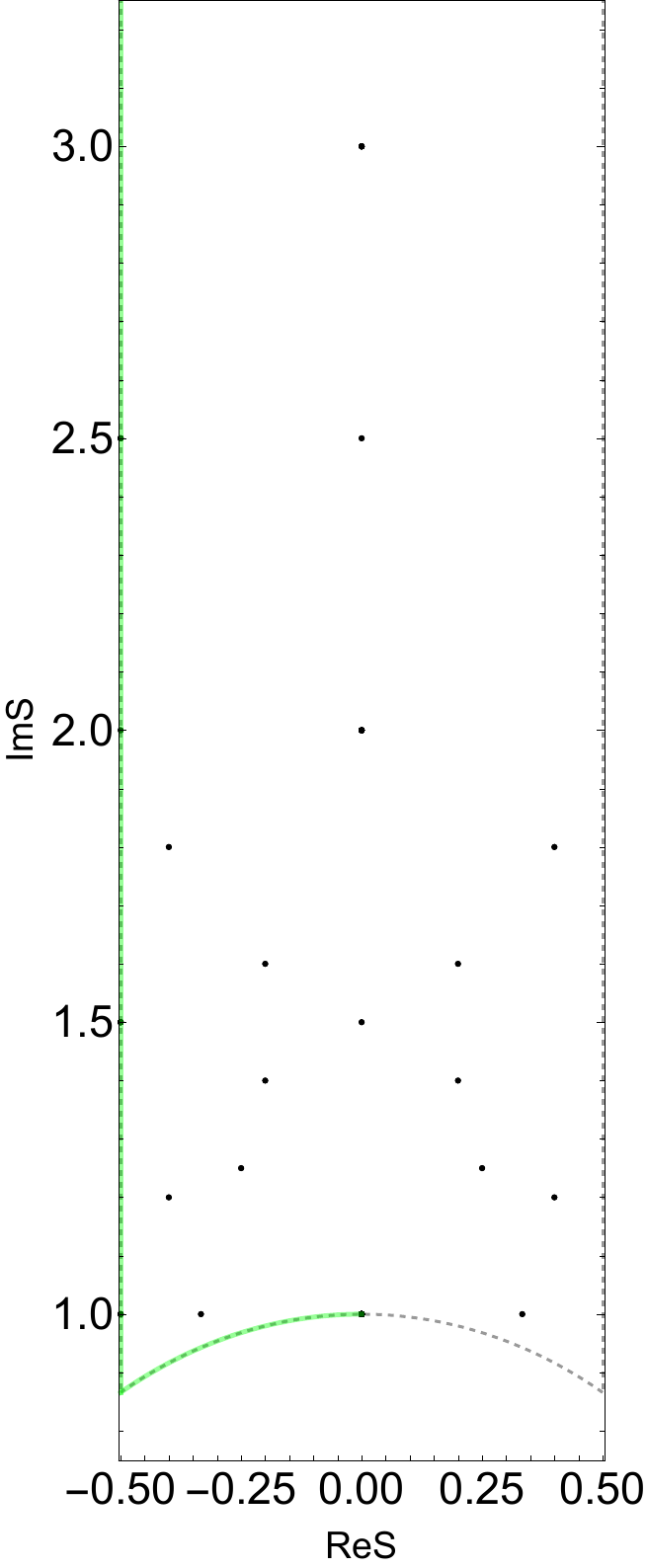}
  \end{minipage}
  \caption{In the left panel, the distribution of $\langle U \rangle$ is illustrated in the case of $N_{\rm flux}^{\rm max} = 32 \times 20$. The green curves denote the boundaries of the fundamental region of $SL(2, \mathbb{Z})$. The right panel is the distribution of $\langle S \rangle$. One can see that they are exactly the same. Both of the figures are cropped to ${\rm Im} (U, S) \geq \frac{\sqrt{3}}{2}$.}
    \label{fig:vevU_vevS_Z2Z4}
\end{figure}

We begin with $N_{\rm flux}^{\rm max} = 32 \times 20$ case. The number of solutions is 123. The solutions which are projected on the $U$-plane and the $S$-plane are illustrated in Fig. \ref{fig:vevU_vevS_Z2Z4}. 
As mentioned, real and imaginary parts of the solutions are rational numbers.
Thus one cannot observe $U = \omega$ for example.

One can find in Fig. \ref{fig:vevU_vevS_Z2Z4} that the distributions of $U$ and $S$ are exactly the same.
This is due to a duality by which $U$ and $S$ are interchanged in the superpotential.
It leads to the flux transformation $\{c^1, c^2, d_1, d_2, a^1, a^2, b_1, b_2\} \rightarrow \{2b_1, c^2, d_1, -a^2 + 2b_1, a^1, c^1 - d_2, \frac{1}{2}c^1, b_2\}$, and it is consistent with the flux quantization (\ref{eq:fluxquantizationcondition-Z2Z4}).
Moreover, $N_{\rm flux}$ is invariant under the transformation.
This duality cannot be expressed as matrix multiplications and linear combinations of $F_3$ and $H_3$.
As discussed in \ref{sec:dualities}, we focus on only the dualities which are understood as basis transformations of the cycles.
Hence, we do not mod out the Landscape by it, and the distributions of $U$ and $S$ become exactly the same.
We simply omit the projection on $S$-plane in the following.

\begin{figure}[H]
\centering
\includegraphics[width = 0.8 \linewidth]{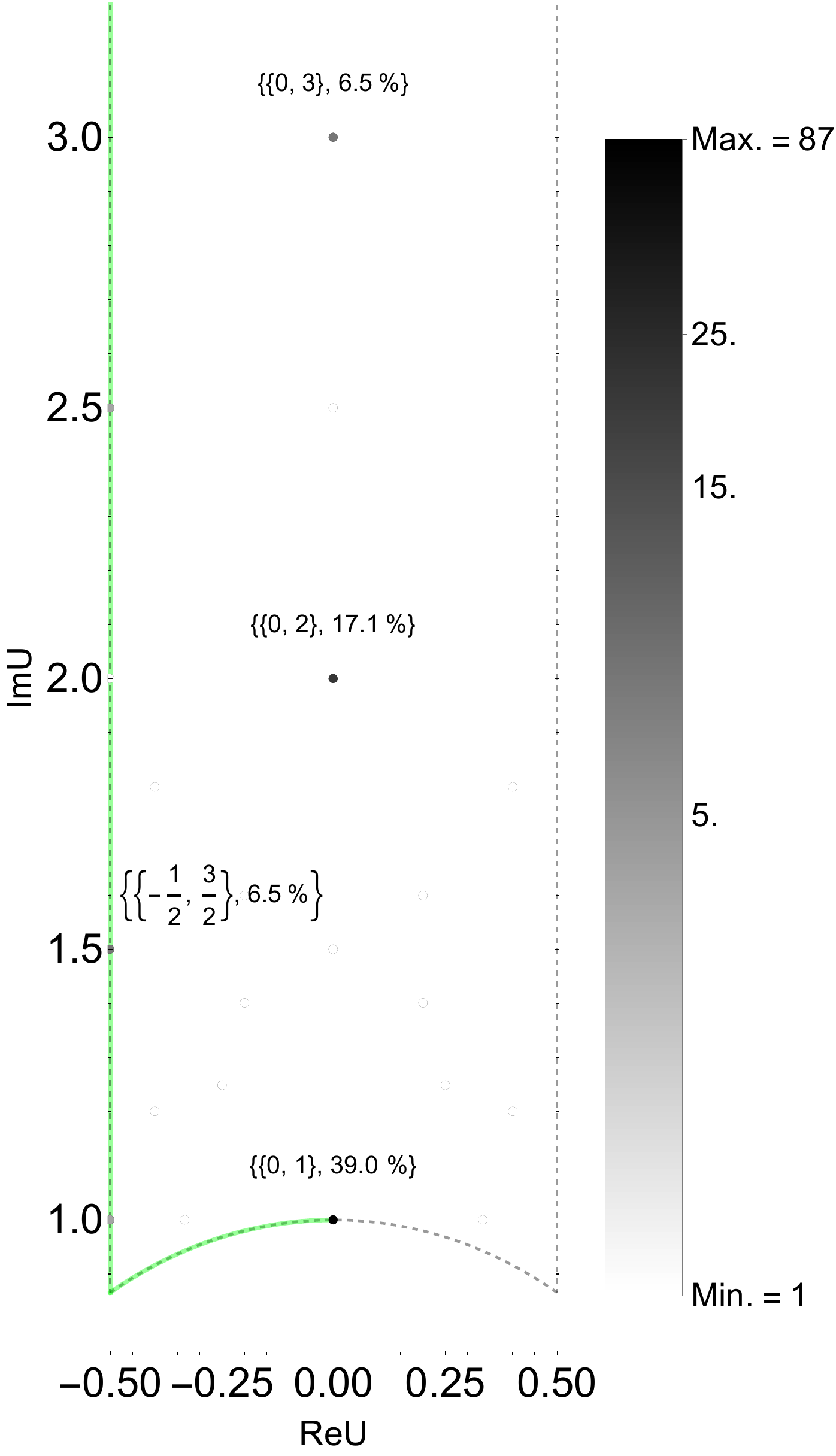}
\caption{The distribution of physically-distinct solutions on $\langle U \rangle$-plane with the figure cropped to ${\rm Im} U \leq 3.25$ being illustrated. Here, $N_{\rm flux}^{\rm max}$ is  $32 \times 20$. The colors are determined by the associated number of degeneracies among the physically-distinct solutions. Here, we additionally listed the positions and ratios of the points whose ratio $> 5.0\%$.}
\label{fig:vevU_enlarge_count_Z2Z4}
\end{figure}

Next, we illustrate the distributions and degeneracies which are projected on the $U$-plane. The result is summarized in Table \ref{tab:RatiovsU-NMAX20-Z2Z4} and Fig. \ref{fig:vevU_enlarge_count_Z2Z4}.
The most favorable point is $U = i$, which is the usual elliptic point of $SL(2, \mathbb{Z})$.
Since $U = \omega$ does not exist on the $\mathbb{Z}_2 \times \mathbb{Z}_4$ orientifold and exact solutions on $U = \infty$ are not achievable in this framework, one can consider that $U = i$ is exclusive among the fixed points.

\begin{figure}[H]
 \centering
  \includegraphics[width=11cm]{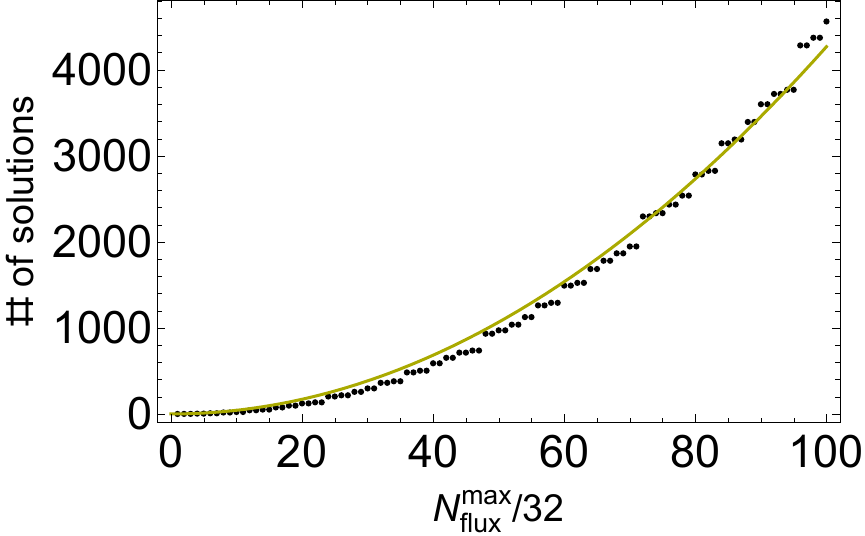}
  \caption{The number (\#) of the whole solutions with $N_{\rm flux} \leq N_{\rm flux}^{\rm max}$. The yellow line is a fitted curve with $\# = 0.427 (N_{\rm flux}^{\rm max}/32)^2.$} 
\label{fig:Nfluxvsnumsolutions-Z2Z4}
\end{figure}

We also vary $N_{\rm flux}^{\rm max}$. 
As in the previous case, small $N_{\rm flux}^{\rm max}$ is quite restrictive, and the number of solutions becomes very small.
The number of solutions for various $N_{\rm flux}^{\rm max}$ is summarized in Table \ref{tab:Number-Z2Z4} and Fig. \ref{fig:Nfluxvsnumsolutions-Z2Z4}.
The number of solutions is well approximated by a quadratic function of $N_{\rm flux}$, as in the previous case.
The minimum value of $N_{\rm flux}$ with which a SUSY Minkowski solution exists is 64.
There, $U = i$ is the only allowed VEV, but $S = i$ thus it is not considered to be in the weak-coupling regime.
We will show correlations between $U$ and $S$ later.
\begin{table}[H]
\centering
\begin{tabular}{|c|cccc|}
\hline
Ratio                 &$39.0\%$&$17.1\%$&$6.50\%$&$3.25\%$\\ \hline
\multirow{4}{*}{$U$} &$i$&$2i$&$3i$&$4i$ \\
                      &  &  &$-\frac{1}{2} + \frac{3}{2}i$&$-\frac{1}{2} + i$\\
                      &  &  &  &$5i$\\
                      &  &  &  &$-\frac{1}{2} + \frac{5}{2}i$\\\hline
\end{tabular}
\caption{The ratios (probabilities) which are associated with each VEV on $U$-plane. $N^{\rm max}_{\rm flux} = 32 \times 20$, and we listed the ratios up to the fourth largest.} 
\label{tab:RatiovsU-NMAX20-Z2Z4}
\end{table}

\begin{table}[H]
\centering
\begin{tabular}{|c|c c c c c c c|} \hline
$N_{\rm flux}^{\rm max} $ & $32 \times 2$  & $32 \times 4$ & $32 \times 10$ & $32 \times 20$ & $32 \times 40$ & $232 \times 80$ & $32 \times 100$ \\ \hline
$\#$ of solutions & 1 & 4 & 23 & 123 & 589 & 2784 & 4562\\ \hline
\end{tabular}
\caption{Number of solutions for each bound $N_{\rm flux}^{\rm max}$.} 
\label{tab:Number-Z2Z4}
\end{table}
\begin{figure}[H]
 \centering
  \includegraphics[height=8cm]{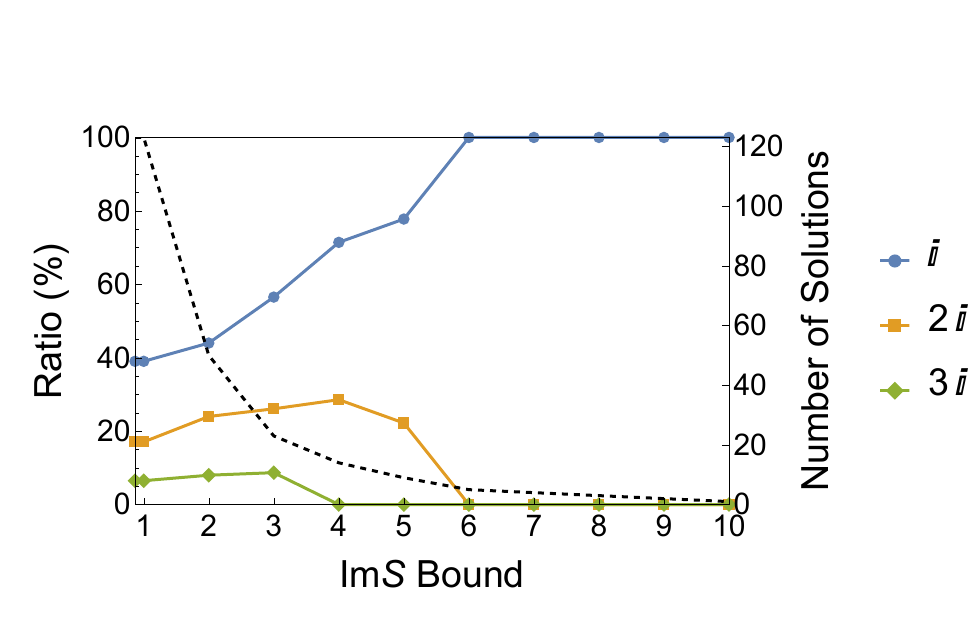}
  \caption{The solid lines present the ratios of the dominant (>5\%) VEVs of $U = i, 2i$ and $3i$ with bounded ${\rm Im}S_{\rm bound}$ and $N_{\rm flux}^{\rm max} = 32 \times 20$. The lines start with ${\rm Im}S_{\rm bound} = \frac{\sqrt{3}}{2}$ (all region). The dashed line presents the number of solutions above certain ${\rm Im}S_{\rm bound}$.} 
\label{fig:Ratio-of-Dominant-VEVs-Bounded-Z2Z4}
\end{figure}
As in the $\mathbb{Z}_{6 - \rm{\greekii}}$ case, we focus on some particular regions on $S$-plane, and study correlations between $U$ and $S$.
Due to the interchanging duality, the VEV on $U = i$ tends to the solutions not in the weak-coupling regime.
We summarize dominant VEVs of $U$ with some bounds on ${\rm Im}S$ in Fig. \ref{fig:Ratio-of-Dominant-VEVs-Bounded-Z2Z4}. One can see if stabilizations in the weak-coupling regime are naturally realized, one obtains $U = i$ exclusively. 
However, in general, it is not natural because the number of solutions decreases as ${\rm Im}S_{\rm bound}$ increases.
Considering some specific values of $S$ leads to the same discussion.
From a perspective of modular flavor symmetric models, stabilizations around $U = i$ is of particular interest.
We leave finding a mechanism to achieve such stabilizations for future work. 
In principle, we can only treat the flux quanta as free parameters at this stage.
We may have to consider different orientifolds or additional fields that enter the scalar potential to make $|D|^2$ in Eq. (\ref{eq:NfluxandVEVs}) small naturally.

\begin{figure}[H]
\centering
\includegraphics[height = 12cm]{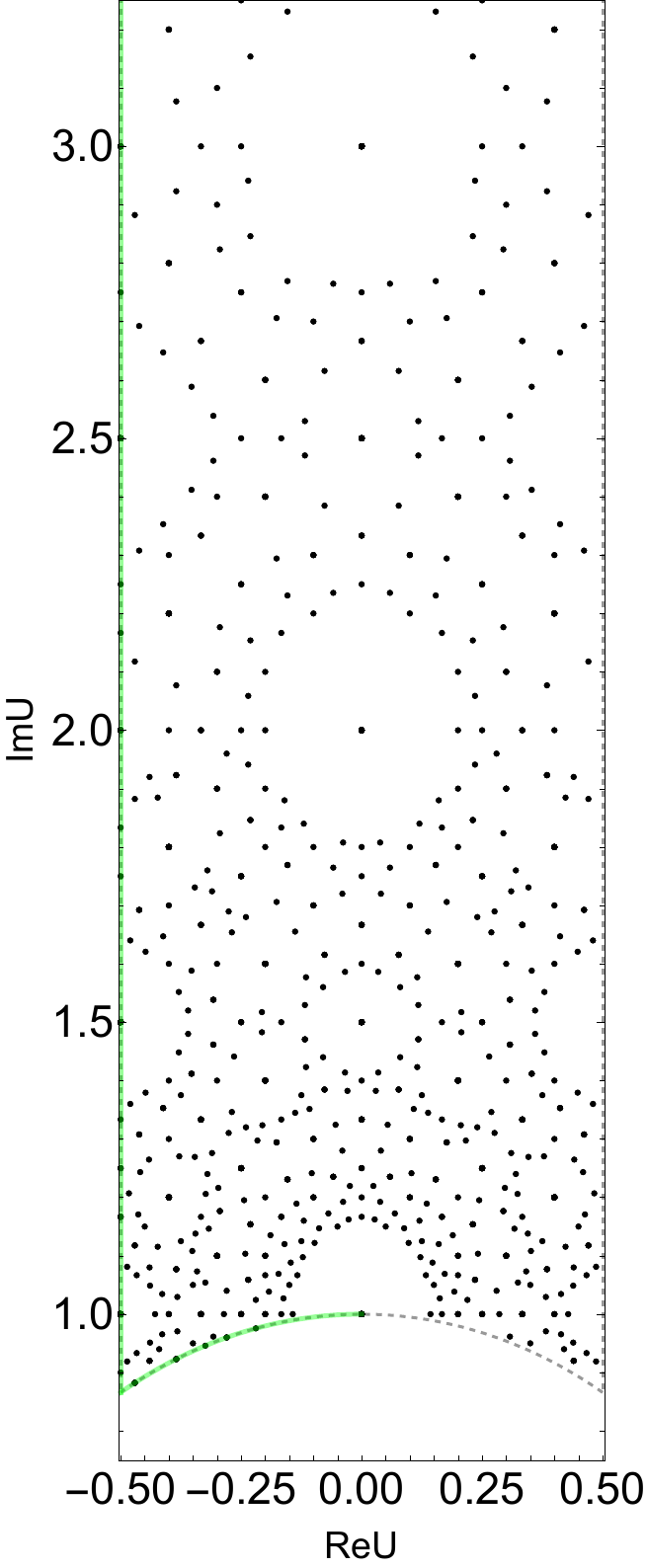}
\caption{The distribution of $\langle U \rangle$ with $N_{\rm flux}^{\rm max} = 32 \times 100$. The plot region is cropped into ${\rm Im}U \leq 3.25$. Some circles are observed, and their centers have high degeneracies. $U = i$ is one of them.}
\label{fig:plot-vevU-100-Z2Z4}
\end{figure}

Let us mention that the void structure also exists in the present case.
We illustrate the distribution of $U$ with $N_{\rm flux}^{\rm max} = 32 \times 100$ in Fig. \ref{fig:plot-vevU-100-Z2Z4}.
Some circles that center at VEVs with high degeneracies are observed.
In this case, $U = i$, which is an elliptic point, is one of the centers.
Together with the previous case, we will report some analysis on this point in the upcoming work.
\section{Conclusions and discussions}
\label{sec:con}
We investigated the flux Landscape of $T^6/{\mathbb{Z}_{6 - \rm{\greekii}}}$ orientifold with ${\rm SU}(6) \times {\rm SU}(2)$ lattice and $T^6/{\mathbb{Z}_2 \times \mathbb{Z}_4}$ orientifold with $({\rm SU}(2))^2 \times ({\rm SO}(5))^2$ lattice respectively, in the manner of our previous studies \cite{Ishiguro:2020tmo} on $T^6/{\mathbb{Z}_2 \times \mathbb{Z}'_2}$.
On the $T^6/{\mathbb{Z}_{6 - \rm{\greekii}}}$ orientifold, the complex-structure modulus does not enjoy $SL(2, \mathbb{Z})$, but its congruence subgroup $\Gamma_0(3)$.
Studying degrees of freedom in reparametrizations of the modulus carefully, it turned out that the actual duality group in the complex-structure modulus space is an outer semi-direct product $\overline{\Gamma}_0(3) \rtimes S_{(3)}$ with an outer automorphism group $S_{(3)}$.
$S_{(3)}$ is a subgroup of $SL(2, \mathbb{R})$, and it is also classified as an elliptic transformation as $S$ is.
It is also checked that this group is the most general duality which is consistent with the symplectic basis in the third (co-)homology.
Breaking into subgroups implies that the fundamental region enlarges, and we explicitly obtain the finite Landscape on such an enlarged region.
Usually, the elliptic points of $SL(2, \mathbb{Z})$ are favored in Landscape, as one finds in the $T^6/{\mathbb{Z}_2 \times \mathbb{Z}'_2}$ case.
However, on the $\mathbb{Z}_{6 - {\rm \greekii}}$ Landscape, the most favored point is found not to be the elliptic points of $SL(2, \mathbb{Z})$.
Instead, it is at the elliptic point of $S_{(3)}$, where $\mathbb{Z}_2$ enhanced symmetry realizes.

In the $T^6/{\mathbb{Z}_2 \times \mathbb{Z}'_2}$ case, a strong correlation between the VEVs of $U$ and $S$ is observed, and it enables the specific VEVs to be exclusive by stabilizing $S$ appropriately.
However, in the present case, we do not find such a correlation. 
It is instead observed that when we move to weakly-coupling limit, the probability associated with the $S_{(3)}$ elliptic point becomes higher.
In this sense, we can conclude that the $S_{(3)}$ elliptic point is the most favored in the most reliable situation.
Furthermore, we investigated the flux Landscape of the $\mathbb{Z}_2 \times \mathbb{Z}_4$ orientifold. 
Opposite to the $\mathbb{Z}_{6 - {\rm \greekii}}$ case, $SL(2, \mathbb{Z})$ survives in the complex-structure modulus space, and it is the most general duality by which we mod out the Landscape.
Then, as in the $T^6/{\mathbb{Z}_2 \times \mathbb{Z}'_2}$ case, the elliptic point of $SL(2, \mathbb{Z})$ is the most favored one.
Due to the explicit form of the period vector, $U = \omega$ is not allowed, and $U = i$ is exclusively realized.
As in the previous case \cite{Ishiguro:2020tmo}, we can choose the elliptic points of $SL(2, \mathbb{Z})$ as VEVs of the complex-structure modulus in the most reliable situation. 

In both cases, the resultant Landscape structures are different from that of the $T^6/{\mathbb{Z}_2 \times \mathbb{Z}'_2}$ case.
From a phenomenological perspective, if one assumes that the modular flavor symmetric models can be embedded in our framework, it suggests that candidate backgrounds are not chosen arbitrarily.
For example, we may consider models with $U \sim i$ are more likely embedded into $T^6/{\mathbb{Z}_2 \times \mathbb{Z}_4}$ rather than $T^6/{\mathbb{Z}_2 \times \mathbb{Z}'_2}$ and $T^6/{\mathbb{Z}_{6 - {\rm \greekii}}}$ as we expected such the phenomenon in the Introduction \ref{sec:intro}.
It is also interesting that $T^6/{\mathbb{Z}_{6 - {\rm \greekii}}}$ leads $U = \frac{i}{\sqrt{3}}$ unlike the usual case.
We expect two directions in our future work:
\begin{itemize}
    \item Systematic classification of dualities in Landscape\\
    These differences in favored elliptic points seem to have originated from the geometric structure. 
    Especially, the dualities which are consistent with the symplectic basis are determined before the flux quanta are turned on, and thus the orbifold group and lattice should determine the structures of Landscape well.
    Since the geometries of toroidal orientifolds are already systematically classified, we will report on the classification of dualities in future work.
    \item Voids and distributions of solutions\\
    Besides the tadpole cancellation condition and dualities, an exact distribution of degeneracy is also important when doing phenomenology.
    The void structure is also observed in the Landscape of $\mathbb{Z}_{6 - {\rm \greekii}}$ where $U = \frac{i}{\sqrt{3}}$ is a center of large circle. 
    Without no quantum corrections, we cannot stabilize moduli inside the voids except for their centers.
    The voids seem to be centered at points with high degeneracy, and their radii shrink as $N_{\rm flux}$ increases.
    We will revisit the origin of the void structure and discuss the correlation between its dependencies on the degeneracy and the tadpole cancellation condition in the upcoming paper. 
\end{itemize}

\appendix

\section{Finite physically-distinct vacua on $T^6/\mathbb{Z}_{6-{\rm \greekii}}$}
\label{app:algorhithm}

In this section, we present a procedure to obtain all physically-distinct solutions of the F-term equations (\ref{eq:ftermeq}) with 
\begin{align}
    N_{\rm flux} \leq N^{\rm max}_{\rm flux}.
\end{align}
$N^{\rm max}_{\rm flux}$ is the maximum value of $N_{\rm flux}$ which we assume that our solutions\footnote{For a while, we call solutions of the F-term equations "solutions". Though it is physically unacceptable in type IIB effective theory, they may have a large $N_{\rm flux}$ and violate the tadpole cancellation condition or induce some backreactions. For our purpose, we simply ignore the problems.} have.
Two elements which are independent under the identification (\ref{eq:identificationmanner}) are called physically-distinct. As a result of the procedure we will show, the Landscape turns out to be finite.

The physically-distinct Landscape can be obtained by fixing the underlying $PSL(2, \mathbb{Z})_S \times \overline{\Gamma}_0(3)_{U} \rtimes_{\varphi(S_{(3})} \mathbb{Z}_2$ transformations, following the method in \cite{Kachru:2002he} for $PSL(2, \mathbb{Z})_S \times PSL(2, \mathbb{Z})_U$ case.
We summarize the T-transformation of the three-form fluxes in Section \ref{sec:dualities} in Table \ref{tab:T-transformation}. 
The $T$-transformation in $\Gamma_0(3)_U$ can be fixed by choosing
\begin{align}
    \frac{1}{2} d_0 &= 0, 1, \dots, \frac{1}{2}|c^0| -1, \\
    \frac{1}{4} d_1 &= 0, 1, \dots, \frac{1}{4}|2c^1| - 1 \quad (c^0 = 0),
\end{align}
where the second choice is for $c^0 = 0$ case. Later, we will choose $c^0 > 0$ or $c^0 = 0, c^1 > 0$ without loss of generality. Note that $c^0 = c^1 = 0$ is forbidden due to Eq. (\ref{eq:ftermsolution}). 
Similarly, the $T$-transformation in $SL(2, \mathbb{Z})_S$ is fixed by choosing
\begin{align}
    \frac{1}{6}a^0 &= 0, 1, \dots, \frac{1}{6}|c^0| - 1, \\
    \frac{1}{2}a^1 &= 0, 1, \dots, \frac{1}{2}|c^1| - 1 \quad (c^0 = 0).
\end{align}

\begin{table}[h]
    \centering
    \begin{tabular}{ccc|c}
    fluxes & $T^q \in SL(2, \mathbb{Z})_S $ &   $T^q \in \Gamma_0(3)_U $  & $m \mathbb{Z}$\\ 
    \hline
    $a^0$ & $a^0 + q c^0$ & $a^0$ & 6 \\
    $a^1$ & $a^1 + q c^1$ & $a^1$ & 2 \\
    $b_0$ & $b_0 + q d_0$ & $b_0 - q a^0$ & 2 \\
    $b_1$ & $b_1 + q d_1$ & $b_1 - q \left( \frac{2}{3} a^0 + 2 a^1 \right)$ & 4 \\
    $c^0$ & $c^0$ & $c^0$ & 6 \\
    $c^1$ & $c^1$ & $c^1$ & 2 \\
    $d_0$ & $d_0$ & $d_0 - q c^0$ & 2 \\
    $d_1$ & $d_1$ & $d_1 - q \left( \frac{2}{3} c^0 + 2 c^1 \right)$ & 4
    \end{tabular}
    \caption{Fluxes under the T-transformation. The fluxes are quantized to be $m \mathbb{Z}$.}
    \label{tab:T-transformation}
\end{table}

\begin{figure}[H]
  \centering
  \includegraphics[width=0.5\textwidth]{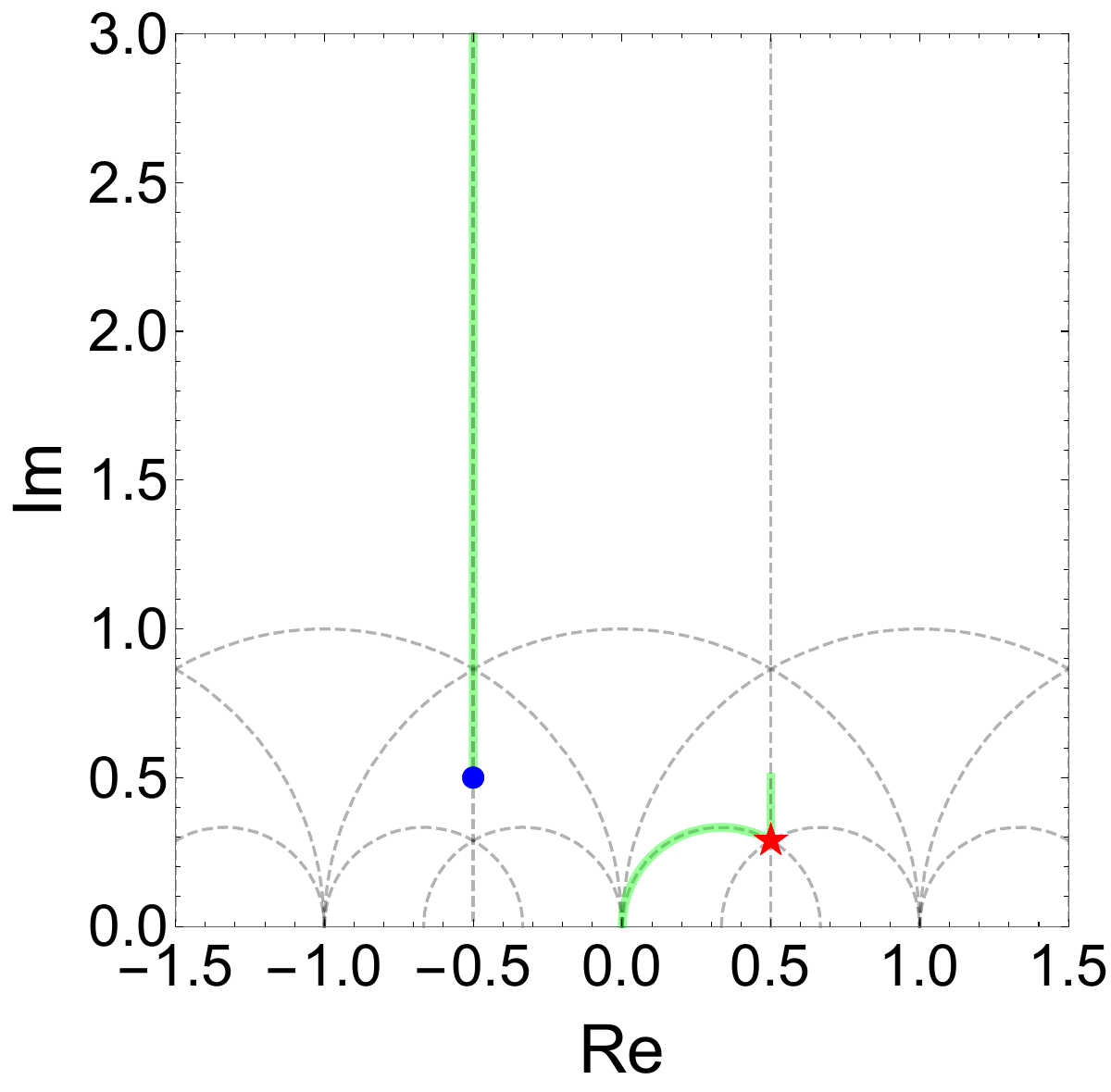}
  \caption{The fundamental region of $\Gamma_0(3)$ (more precisely, $\overline{\Gamma}_0(3)$). The boundary is indicated as the green curves. The blue disk represents $-\frac{1}{2} + i \frac{1}{2}$, which is the image of $i$ under $S$. The red star represents $\frac{1}{2} + i \frac{\sqrt{3}}{6}$, which is the fixed point under $S'$. Since $S'^3 = -1 \simeq 1$ in $\overline{\Gamma}_0(3)$, there is a $\mathbb{Z}_3$ enhanced symmetry.$\frac{1}{2} + i \frac{1}{2}$ is not included in the fundamental region. The green arc is given by $|U - \frac{1}{3}| = \frac{1}{3}$ with $0 < {\rm Re} U \leq \frac{1}{2}$.}
  \label{fig:fundamentalregionG03}
\end{figure}

Since there always exist $S$-transformations of $SL(2, \mathbb{Z})_{S}$ on the region considered, we can simply ignore the flux region which is connected with the fundamental region of $S$ by the single $S$-transformation. 
We choose the fundamental region $\mathcal{F}_{SL(2, \mathbb{Z})_S}$ as
\begin{align}
{\cal F}_{SL(2, \mathbb{Z})_S} = \left\{S \middle|-\frac{1}{2} \leq {\rm Re}S \leq 0, |S|^2 \geq 1 \right\} \bigcup \left\{S \middle| 0 < {\rm Re}S < \frac{1}{2}, |S|^2 > 1 \right\}.
\end{align}
Furthermore, there also exists the fundamental region $\mathcal{F}_{\Gamma_0(3)_U}$ for $\Gamma_0(3)$. At present the $S$-transformation is broken, and there are four\footnote{More generally, it is known in the mathematical literature that the index of $\Gamma_0(n)$ is given by $[PSL(2, \mathbb{Z}) : \Gamma_0(n)] = n \prod_{p|n} \left( 1 + \frac{1}{p}\right)$, where $p$ is prime (for more details, see \cite{diamond2005first}).} independent regions under $\mathcal{F}_{\Gamma_0(3)_U}$. We can obtain those regions by transforming the $SL(2, Z)$ fundamental region via $1, S, ST, ST^{-1}$, and $\mathcal{F}_{\Gamma_0(3)_U}$ is their union:
\begin{align}
    \mathcal{F}_{\Gamma_0(3)_U} = {\cal F}_{SL(2, \mathbb{Z})_U} \bigcup S {\cal F}_{SL(2, \mathbb{Z})_U} \backslash \{-\bar{\omega}\} \bigcup (ST) {\cal F}_{SL(2, \mathbb{Z})_U} \bigcup (ST^{-1}) {\cal F}_{SL(2, \mathbb{Z})_U},
\end{align}
where we have to exclude the point $U = -\bar{\omega} \equiv \frac{1}{2} + i \frac{\sqrt{3}}{2}$ because $T \in \Gamma_0(3)$ and $S\omega = T \omega = - \bar{\omega}$. 
We illustrate the region in Fig. \ref{fig:fundamentalregionG03}. The cusps are $0$ and $i \infty$, and ${\rm Im} U \ll 1$ is necessarily included.

\begin{figure}[H]
  \begin{minipage}[b]{0.45\linewidth}
 \centering
  \includegraphics[height=12.0cm]{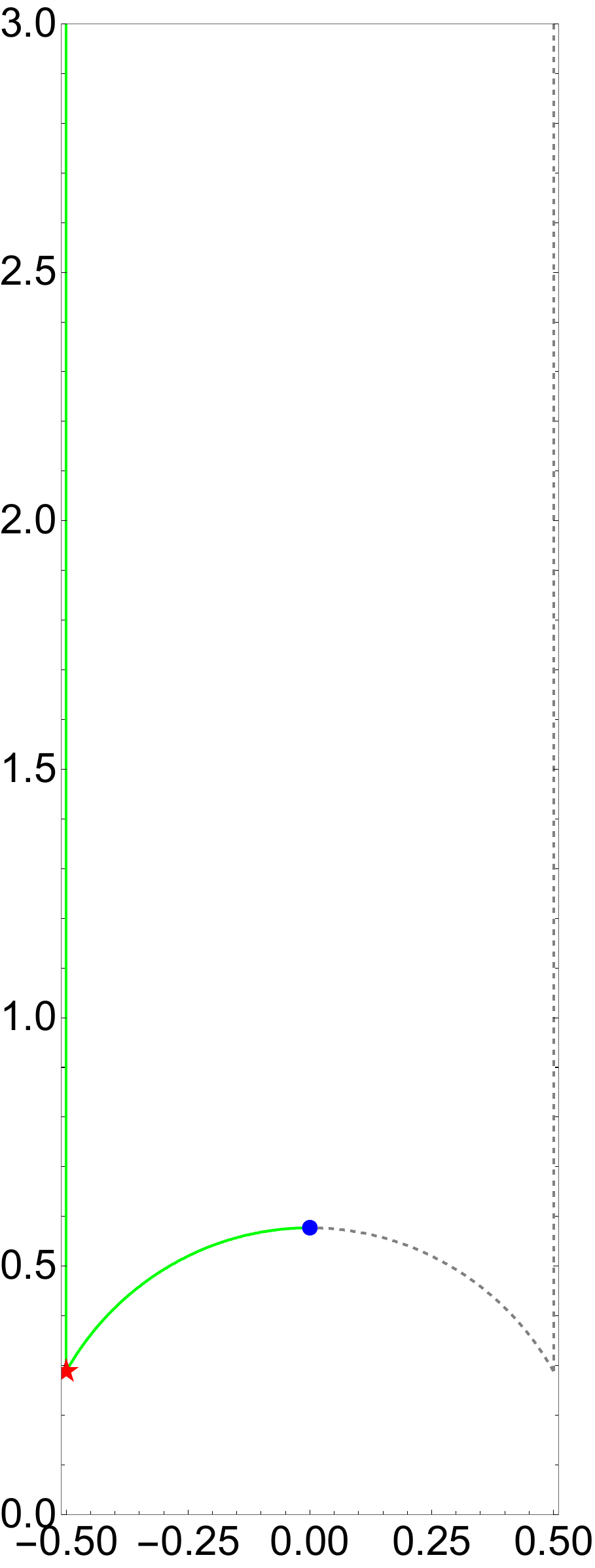}
  \end{minipage}
  \begin{minipage}[b]{0.45\linewidth}
 \centering
  \includegraphics[height=12.0cm]{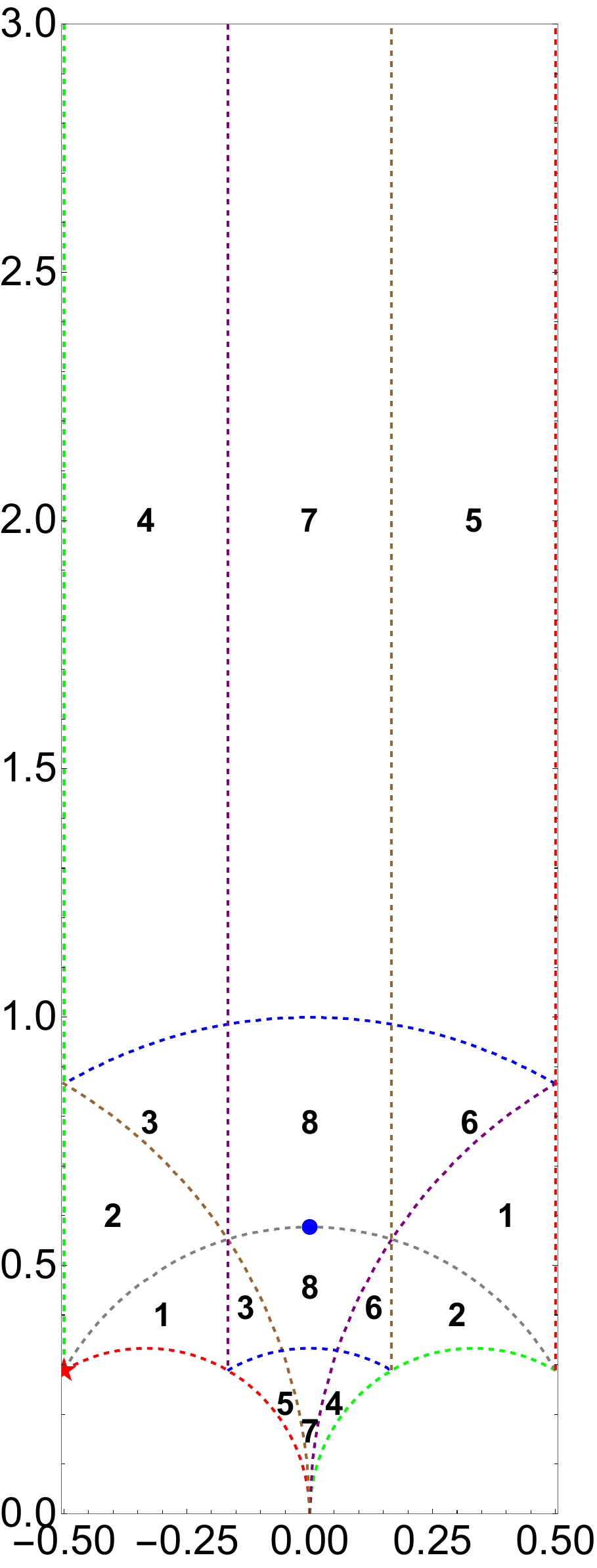}
  \end{minipage}
  \caption{The boundary which we chose for the fundamental region of the total group, $\overline{\Gamma}_0(3) \rtimes \mathbb{Z}_2$. The blue disk represents $\frac{i}{\sqrt{3}}$, which is the fixed point under $U \rightarrow - \frac{1}{3U}$. The red star represents $-\frac{1}{2} + i \frac{\sqrt{3}}{6}$, which is the fixed point under $S'$. There are $\mathbb{Z}_2$ and $\mathbb{Z}_3$ enhanced symmetries on those points, respectively. We also illustrate the correspondences between regions under $U \rightarrow -\frac{1}{3U}$ in the right panel. Two regions with identical numbers written on them are equivalent under the identification. Here, the colors of the disk and the star have no meaning, but two (dashed) lines with the same color correspond to each other.}
    \label{fig:TotalFR}
\end{figure}

As mentioned in \ref{sec:dualities} and will be proved in \ref{app:generalreparametrization}, there is also the scaling duality, $U \rightarrow - \frac{1}{3U}$. 
Let us mod out the fundamental region by this duality.
Circles which are fixed under $U \rightarrow - \frac{1}{3U}$ are given by two one-parameter families\footnote{The first one is fixed because of $a + d = 0$. There is only one family of fixed circles under general elliptic transformations (for more details, see \cite{ford1929automorphic}).};
\begin{align}
    ({\rm Re} U - \alpha)^2 + {\rm Im} U^2 &= \frac{1}{3} + \alpha^2,\\
    {\rm Re}U^2 + ({\rm Im} U - \alpha)^2 &= -\frac{1}{3} + \alpha^2,
\end{align}
with $\alpha$ being an arbitrary real number. The cycle with $\alpha = 0$, which is given by $|\sqrt{3}U| = 1$,  is called the isometric circle since the length of a line element is multiplied by $|c U  + d|^2$ under a linear fractional transformation.
One can choose the isometric circle as a boundary for the resultant fundamental region.
As a result, we can exclude ${\rm Im} U \simeq 0$ region. 
We illustrate the resultant region in Fig. \ref{fig:TotalFR}.

The imaginary parts of $S$ and $U$ are given by
\begin{align}
    {\rm Im} S = - \frac{{\rm Im} ( C \overline{D} )}{|D|^2}, \quad {\rm Im} U = - \frac{{\rm Im} ( B \overline{D} )}{|D|^2}
    \label{eq:imaginarypartVEVsinflux}
\end{align}
in terms of $A, B, C$ and $D$ which is defined in Eq. (\ref{eq:ABCDinrealfluxes}).
It is easier to consider the region
\begin{align}
    {\rm Im} S \geq \frac{\sqrt{3}}{2}
    \label{eq:regionfixingS-transformation}
\end{align}
than fixing $S$-transformation strictly at this level. The reason why it works is the following. Assume that we have all $T$-independent solutions on the $S$-plane with ${\rm Im}S \geq \frac{\sqrt{3}}{2}$. If there is a solution with ${\rm Im}S < \frac{\sqrt{3}}{2}$ which we do not consider, it can be mapped to a solution with ${\rm Im}S \geq \frac{\sqrt{3}}{2}$ by a possible combination of $S$-transformations and $T$-transformations. Since we have all $T$-independent solutions with ${\rm Im}S \geq \frac{\sqrt{3}}{2}$, it is not a new physically-distinct solution, and our set of solutions includes it already. As for a solution with ${\rm Im}S \geq \frac{\sqrt{3}}{2}$ and can be mapped to a solution with $|S| < 1, -\frac{1}{2} \leq {\rm Re}S < \frac{1}{2}$, we have the corresponding solution which is given by $S$-transformation on it. We can simply drop solutions which are inside the unit circle.   

On the other hand, unless taking account of $U \rightarrow -\frac{1}{3U}$, there is no restriction on ${\rm Im}U$ itself, and it obviously leads to necessary infinite solutions even if $T$-transformation is fixed since there are an infinite number of regions which are $\overline{\Gamma}_0(3)$ images of the fundamental region ${\cal F}_{SL(2, \mathbb{Z})_U}$ within $|{\rm Re}U| \leq \frac{1}{2}$.
However, owing to the scaling duality, we can restrict ${\rm Im} U$ to be within 
\begin{align}
    {\rm Im}U \geq \frac{1}{2\sqrt{3}}.
\end{align}
Thus, if the number of solutions inside ${\cal F}_{SL(2, \mathbb{Z})_U}$ is finite, we certainly find finite solutions.

Let us consider only the region ${\cal F}$ at first. 
It implies the positive integer $|D|^2 \leq N_{\rm flux}$ via Eq. (\ref{eq:NfluxandVEVs}).
Then $c^0$ and $c^1$ are respectively multiples of 6 and 2 within
\begin{align}
   6 \left\lceil - \frac{1}{6} \sqrt{N_{\rm flux}} \right\rceil  \leq &c^0 \leq 6 \left\lfloor \frac{1}{6} \sqrt{N_{\rm flux}} \right\rfloor , \\
    \begin{split}
         2 \left\lceil -\frac{1}{2} \frac{\sqrt{3}\sqrt{N_{\rm flux} - (c^0)^2} + c^0}{3} \right\rceil  \leq & c^1  \leq 2 \left\lfloor \frac{1}{2} \frac{\sqrt{3}\sqrt{N_{\rm flux} - (c^0)^2} - c^0}{3} \right\rfloor,
    \end{split}
\end{align}
where $c^0$ in the second and third lines should be understood a fixed value which is determined in the first line.

One can see that $-\frac{1}{\sqrt{3}} {\rm Im}(C \overline{D})$ and  $-2\sqrt{3} {\rm Im}(B \overline{D})$ are integers and should be positive in the region ${\cal F}$.
From
\begin{align}
 \frac{3 |D|^2 N_{\rm flux}}{6} = \left( - \frac{1}{\sqrt{3}} {\rm Im} (C \overline{D}) \right) \left( - 2 \sqrt{3} {\rm Im} (B \overline{D}) \right),
 \label{eq:Nfluxintegerdivisor}
\end{align}
$B$ is restricted as
\begin{align}
    3 |D|^2 \leq {\rm Im}D (2\sqrt{3} {\rm Re}B) - {\rm Im}B (2\sqrt{3} {\rm Re}D) \leq 3|D|^2 \frac{N_{\rm flux}}{6}.
\end{align}
After that $B$ is determined, $C$ can be fixed via Eq. (\ref{eq:Nfluxintegerdivisor}). It is needed to check that $-2\sqrt{3} {\rm Im}(B \overline{D})$ is in the divisors of $\frac{3|D|^2 N_{\rm flux}}{6}$, which is an integer due to the quantization condition in Table \ref{tab:T-transformation}. It is also required that ${\rm Im}S$ reside in the region (\ref{eq:regionfixingS-transformation}). The result is as follows;
\begin{itemize}
    \item $c^0 > 0$ case\\
    $d_0$ is already fixed and
    \begin{align}
        4 \left \lceil \frac{1}{4} \frac{|D|^2 + {\rm Im}B 2 \sqrt{3} {\rm Re}D}{3 {\rm Im}D} \right \rceil \leq d_1 \leq 4 \left \lfloor \frac{1}{4} \frac{3|D|^2 \frac{N_{\rm flux}}{6} + {\rm Im}B 2\sqrt{3} {\rm Re}D}{3 {\rm Im}D} \right \rfloor.
    \end{align}
    Since $a_0$ is already fixed, we obtain
    \begin{align}
        a^1 = - \frac{1}{c^0} \left( \frac{3 |D|^2 N_{\rm flux}}{6} \frac{1}{-2\sqrt{3}{\rm Im}(B \overline{D})}  - a^0 c^1 \right),
    \end{align}
    where we have to check that $a^1 \in 2 \mathbb{Z}$. 

    \item $c^0 = 0$, $c^1 >0$ case\\
    $d_1$ is already fixed and
    \begin{align}
        2 \left \lceil -\frac{1}{2} \frac{3 |D|^2 \frac{N_{\rm flux}}{6} - 2 \sqrt{3} {\rm Im}D {\rm Re}B}{2\sqrt{3}{\rm Re}D} \right \rceil \leq d_0 \leq 2 \left \lfloor \frac{1}{2} \frac{- |D|^2 + 2\sqrt{3} {\rm Im} D {\rm Re}B }{2 \sqrt{3} {\rm Re}D}  \right \rfloor.
    \end{align}
    $a^1$ is already fixed, and we obtain
    \begin{align}
        a^0 = \frac{1}{c^1} \left( \frac{|D|^2 N_{\rm flux}}{6} \frac{1}{-2\sqrt{3} {\rm Im}(B \overline{D}) } + a^1 c^0 \right),
    \end{align}
    where we have to check that $a^0 \in 6 \mathbb{Z}$.
\end{itemize}
Here, we make $c^0$ (or, $c^1$) positive to cancel the degree of freedom in the flux sign. 
In the end, we can fix $A$ via $AD - BC = 0$ in Eq. (\ref{eq:ftermsolution}):
\begin{align}
    b_0 &= - {\rm Im} \frac{CB}{D} = \frac{-1}{|D|^2} \left( {\rm Im}B {\rm Im}C {\rm Im}D - {\rm Re}B {\rm Re} C {\rm Im} D + {\rm Re}B {\rm Im}C {\rm Re}D + {\rm Im}B {\rm Re}C {\rm Re}D\right), \\
    b_1 &= - \frac{2}{\sqrt{3}} {\rm Re} \frac{CB}{D} 
    \nonumber\\
    &= \frac{-2}{\sqrt{3}|D|^2} \left( {\rm Re}B {\rm Im}C {\rm Im}D + {\rm Im}B {\rm Re}C {\rm Im}D - {\rm Im}B {\rm Im}C {\rm Re}D + {\rm Re}B {\rm Re}C {\rm Re}D \right),
\end{align} 
where we have to check that $b^0 \in 2 \mathbb{Z}$ and $b^1 \in 4 \mathbb{Z}$.

We have obtained the whole solutions with given $N_{\rm flux}$, which are independent under $T \in \Gamma_0(3)$. Solutions in the region $|U|<\frac{1}{3}$ are inequivalent with solutions in the region $|U| \geq \frac{1}{3} \land |U| \leq \frac{1}{2}$ unless one consider $U \rightarrow - \frac{1}{3U}$.
Modding out the Landscape by the scaling duality, we can simply discard solutions inside of the isometric circle. 

Let us note that one can adopt another procedure, which is inefficient for large $N_{\rm flux}$, to find whole solutions for reference. First, one obtain whole solutions inside $\mathcal{F}_{SL(2, \mathbb{Z})_U}$ with given $N_{\rm flux}$. This can be done in the manner of the procedure which we adopted above. As mentioned, regarding the regions $S \mathcal{F}, ST\mathcal{F}$ and $ST^{-1}\mathcal{F}$, the procedure for the $SL(2, \mathbb{Z})$ case should fail since there is no constraint on ${\rm Im}U$ itself. Nevertheless, the finiteness of the solutions is ensured. For arbitrary fluxes, the $S$-transformation on (fluxes, $U$) may or may not be broken. Indeed, for fluxes where $S$-transformation survives, the corresponding solutions must reside in $\mathcal{F}$, and we have already found it. Then we should only consider the broken $S$-transformation. Under the $S$-transformation, where we virtually impose although it is broken, the fluxes become fractional ones. However, we can always associate them with solutions in $\mathcal{F}$ with $k^2 N_{\rm flux}$, where $k (F_3, H_3)$ have integer coefficients. Since we proved the number of physically-distinct solutions in $\cal{F}$ with a finite $N_{\rm flux}$ is finite, we conclude the number of solutions in ${S \mathcal{F}}$ should be finite. We can argue analogously for the other two regions.
The explicit procedure to obtain solutions in $S{\cal F}$ is as follows. Let us assume that all solutions in ${\cal F}$ are given for arbitrary $N_{\rm flux}$ by the procedure above with possible $T$-transformations. 
The $S$-transformation, which is broken but can be imposed virtually, should map solutions in $S {\cal F}$ to corresponding ones in ${\cal F}$.   
Recalling the action of $S$ on the fluxes \ref{eq:fluxtrf-S}, we observe that the rescaling $(F, H) \rightarrow 3(F, H)$ after $S$-transformation cancels out the denominator to ensure the quantization condition. 
Since this composed map is bijective, we can trace it back and obtain solutions in $S {\cal F}$. 
Concretely, for each $N_{\rm flux}$, we search the solutions with $9N_{\rm flux}$ and whose fluxes are quantized as
\begin{align}
    (b_0, d_0) \in 18 \mathbb{Z},\quad (b_1, d_1) \in 12 \mathbb{Z},\quad (a_0, c_0) \in 6 \mathbb{Z},\quad  (a_1, c_1) \in 2 \mathbb{Z}.
    \label{eq:quantizationconditionofSFwith9Nflux}
\end{align}
For those solutions, there may exist the corresponding solutions in $S {\cal F}$. Since the condition (\ref{eq:quantizationconditionofSFwith9Nflux}) is not a sufficient but necessary condition, we have to check that fluxes are quantized as Eq. (\ref{eq:fluxquantizationcondition}) after dividing them by three.
Other solutions in $ST{\cal F}$ and $ST^{-1}{\cal F}$ are obtained via the same procedures. In the real implementation, this algorithm is quite inefficient, since the flux ranges will broaden as $N_{\rm flux}$ increases, as the maximum values of $c^0, c^1$ scales roughly $\sim \sqrt{N_{\rm flux}}$ and that of $d_1$ does $\sim N_{\rm flux}^{3/2}$.

Either procedure may be used, and we obtain the whole physically-distinct Landscape. 
Since the number of solutions is finite, counting them is meaningful. Keep in mind that when one counts the number of solutions, one should divide the number of solutions on the elliptic points of $SL(2, \mathbb{Z})$ and $\Gamma_0(3)$ by corresponding factors. 
There are two elliptic points for $SL(2, \mathbb{Z})$, and two elliptic points for $\overline{\Gamma}_0(3) \rtimes \mathbb{Z}_2$, where by definition stabilizer groups become non-trivial. 
We denote a stabilizer group $G$ on a point $x$ by $G_x$.
$\mathbb{Z}_{3, \omega}$ on $\omega$ and $\mathbb{Z}_{2, i}$ on $i$ are the stabilizer groups for $PSL(2, \mathbb{Z})$, and $\mathbb{Z}_{3, \omega'}$ on $\omega' \equiv \frac{1}{2} + i \frac{\sqrt{3}}{6}$ is for $\overline{\Gamma}_0(3)$. 
$\mathbb{Z}_{2, \frac{i}{\sqrt{3}}}$ is realized on $U = \frac{i}{\sqrt{3}}$, and it is the other elliptic point of $\overline{\Gamma}_0(3) \rtimes \mathbb{Z}_2$.
Note that $\omega$ and $\omega'$ are $SL(2, \mathbb{Z})$ equivalent as $\omega' =  S T^{-1} \omega$. In addition, $S = i$ is turned out not to be realized since $N_{\rm flux}$ must be an integer. 
Then, one find non-trivial $\mathbb{Z}_3$ orbits of three sets (fluxes, $S$, $U$) at $S = \omega$ or $U = \omega'$. 
We have to divide the number of solutions on $(S, U) =(\omega, \cdot)$, $(\cdot, \omega')$ by three. 
We have to divide $(\omega, \omega')$ by three, not nine. As mentioned, $\omega$ and $\omega'$ are $SL(2, \mathbb{Z})$ equivalent. 
Then, $\langle W \rangle$ on the solution at $(\omega, \omega')$ 
\begin{align}
    \langle W \rangle = A + B \omega + \omega' [C + D \omega] = 0
\end{align}
can be mapped into 
\begin{align}
    \langle W \rangle = A' + B' \omega' + \omega' [C + D \omega'] = 0.
\end{align}
There is a symmetry of $B' \leftrightarrow C'$. Under the element $g$ of the stabilizer group at $\omega'$: $g \in \Gamma_0(3)_{\omega'} \simeq \mathbb{Z}_3$, the fluxes must transform as 
\begin{align}
    \begin{pmatrix}
        A' \\  C'
    \end{pmatrix} \rightarrow \rho(g) \begin{pmatrix}
        A' \\  C'
    \end{pmatrix},\quad
    \begin{pmatrix}
        B' \\  D'
    \end{pmatrix} \rightarrow \rho(g) \begin{pmatrix}
        B' \\  D'
    \end{pmatrix},
\end{align}
where $\rho(g)$ is fixed in Eq. (\ref{eq:fluxesABCDtransformation-sl2z-cs}). Since $\rho(g)$ is also a representation of $SL(2, \mathbb{Z})_S$ and there exists $B' \leftrightarrow C'$ symmetry, we can embed the $\Gamma_0(3)_{\omega'}$ transformation in the $SL(2, \mathbb{Z})_{\omega'}$ transformation
\begin{align}
    \begin{pmatrix}
        A' \\  B'
    \end{pmatrix} \rightarrow \rho(g) \begin{pmatrix}
        A' \\  B'
    \end{pmatrix},\quad
    \begin{pmatrix}
        C' \\  D'
    \end{pmatrix} \rightarrow \rho(g) \begin{pmatrix}
        C' \\  D'
    \end{pmatrix}.
\end{align}
$\Gamma_0(3)_{\omega'} \simeq SL(2, \mathbb{Z})_{\omega'}$ implies they are in one-to-one correspondence. 
Thus, we divide them by three. Explicitly, one can show that $S'_{\Gamma_0(3)_U} = (T^{-1} S)_{SL(2, \mathbb{Z})_S}$ holds at $(\omega, \omega')$ up to sign. 
Similar symmetry happens when the superpotential is a polynomial of the moduli, and the VEVs of $S$ and $U$s are $SL(2, \mathbb{Z})$ equivalent. 

On the other hand, the elliptic point $\frac{i}{\sqrt{3}}$ is not equivalent to other elliptic points. 
Thus, the above phenomenon cannot occur. However, one can explicitly show that $\mathbb{Z}_{2, \frac{i}{\sqrt{3}}}$ and the flux sign-flip $(F_3, H_3) \rightarrow -(F_3, H_3)$ are equivalent on the elliptic point $U = \frac{i}{\sqrt{3}}$. Thus, the multiplicity of $U \rightarrow -\frac{1}{3U}$ on $U = \frac{i}{\sqrt{3}}$ is not two, but one since we have already modded out the theory by the sign-flip.

Let us end this section with a comment that $N_{\rm flux}$ (\ref{eq:Nfluxintegerdivisor}) can easily exceed the tadpole cancellation because of the quantization condition in Table \ref{tab:T-transformation}. Indeed, it implies $N_{\rm flux} \in  24 \mathbb{Z}$, and we found there are only solutions with $N_{\rm flux} \geq 48$\footnote{The solutions which we obtained with $N_{\rm flux} = 48$ does not contain the solutions with the same $N_{\rm flux}$ listed in \cite{Font:2004cy}, as those lead to AdS, not Minkowski, which we search in this paper. }. Since $N_{\rm flux} > 0$ is required by the ISD condition and $N_{\rm flux} = 0$ does not lead to viable solutions with ${\rm Im} S > 0$ and ${\rm Im}U > 0$, it will exceed the limit determined by the tadpole cancellation condition. Furthermore, for the large flux contribution, we have to take backreactions into account. The former may be tamed if one considers an F-theory uplifting while we simply postpone the latter to future work. In the present paper, we concentrate on studying the behavior of solutions of the F-term equation.

\section{The most general duality in reparametrizations of $U$}
\label{app:generalreparametrization}
The period matrix is transformed by the $Sp(4, \mathbb{Z})$ matrices (\ref{eq:period-under-sl2z}) under the $\Gamma_0(3)$ transformation of $U$. 
We also found that $U \rightarrow - \frac{1}{3U}$ corresponds to the symplectic transformation (\ref{eq:periodtrfscaling}). 
Thus, we need to classify all possible dualities, which can be interpreted as basis transformations of the symplectic basis. 

First, let us outline the condition for identifying reparametrizations of $U$ with the basis transformations. 
Obviously, the action is invariant under the basis transformation and the K\"{a}hler transformation. 
We can decompose pairs $(G_3, \Pi)$ into orbits under these transformations, $(G_3, \Pi) \rightarrow (M G_3, k M\Pi)$, where $k=k(U)$ is a coordinate-independent factor and $M \in Sp(4, \mathbb{Z})$. Without loss of generality, we can choose their representatives are ones with $\Pi$.
Then, two elements $(G_3, \Pi)$ and $(M G_3, \Pi)$ with $M \in Sp(4, \mathbb{Z})$ are in different classes from each other.
However, those two classes may be identified under a reparametrization of $U$, $U' = f(U)$. Indeed, the existence of $\Gamma_0(3)$ and $S_{(3)}: U \rightarrow - \frac{1}{3U}$ dualities suggests that
\begin{align}
    {}^\exists f: U' = f(U), \left( {}^\exists M \in Sp(4, \mathbb{Z}),  {}^\exists k = k(U) \in \mathbb{C} ~\text{s.t.}~ \Pi(U') = \Pi\left(f(U)\right) = k(U) M \Pi(U) \right).
    \label{eq:def-reparametrization-duality}
\end{align}
In that cases, the two orbit classes $[G_3, \Pi(U)]$ and $[M G_3, \Pi(U')]$ are identified under the reparametrization.
In terms of solutions for the F-term equations, sets of the solutions are mapped into each other by the function, which connects $U$ with $U'$.
Note that all elements of the $Sp(4, \mathbb{Z})$ may not appear, even if we consider arbitrary nontrivial $f$ in (\ref{eq:def-reparametrization-duality}). The statement is that, if there are two solutions with fluxes $G_3, G_3'$ being related by $G_3' = M G_3$, there may be a reparametrization $f$ behind those two solutions.

The allowed functional form of $f$ is linear fractional transformation. In general, the complex-structure modulus $U$ is defined as
\begin{align}
    U \propto \frac{\displaystyle\int_B \Omega}{\displaystyle\int_A \Omega}.
\end{align}
On the other hand, the basis transformation of the symplectic basis will act on the periods as
\begin{align}
    \begin{pmatrix}
        \int_A' \Omega \\
        \int_B' \Omega
    \end{pmatrix} = 
    M
    \begin{pmatrix}
        \int_A \Omega \\
        \int_B \Omega
    \end{pmatrix},
\end{align}
where $M \in Sp(4, \mathbb{Z})$. Thus $U$ is transformed as
\begin{align}
    U \rightarrow \frac{a U + b}{c U + d},
\end{align}
which is a linear fractional transformation with the integer coefficients\footnote{Without loss of generality, we can redefine real $a, b, c, d$ to make the correspondent matrix be element of $SL(2, \mathbb{R})$. This enables us to consider the transformation group, however, in the following discussion, we continue to consider them as integers to simplify the problem.}.

Then, all we need to do is to find constraints on $a, b, c, d$ by imposing $M$ as an element of $Sp(4, \mathbb{Z})$.
Since the linear fractional transformation is defined all over the modulus space, we can consider the constraints as identities of $U$. 

First, one can show that if we factorize $k$ as $k = (cU + d)^{-1} k'$, $k'$ is independent of $U$. For the period matrix $\Pi = (1 ~ -\sqrt{3}iU ~ 3U ~ \sqrt{3}i)^T$, $k \equiv k'$ satisfies 
\begin{align}
    \begin{pmatrix}
        c U + d \\ - i \sqrt{3} ( a U + b ) \\ 3 ( a U + b ) \\ i \sqrt{3} (c U + d)
    \end{pmatrix}
    = k 
    \begin{pmatrix}
    x_{11} + i \sqrt{3} x_{14} + U (- x_{12} i \sqrt{3} + 3 x_{13}) \\
    x_{21} + i \sqrt{3} x_{24} + U (- x_{22} i \sqrt{3} + 3 x_{23}) \\
    x_{31} + i \sqrt{3} x_{34} + U (- x_{32} i \sqrt{3} + 3 x_{33}) \\
    x_{41} + i \sqrt{3} x_{44} + U (- x_{42} i \sqrt{3} + 3 x_{43})
    \end{pmatrix},
    \label{eq:consistency-symplectictrf-reparametrization}
\end{align}
where $X \equiv (x_{ij})$ is chosen to be an element of $Sp(4, \mathbb{Z})$ without loss of generality, if $f(U)$ is a duality of the solutions. Since the left-hand side is linear in $U$, it is sufficient to consider $k \sim \frac{1}{U}$ or $k \sim U$ as the dependence of $k$ on $U$, if the reparametrization domain is the whole space. Then, it is respectively required that $\forall i, x_{i1} + i \sqrt{3} x_{i4} = 0$ and $-x_{i2} i\sqrt{3} + 3 x_{i3} = 0$ for each dependence. However, both conditions lead to $\det X = 0$, and it is a contradiction. Thus, $k$ is independent of $U$.

Then, Eq. (\ref{eq:consistency-symplectictrf-reparametrization}) is reduced to 
\begin{align}
\begin{alignedat}{2}
x_{14} &= \frac{1}{\sqrt{3}k} \left(-i (cU + d) + ik (x_{11} + 3 U x_{13}) + \sqrt{3}k U x_{12}\right), \\
x_{24} &= \frac{1}{3k} \left( -3(aU + b) + i k \sqrt{3} (x_{21} + U x_{23}) + 3k U x_{22} \right),\\
x_{34} &= \frac{1}{3} \left( 3(x_{21} + 3 U x_{23} + U x_{32}) + i \sqrt{3} (x_{31} - 3 U x_{22} + 3 x_{24} + 3 U x_{33}) \right),\\
x_{44} &= \frac{1}{\sqrt{3}} \left( \sqrt{3} (x_{11} + 3 U x_{13} + U x_{42}) + i (x_{41} + 3x_{14} - 3 U x_{12} + 3 U x_{43}) \right).
\end{alignedat}
\end{align}
As a set of identities which holds independent of $U$, the last two identities imply that $X$ takes the form of
\begin{align}
    X = 
    \begin{pmatrix}
      x_{44} & x_{43} & -\frac{x_{42}}{3} & -\frac{x_{41}}{3} \\
      x_{34} & x_{33} & -\frac{x_{32}}{3} & -\frac{x_{31}}{3} \\
      x_{31} & x_{32} & x_{33} & x_{34} \\
      x_{41} & x_{42} & x_{43} & x_{44}
    \end{pmatrix},
\end{align}
while the first two identities impose that
\begin{align}
    \begin{alignedat}{2}
        -\frac{\sqrt{3}x_{12}}{c}  &= \frac{\sqrt{3}x_{14}}{d} = \frac{x_{21}}{\sqrt{3}b}  = \frac{\sqrt{3}x_{23}}{a} = {\rm Im}\frac{1}{k}, \\
        \frac{3x_{13}}{c} &= \frac{x_{11}}{d} = - \frac{x_{24}}{b} = \frac{x_{22}}{a} = {\rm Re}\frac{1}{k}.
    \end{alignedat}
\end{align}
Here, $a, b, c, d$ are assumed to be nonzero. If a denominator vanishes, then $x_{ij}$ in the corresponding numerator must vanish.
\paragraph{Case I. $a, b, c, d \neq 0$} \mbox{}\\
For non-vanishing $a, b, c, d$, $X$ is given by
\begin{align}
    X = 
    \begin{pmatrix}
        \frac{d}{a} x_{33} & \frac{c}{3a} x_{32} & \frac{c}{3a} x_{33} & - \frac{d}{3a}x_{32} \\
        - \frac{b}{a}x_{32} & x_{33} & - \frac{x_{32}}{3} & - \frac{b}{a} x_{33} \\
        \frac{3b}{a} x_{33} & x_{32} & x_{33} & - \frac{b}{a} x_{32} \\
        \frac{d}{a} x_{32} & - \frac{c}{a} x_{33} & \frac{c}{3a} x_{32} & \frac{d}{a} x_{33}
    \end{pmatrix}.
\end{align}
$X$ should be an integral matrix. 
The condition $X^T \Sigma X = X$ is reduced to 
\begin{align}
    (ad-bc) \frac{x_{32}^2 + 3 x_{33}^2}{3a^2} = 1,
    \label{eq:nonzero-abcd-symplectic-condition}
\end{align}
which implies $ad - bc \geq 1$. 

If $ad - bc = 1$, then the condition $X$ is an integral matrix leads $x_{32} = \frac{3a}{c} m, x_{33} = \frac{3a}{c}n ~ (m, n \in \mathbb{Z})$ and 
\begin{align}
    \frac{3 (m^2 + 3n^2)}{c^2} = 1,
\end{align}
which implies $c \in 3\mathbb{Z}$ restricts $f(U)$ to be an element of $\Gamma_0(3)$. Since it is explicitly checked that $\Gamma_0(3)$ is consistent with the symplectic transformation in Eq. (\ref{eq:period-under-sl2z}), the whole $\Gamma_0(3)$ corresponds to this case.

If $ad - bc > 1$, by the similar way, we obtain
\begin{align}
    \frac{3(m^2 + 3n^2)}{c^2} = \frac{1}{ad - bc} < 1.
\end{align}
This also implies $c \equiv 0 ~ ({\rm mod} 3)$.
Next, assume that $x_{33} \neq 0$. Let us modify (\ref{eq:nonzero-abcd-symplectic-condition}) as
\begin{align}
    x_{32}^2 + 3x_{33}^2 
        &= \frac{3a^2}{ad - bc} \nonumber \\
        &= \frac{3}{\frac{d}{a} - (-\frac{b}{a}) 3(-\frac{c}{3a})} \nonumber \\
        &= \displaystyle\frac{3 x_{33}^2}{\left(\frac{d}{a} x_{33}\right) x_{33} - \left(- \frac{b}{a} x_{33}\right)3\left(- \frac{c}{3a} x_{33}\right)} ~ (x_{33} \neq 0).
\end{align}
Since $X$ must be an integral matrix, $e = \frac{d}{a}x_{33}, f = - \frac{b}{a}x_{33}, g = -\frac{c}{3a}x_{33}$ are all integers. Then, the above equation implies
\begin{align}
    e x_{33} - 3 f g = \frac{3 x_{33}^2}{x_{32}^2 + 3 x_{33}^2}
\end{align}
must be an integer. Hence it leads $x_{32}^2 + 3 x_{33}^2 \leq 3 x_{33}^2$, and $x_{32} = 0$. By a similar calculation, we have $x_{33} = 0$ when $x_{32} \neq 0$ is assumed. Therefore $ad - bc > 1$ case is split into two cases; (i) $a d - b c > 1$ and $x_{33}\neq 0, x_{32} = 0$ and (ii) $a d - b c > 1$ and $x_{32}\neq 0, x_{33} = 0$.

\subparagraph{Case I-i. $a d - b c > 1$ and $x_{33}\neq 0, x_{32} = 0$} \mbox{}\\
In this case, $X$ becomes
\begin{align}
    X = x_{33}
    \begin{pmatrix}
        \frac{d}{a} & 0& \frac{c'}{a} & 0 \\
        0 & 1 & 0 & - \frac{b}{a} \\
        \frac{3b}{a} & 0 & 1 & 0 \\
        0 & - \frac{3c'}{a} & 0 & \frac{d}{a}
    \end{pmatrix},
\end{align}
and $X^T \Sigma X = \Sigma$ leads
\begin{align}
    |x_{33}| = \frac{|a|}{\sqrt{a d - b c}},
\end{align}
where $c' = \frac{c}{3}$ is an integer. Since $X$ is an integral matrix, it is required that there exist $\tilde{a}, \tilde{b}, \tilde{c}, \tilde{d} \in \mathbb{Z}$ such that
\begin{align}
    \tilde{a} = \frac{a}{\sqrt{a d - b c}}, \tilde{b} = \frac{b}{\sqrt{a d - b c}}, \tilde{c}' = \frac{c'}{\sqrt{a d - b c}}, \tilde{d}' = \frac{d}{\sqrt{a d - b c}}.
\end{align}
It leads $\tilde{a} \tilde{d} - 3 \tilde{b} \tilde{c}' = 1$. If such  $\tilde{a}, \tilde{b}, \tilde{c}', \tilde{d} \in \mathbb{Z}$ exist, $\gcd (a, b, c, d) = \sqrt{a d - b c} \neq 1$ which indicates that this case is reduced to the $ad-bc=1$ case.

Since the existence of $\tilde{a}, \tilde{b}, \tilde{c}', \tilde{d} \in \mathbb{Z}$ is the necessary condition for $X \in Sp(4, \mathbb{Z})$ exists under the reparametrization $f(U)$, we can conclude that $x_{33} = 0$ leads a $\Gamma_0(3)$ transformation with $ad - bc = 1$.

\subparagraph{Case I-ii. $a d - b c > 1$ and $x_{32}\neq 0, x_{33} = 0$} \mbox{}\\
$X$ becomes
\begin{align}
    X = 
    \frac{x_{32}}{3}
    \begin{pmatrix}
        0 & \frac{c}{a} & 0 & -\frac{d}{a} \\
        -\frac{3b}{a} & 0 & - 1 & 0 \\
        0 & 3 & 0 & -\frac{3b}{a} \\
        \frac{3d}{a} & 0 & \frac{c}{a} & 0 
    \end{pmatrix},
\end{align}
and $X^T \Sigma X = \Sigma$ leads
\begin{align}
    |x_{32}| = \frac{\sqrt{3}|a|}{\sqrt{ad-bc}}.
\end{align}
Then, define $m_1 \equiv \frac{c}{3a}x_{32}, m_2 \equiv \frac{d}{3a}x_{32}, m_3 \equiv \frac{b}{a}x_{32}, m_4 = \frac{x_{32}}{3}$ with $m_1, m_2, m_3, m_4 \in \mathbb{Z}$. Taking into account $X^T \Sigma X = \Sigma$, we obtain
\begin{align}
    m_4 = \frac{1+ m_1 m_3}{3m_2}, b = \frac{a}{3m_4}, c = -  3(b m_1^2 - a m_1 m_2), d = - 3 (b m_1 m_2 - a m_2^2).
\end{align}
Further, $a^2 = 3 m_4^2 (a d - b c)$ also holds. Thus, we see $a, c, d$ must be multiples of three.
Hence we can define $a = 3a', c = 3c', d = 3d'$, and $X$ is given by
\begin{align}
    X = \frac{1}{\sqrt{3 a' d' - b c'}} 
    \begin{pmatrix}
        0 & c' & 0 & -d' \\
        -b & 0 & -a' & 0 \\
        0 & 3a' & 0 & -b \\
        3d' & 0 & c' & 0
    \end{pmatrix}.
\end{align}
$X$ should be an integral matrix, and there should exist $\tilde{a}', \tilde{b}, \tilde{c}', \tilde{d}' \in \mathbb{Z}$ such that
\begin{align}
    \tilde{a}' = \frac{a'}{\sqrt{3 a' d' - b c'}}, \tilde{b} = \frac{b}{\sqrt{3 a' d' - b c'}}, \tilde{c}' = \frac{c'}{\sqrt{3 a' d' - b c'}}, \tilde{d}' = \frac{d'}{\sqrt{3 a' d' - b c'}}.
\end{align}
It follows that $3 \tilde{a}' \tilde{d}' - \tilde{b}\tilde{c}' = 1$. If such $\tilde{a}', \tilde{b}, \tilde{c}', \tilde{d}' \in \mathbb{Z}$ exist, $X$ is just rewritten as 
\begin{align}
    X =
    \begin{pmatrix}
        0 & \tilde{c}' & 0 & -\tilde{d}' \\
        -\tilde{b} & 0 & - \tilde{a}' & 0 \\
        0 & 3 \tilde{a}' & 0 & -\tilde{b} \\
        3 \tilde{d} & 0 & \tilde{c}' & 0,
    \end{pmatrix}
    ,
\end{align}
which is a symplectic matrix with the integral coefficients.  
Recalling the symplectic matrices for the modular $\Gamma_0(3) \ni \begin{pmatrix}
    v & w \\
    3x & y
\end{pmatrix} ~ (v, w, x, y \in \mathbb{Z})$ and the scaling $U \rightarrow -\frac{1}{3U}$, consecutive transformations are given by
\begin{align}
    \begin{pmatrix}
        y & 0 & x & 0 \\
        0 & v & 0 & -w \\
        3w & 0 & v & 0 \\
        0 & -3x & 0 & y
    \end{pmatrix}
    \begin{pmatrix}
        0 & 1 & 0 & 0 \\
        1 & 0 & 0 & 0 \\
        0 & 0 & 0 & 1 \\
        0 & 0 & 1 & 0
    \end{pmatrix} = 
    \begin{pmatrix}
        0 & y & 0 & x \\
        v & 0 & -w & 0 \\
        0 & 3w & 0 & v\\
        -3x & 0 & d & 0
    \end{pmatrix},
\end{align}
or
\begin{align}
    \begin{pmatrix}
        0 & 1 & 0 & 0 \\
        1 & 0 & 0 & 0 \\
        0 & 0 & 0 & 1 \\
        0 & 0 & 1 & 0
    \end{pmatrix}
    \begin{pmatrix}
        y & 0 & x & 0 \\
        0 & v & 0 & -w \\
        3w & 0 & v & 0 \\
        0 & -3x & 0 & y
    \end{pmatrix} = 
    \begin{pmatrix}
        0 & v & 0 & -w \\
        y & 0 & x & 0 \\
        0 & -3x & 0 & y \\
        3w & 0 & v & 0
    \end{pmatrix}.
\end{align}
One can see these are quite similar to $X$. Indeed, for the first transformation, there exist
\begin{align}
    v = - \tilde{b}, w = \tilde{a}', x = -\tilde{d}', y = \tilde{c}'
\end{align}
which satisfies $vy - w(3x) = 1$. Thus, if $\tilde{a}', \tilde{b}, \tilde{c}', \tilde{d}' \in \mathbb{Z}$ exists, we see that the result $X$ is represented as a product of a $\Gamma_0(3)$ matrix and the scaling duality. 

Hence, the independent dualities are the modular $\Gamma_0(3)$ and the scaling $U \rightarrow - \frac{1}{3U}$.
The second transformation can also be related similarly. Hence the scaling duality is realized as an outer automorphism of $\Gamma_0(3)$. 
\paragraph{Case \greekii. $a, d = 0, b, c \neq 0$} \mbox{}\\
In this case, $X$ takes the form
\begin{align}
    X = \begin{pmatrix}
        0 & -\frac{c}{3b}x_{34} & \frac{c}{9b}x_{31} & 0 \\
        x_{34} & 0 & 0 & -\frac{x_{31}}{3} \\
        x_{31} & 0 & 0 & x_{34} \\
        0 & -\frac{c}{3b}x_{31} & -\frac{c}{3b}x_{34} & 0
    \end{pmatrix}.
\end{align}
One can see that $x_{31} = \frac{9b}{c}m, x_{34} = \frac{3b}{c}n, ~ (m, n \in \mathbb{Z})$. The condition $X^T \Sigma X  = \Sigma$ is reduced to 
\begin{align}
    -3b (3m^2 + n^2) = c,
\end{align}
thus $c \equiv 0 ~ (\rm{mod} 3)$ and $bc < 0$. Defining $\frac{3b}{c} = q^{-1}$, $q = - (3m^2 + n^2) \in - \mathbb{N}$. 
It leads $x_{31} = \frac{3m}{q}, x_{34} = \frac{n}{q}$. Taking into account $x_{31} \in 3 \mathbb{Z}$ and $x_{34} \in \mathbb{Z}$, $m = 0, n = 1$ is only possible and $b = -\frac{c}{3}$. Thus, the reparametrization is the scaling duality, $f(U) = - \frac{1}{3U}$. It also implies that the $S$-transformation $U \rightarrow - \frac{1}{U}$ is not allowed.

\paragraph{Case I\hspace{-1.2pt}I\hspace{-1.2pt}I. $b, c = 0$ , $a, d  \neq 0$} \mbox{}\\
In this case, $X$ takes the form
\begin{align}
    X = 
    \begin{pmatrix}
        \frac{d}{a} x_{33} & 0 & 0 & - \frac{d}{3a} x_{32} \\
        0 & x_{33} & - \frac{x_{32}}{3} & 0 \\
        0 & x_{32} & x_{33} & 0 \\
        \frac{d}{a} x_{32} & 0 & 0 & \frac{d}{a} x_{33}
    \end{pmatrix}.
\end{align}
Defining $x_{32} = -\frac{3a}{d} m$ and $x_{33} = \frac{a}{d}n$, $X^T \Sigma X = \Sigma$ reduces to 
\begin{align}
    \frac{a}{d} (3m^2 + n^2) = 1.
\end{align}
Defining $\frac{a}{d} = q^{-1}$, we find $q = 3m^2 + n^2 \in \mathbb{N}$. Similar to the previous case, taking into account $x_{32} \in 3\mathbb{Z}$ and $x_{33} \in \mathbb{Z}$ gives $m =0, n=1$. Thus we obtain the trivial reparametrization $U \rightarrow U$, and $X = \mathbf{1}_{4\times4}$. 

\paragraph{Case I\hspace{-1.2pt}V. $a = 0$, $b, c, d \neq 0$} \mbox{}\\
\begin{align}
X =
\begin{pmatrix}
\frac{d}{3b}x_{31} & \frac{c}{3d} x_{41} & -\frac{c}{9b}x_{31} & -\frac{x_{41}}{3} \\
-\frac{b}{d}x_{41} & 0 & 0 & -\frac{x_{31}}{3} \\
x_{31} & 0 & 0 & -\frac{b}{d} x_{41} \\
x_{41} & \frac{c}{3b} x_{31} & \frac{c}{3d} x_{41} & \frac{d}{3b}x_{31}
\end{pmatrix}.
\end{align}
The condition $X^T \Sigma X = \Sigma$ reduces to
\begin{align}
\begin{alignedat}{2}
     x_{31}x_{41} &= 0, \\
    \frac{c}{9b} x_{31}^2 - \frac{bc}{3d^2} x_{41}^2 &= 1.
\end{alignedat}
\end{align}
Thus, either $x_{31} = 0$ or $x_{41} = 0$. If we choose $x_{41} = 0$, it can be represented as a multiplication of two integers as
\begin{align}
    \frac{c}{9b}x_{31}^2 = \left(-\frac{c}{3b}x_{31}\right) \left(-\frac{x_{31}}{3}\right) = 1.
\end{align}
However, since $\frac{c}{3b} x_{31}$ must be an integer, this cannot be satisfied. Hence, $x_{31} = 0$ is only allowed. 
\begin{align}
    -\frac{b c}{3d^2}x_{41}^2 = \left(\frac{c}{3d}x_{41}\right)\left(-\frac{b}{d}x_{41}\right) = 1
\end{align}
implies $b = - \frac{c}{3}$. Furthermore, $d \in 3\mathbb{Z}$ because of $x_{41} \in 3\mathbb{Z}$ and $d^2 = b^2 x_{41}^2$. Thus, the reparametrization is restricted to $U \rightarrow -\frac{m}{3(m U + n )}$, where $c = 3m$ and $d = 3n$, $m, n \in \mathbb{Z}$.
We further obtain $\frac{n}{m} \in \mathbb{Z}$, since $(-\frac{x_{41}}{3})^2 = \frac{d^2}{c^2} \in \mathbb{Z}$. Then, the reparametrization is reduced to 
\begin{align}
    U \rightarrow U = - \frac{1}{3(U + q)} \quad q \in \mathbb{Z},
\end{align}
and it is a composition of $T$-transformation and the scaling duality $U \rightarrow - \frac{1}{3U}$.

\paragraph{Case V. $b=0$, $a, c, d \neq 0$} \mbox{}\\
\begin{align}
    X = 
\begin{pmatrix}
    \frac{d}{a} x_{33} & \frac{c}{3a} x_{32} & \frac{c}{3a} x_{33} & - \frac{d}{3a} x_{32} \\
    0 & x_{33} & - \frac{x_{32}}{3} & 0 \\
    0 & x_{32} & x_{33} & 0 \\
    \frac{d}{a} x_{32} & - \frac{c}{a}x_{33} & \frac{c}{3a} x_{32} & \frac{d}{a} x_{33}
\end{pmatrix},
\end{align}
and $X^T \Sigma X = \Sigma$ gives 
\begin{align}
    U \rightarrow \frac{U}{3q U + 1} \quad q \in \mathbb{Z},
\end{align}
in a similar way to the case I\hspace{-1.2pt}I\hspace{-1.2pt}I. Thus, $f(U)$ is an element of the modular duality $\Gamma_0(3)$. 

\paragraph{Case V\hspace{-1.2pt}I. $c=0$, $a, b, d \neq 0$} \mbox{}\\
Similarly, one can obtain $f(U) = U + q, ~ q \in \mathbb{Z}$. Thus, $f(U)$ is the usual $T$-transformation.

\paragraph{Case V\hspace{-1.2pt}I\hspace{-1.2pt}I. $d=0$, $a, b, c \neq 0$} \mbox{}\\
The condition implies $f(U) = \frac{3q U - 1}{3q} = - \frac{1}{3U} + q, ~ q \in \mathbb{Z}$. Thus, $f(U)$ is a composition of the scaling duality $U \rightarrow - \frac{1}{3U}$ and the $T$-transformation $U \rightarrow U + q$.

\acknowledgments

This work was supported by JSPS KAKENHI Grant Numbers JP20K14477 (H. O.), JP22J12877 (K. I.), JP23H04512 (H. O.) and JP23K03375 (T. Kobayashi). We would like to thank T. Oikawa for useful comments.

\bibliography{references}{}
\bibliographystyle{JHEP} 

\end{document}